\documentclass[10pt]{iopart}
\usepackage{graphicx,iopams,upgreek}
\usepackage{color,nameref}
\renewcommand{\etal}[0]{\textit{et al.}~}
\usepackage{cite}
\begin{document}
\topical[Emergent behavior in active colloids]{Emergent behavior in active colloids}
\author{Andreas Z\"{o}ttl$^{1,2}$ and Holger Stark$^2$}
\address{$^1$ The Rudolf Peierls Centre for Theoretical Physics, University of Oxford,\newline   1 Keble Road, OX1 3NP, United Kingdom}
\address{$^2$ Institut f\"{u}r Theoretische Physik, Technische Universit\"{a}t Berlin, Hardenbergstrasse 36, 10623 Berlin, Germany}
\ead{andreas.zoettl@physics.ox.ac.uk}

\begin{abstract}
Active colloids are microscopic particles, which self-propel through viscous fluids
by converting energy extracted from their environment into directed motion.
We first explain how
articial microswimmers 
move forward
by 
generating
near-surface 
flow fields
via 
self-phoresis
or 
the
self-induced
Marangoni 
effect.
We 
then
discuss 
generic  features of the dynamics of 
single
active colloids
in bulk and in confinement, as well as in the presence of 
gravity, field gradients, and fluid flow.
In the third part, we
review the emergent 
collective
behavior of 
active colloidal suspensions
focussing on their structural and 
dynamic
properties.
After summarizing
experimental observations, we 
give an overview 
on
the progress in modeling collectively moving active colloids.
While 
active Brownian 
particles
are heavily used to study 
collective 
dynamics on large scales,
more advanced methods 
are necessary
to explore the importance of hydrodynamic and phoretic 
particle
interactions.
Finally, the relevant physical 
approaches
to quantify the emergent collective behavior 
are
presented.
\end{abstract}
\submitto{\JPCM}
\maketitle
\ioptwocol

\tableofcontents

\vspace{2ex}

\section{Introduction}
\label{sec:intro}

Active particles are self-driven units, which are able 
to
move autonomously,  \emph{i.e.}, in the absence of external forces 
and torques, by converting energy into directed motion \cite{Vicsek2012,Schweitzer2003,Romanczuk2012}.
In nature many living organisms are active, ranging from mammals at the 
macro-scale
to bacteria at the 
micro-scale.

Inspired by nature, 
chemists, physicists, and engineers
have started to create and study the motion of artificial nano- and microswimmers \cite{Kapral2013,Poon2013a}.
The first experimental realizations,
constructed only about ten years ago \cite{Paxton2004b,Fournier-Bidoz2005b},
were
actively moving and spinning 
bimetallic
nanorods,
driven by a catalytic reaction on one of the two metal surfaces. 
Since then, many different experimental realizations of 
\textit{nano-}
and \textit{micromachines} have 
been investigated.
In particular,  spherical active colloids, due
to their simple shape, 
are useful to study novel physical phenomena,
which one expects from the
intrinsic nonequilibrium nature
of autonomous swimmers.
Indeed, 
in experiments they reveal interesting
emergent collective properties \cite{Thutupalli2011,Theurkauff2012,Palacci2013,Bricard2013,Buttinoni2013e,Nishiguchi2015}.

The locomotion of microswimmers is governed by low Reynolds number hydrodynamics and thermal noise.
Biological microswimmers  have to perform periodic, non-reciprocal body deformations 
or wave long thin appendages
in order to swim \cite{Purcell1977}.
In contrast, active colloids
and emulsion droplets
are able to propel themselves through a viscous medium 
by 
creating fluid flow
close to their surfaces
via self-phoresis or 
the
self-induced Marangoni 
effect.
Hence, their 
swimming
direction 
is not determined by external forces, such as in driven nonequilibrium systems,
but is an intrinsic property of the 
individual swimmers.
Simple models of active colloids are active Brownian particles, which move with a 
constant
speed $v_0$ along 
an intrinsic direction
$\mathbf{e}(t)$,
which varies in time, 
\emph{e.g.}, due to rotational Brownian motion.
More detailed models for self-phoretic active colloids are not only able to 
evaluate
$v_0$ from the local fluid-colloid 
interaction based on low Reynolds number hydrodynamics, but also
determine
the surrounding fluid 
flow and chemical concentration fields.
The cooperative motion of self-propelled active colloids, which interact via 
hydrodynamic, phoretic, electrostatic, and other 
forces,
results in various emergent collective behavior, which we discuss in this Topical Review.

Several reviews on self-propelled particles, microswimmers, and active colloids already exist, many of  them with different foci.
A general introduction to collective motion can be found in Ref.\ \cite{Vicsek2012}.
The dynamics of active Brownian particles under various conditions 
is extensively
reviewed in Refs.\ \cite{Schweitzer2003} and \cite{Romanczuk2012}.
Several
reviews on 
the
basic 
fluid mechanical principles of swimming at low Reynolds number are available
\cite{Lauga2009a,Ishikawa2009,Guasto2012,Yeomans2014,Elgeti2015b,Pak2015}.
How interfacial forces 
drive
phoretic motion of particles 
is
discussed and reviewed in Refs.~\cite{Anderson1989,Juelicher2009,Kapral2013,Poon2013a}.
More general 
survey articles
about the physics 
of
active colloidal systems 
are
found in \cite{Aranson2013a,Poon2013a}.
Articles 
on the
fabrication
of active colloids, the involved
chemical processes, 
or their
control  and technical applications 
exist
\cite{Ozin2005,Paxton2006,Hong2010d,Ebbens2010a,Mirkovic2010,Sengupta2012,Wang2013a,Abdelmohsen2014,Guix2014,Wang2015,Wang2015a}.
Nonequilibrium and thermodynamic properties, as well as motility-induced phase separation and active clustering 
are
discussed in Refs.\ \cite{Cates2012a,Cates2015,Bialke2014,Marchetti2015}.
We also 
mention
reviews on continuum modeling of microswimmers \cite{Koch2011,Saintillan2013} or of
generic active fluids  including flocks and active gels \cite{Ramaswamy2010,Marchetti2013}.

Here we review the emergent behavior in active colloidal systems by focussing on the basic physical concepts.
In section~\ref{Sec:HowSwim} we explain how active col\-loids are able to move autonomously through a fluid.
We explain how the basic fluid mechanics at low Reynolds number works (section~\ref{Sec:LowRe}),
how near-sur\-face flows (sec\-tion~\ref{Sec:Propulsion}) or the presence of sur\-faces (section~\ref{Sec:RollSurf}) 
ini\-ti\-ate
self-propulsion, and in\-tro\-duce the concept of active Brownian particles (sec\-tion~\ref{Sec:ABP}).
In section~\ref{Sec:Single} 
we discuss generic features of active particles. We
present the motion of a sin\-gle particle  in the presence of noise (section~\ref{Sec:Walk}), gra\-vity (section~\ref{Sec:Gravity}),
taxis (section~\ref{Sec:Taxis}), external fluid flow (section~\ref{Sec:FluidFlow}), near surfaces (section~\ref{Sec:Surface}), and
in complex environments (sec\-tion~\ref{Sec:Complex}).
The col\-lec\-tive dynamics of many in\-ter\-ac\-ting active colloids is discussed in sec\-tion~\ref{Sec:Collective}.
We 
first
review
ex\-peri\-men\-tal observations in sec\-tion~\ref{Sec:Collective:Exp},
and then 
address
the progress in modeling  
systems of
interacting active 
colloids
in section~\ref{Sec:Collective:Simu}.
General physical con\-cepts 
for quantifying
emergent 
collective
behavior of 
ac\-tive particle suspensions are 
presented
in sec\-tion~\ref{Sec:Collective:Phys}.
Finally, we give an outlook in section~\ref{Sec:Outlook}.


\section{How do active colloids swim}
\label{Sec:HowSwim}
\textit{Passive} microscopic particles, such as colloids immersed in a fluid,
move when external forces are applied.
These can either be external body forces,
for example, due to
gravity, or surface forces 
induced by physical or chemical gradients,
which 
then initiate
phoretic particle transport  \cite{Anderson1989}.
For example, the directed motion of colloids  along  temperature gradients $\nabla T$,
chemical gradients $\nabla c$, or 
electric
potential gradients $\nabla \zeta$,
is named \textit{thermophoresis},  \textit{diffusiophoresis}, and \textit{electrophoresis}, respectively \cite{Anderson1989}.
The velocity of the colloids can be calculated 
using
low Reynolds number hydrodynamics.

\textit{Active} colloids are able to move in a fluid in the absence of external forces.
Typically, they create  field gradients 
by themselves
localized around their body
when they consume
fuel \cite{Paxton2004b,Howse2007a} or 
are heated by
laser light \cite{Jiang2010b,Volpe2011}.
This initiates
\textit{self-phoretic}
motion \cite{Golestanian2005a,Golestanian2007a}.
While passive particles move along
an externally set 
field 
gradient,
active colloids 
change the direction of the 
self-generated gradient
and 
thereby
their propulsion direction,
when they experience, for example,
rotational thermal noise \cite{Howse2007a,Golestanian2009a,Jiang2010b,Volpe2011}.

Active emulsion droplets
start to move when gradients in surface tension form at the fluid-fluid interface
by spontaneous symmetry breaking 
\cite{DeBuyl2013a,Schmitt2013,Michelin2013,Schmitt2016}.
They then drive Marangoni flow and thereby propel the droplet
 \cite{Sumino2005,Hanczyc2007,Toyota2009,Thutupalli2011,Herminghaus2014}.

In the following we 
present some basic physical principles of low Reynolds number flow and how
active colloids are able to move in Newtonian fluids.

\subsection{Hydrodynamics at low Reynolds number: Stokes equations and fundamental solutions}
\label{Sec:LowRe}
The motion of particles immersed in an incompressible Newtonian fluid with viscosity $\eta$ and density $\rho$ is governed by 
the Navier Stokes equations \cite{Landau1986}.
At the micron scale inertial effects can be 
neglected
and the Navier Stokes equations for the flow field $\mathbf{v}(\mathbf{r},t)$ 
and pressure field $p(\mathbf{r},t)$ simplify to the Stokes equations \cite{Happel2012},
\begin{equation}
-\boldsymbol{\nabla} p + \eta \nabla^2 \mathbf{v} =   - \mathbf{f}  \enspace , \quad \boldsymbol{\nabla}\cdot  \mathbf{v}  = 0, 
\label{Eq:StokesE}
\end{equation}
where $\mathbf{f}(\mathbf{r},t)$ are 
body
forces acting on the fluid.
Together with appropriate boundary conditions 
at
the surface of the particle, 
equations~(\ref{Eq:StokesE})
are
solved for $\mathbf{v}(\mathbf{r},t)$  and $p(\mathbf{r},t)$,
and one can immediately determine the stress tensor  \cite{Landau1986}
\begin{equation}
\boldsymbol{\upsigma} = -  p \textbf{1}
 + \eta [ \boldsymbol{\nabla} 
 \otimes
  \mathbf{u} + ( \boldsymbol{\nabla} 
  \otimes \mathbf{u})^t  ].
%
%
\label{Eq:stress}
\end{equation}
The hydrodynamic force $\mathbf{F}(t)$  and torque  $\mathbf{M}(t)$ acting on a body 
immersed
in a Newtonian fluid 
are then calculated by integrating the stress tensor along the body
surface $S$, 
\begin{eqnarray}
\label{Eq:FMstress1}
  \mathbf{F}(t) &= \int_S  \boldsymbol{\upsigma}(\mathbf{r},t)\mathbf{n} \,\mathrm{d}S,  \\
\label{Eq:FMstress2}
  \mathbf{M}(t) &= \int_S  \mathbf{r} \times (\boldsymbol{\upsigma}(\mathbf{r},t)\mathbf{n})\,\mathrm{d}S.
\end{eqnarray}
Here, $\mathbf{n}$ is 
a unit vector normal to the surface 
and
pointing into the body.

\begin{figure}
\includegraphics[width=\columnwidth]{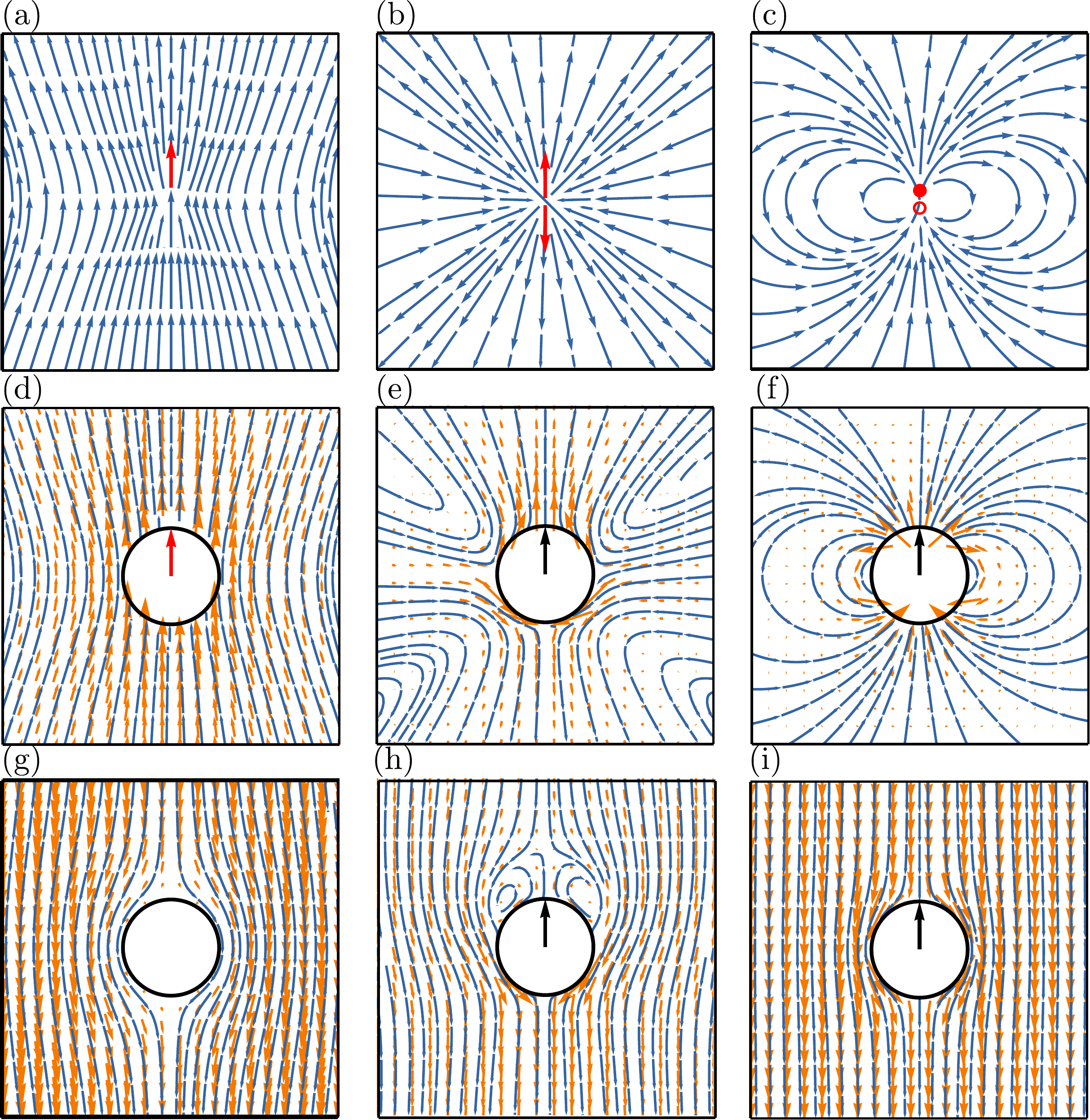}
\caption{
Typical
velocity fields (orange)
and stream lines (blue) of 
colloidal objects that behave like
a stokeslet (a,d,g), 
a force dipole (b,e,h), and 
a source dipole (c,f,i)
in the far field.
The first line shows the pure 
flow singularities. 
The second line shows the flow fields around spherical particles:
(d) driven by an external force, (e) a pusher squirmer with $\beta = -3$, and 
(f) a neutral squirmer with $\beta = 0$.
The third line shows the flow fields in the frame of the moving particles.
The red vectors indicate forces and the black vectors swimming
velocities. The  source dipole is indicated by a filled circle (source) and an open circle (sink).
Two-dimensional slices of the three-dimensional flow fields
are shown.
}
\label{Fig:flow}
\end{figure}

Equations~(\ref{Eq:StokesE}) are linear
in $\mathbf{v}(\mathbf{r},t)$ and $ p(\mathbf{r},t)$
and it is possible to 
solve for 
both fields using
Green's functions 
acting on the inhomogeneity
$\mathbf{f}(\mathbf{r},t)$
in
the Stokes equations.
The
unique
solution is 
formally 
written as  \cite{Dhont1996}  
\begin{eqnarray}
\label{Eq:Green1}
  \mathbf{v}(\mathbf{r},t) &=  \int \textbf{\textsf{O}}(\mathbf{r}-\mathbf{r}') \, \mathbf{f}(\mathbf{r}',t) \,\mathrm{d}^3x',    \\
\label{Eq:Green2}
  p(\mathbf{r},t) &= \int \mathbf{g}(\mathbf{r}-\mathbf{r}') \cdot \mathbf{f}(\mathbf{r}',t) \,\mathrm{d}^3x',
\end{eqnarray}
%
%
where 
$\textbf{\textsf{O}}(\mathbf{r}-\mathbf{r}')$ and $\mathbf{g}(\mathbf{r}-\mathbf{r}')$
are called \textit{Oseen tensor} and \textit{pressure vector}, respectively.
In 
three dimensions
they read \cite{Dhont1996},   
\begin{eqnarray}
\textbf{\textsf{O}}(\mathbf{r}) &= \frac{1}{8\pi\eta}\left( \frac{1}{|\mathbf{r}|}\mathbf{1}
 + \frac{\mathbf{r}\otimes\mathbf{r}}{|\mathbf{r}|^3}   \right),  \\
\mathbf{g}(\mathbf{r}) &= \frac{1}{4\pi}\frac{\mathbf{r}}{|\mathbf{r}|^3}.
\label{Eq:Oseen}
\end{eqnarray}

Now, consider
a static point force
or \textit{force monopole} 
$\mathbf{f}= f\mathbf{e} \delta(\mathbf{r}-\mathbf{r}_0)$
located at position $\mathbf{r}_0$ with strength $f$ and directed along $\mathbf{e}$
in an unbounded fluid, where $\delta(\cdots)$ is the Dirac $\delta$-function.
Using 
equation~(\ref{Eq:Green1}) the re\-sul\-ting flow field, a \textit{stokeslet,} then simply is
\begin{equation}
 \mathbf{v}_S(\mathbf{r}) 
 =  \frac{f}{8\pi\eta r}[
  \mathbf{e}
 + (\hat{\mathbf{r}} \cdot \mathbf{e}) \hat{\mathbf{r}}
] .
\label{Eq:Stokeslet}
\end{equation}
It
decays as $r^{-1}$ with  $r = |\mathbf{r}-\mathbf{r}_0|$ and
$\hat{\mathbf{r}} = (\mathbf{r}-\mathbf{r}_0)/r$ is the radial unit vector. 
The streamlines around a stokeslet are illustrated in figure~\ref{Fig:flow}(a).
The stokeslet 
is the fundamental solution of 
the
Stokes equations. 
It describes the 
flow field
far away from the particle,
when it is forced from outside, for example, by gravity.
Similar as in electrostatics,
one can 
construct 
solutions of the Stokes equations of
higher order in $1/r$
by a multipole expansion of the flow field.
The contributions
are called \textit{force dipole} $ \sim r^{-2}$, \textit{force quadrupole}
 $ \sim r^{-3}$, and so on \cite{Pozrikidis1992,Kim2013,Spagnolie2012,Mathijssen2015}.
The force dipole consists of two point forces, 
$\mathbf{f}=
f\mathbf{e} \delta(\mathbf{r}-\mathbf{r}_0-(l/2)\mathbf{e})$ and 
$ -\mathbf{f} = - f\mathbf{e} \delta(\mathbf{r}-\mathbf{r}_0+(l/2)\mathbf{e})$,
separated by 
a
distance $l$.
At 
$r \gg l$ or in the limit $l \rightarrow 0$ the force dipole flow field reads
\begin{equation}
\mathbf{v}_D(\mathbf{r}) 
 = \frac{p}{8\pi\eta r^2} \left[ -1+3(\mathbf{e}\cdot\hat{\mathbf{r}})^2  \right]\hat{\mathbf{r}},
  \label{Eq:Dipole} 
\end{equation}
where 
$p=fl$
is the 
strength of the
force dipole.
For $p \propto f >0$, we plot the flow field
in figure~\ref{Fig:flow}(b).
Since the two point forces 
point
outwards, the flow field is called \textit{extensile}.
In contrast, 
two point forces pointing towards each other ($p<0$) initiate a
\textit{contractile} 
flow field and the field lines of figure~\ref{Fig:flow}(b) are simply reversed.
The flow fields of microswimmers are often dominated by force dipoles.
Microswimmers
with 
extensile flow 
fields
($p>0$) are called \textit{pushers}, since they
push fluid outwards along their body axis.
Those
with 
contractile flow 
fields
($p<0$) are called \textit{pullers}, 
since
they pull fluid inwards along their body axis \cite{Lauga2009a}.

In addition to force singularities also source 
singularities
exist. Since they solve the Stokes equations for constant pressure, which gives the 
Laplace equation, they are potential flow solutions. Combinations of
sources and sinks in the fluid 
are named
 \textit{source monopole} $ \sim r^{-2}$,  \textit{source dipole}  $\sim r^{-3}$,
 \textit{source quadrupole}  $ \sim r^{-4}$, and so on.
In figure~\ref{Fig:flow}(c) we show the source-dipole 
flow
field,
\begin{equation}
\mathbf{v}_{SD}(\mathbf{r}) 
 = \frac{q}{8\pi\eta r^3} \left[ -\mathbf{e} + 3(\mathbf{e}\cdot\hat{\mathbf{r}})\hat{\mathbf{r}}  \right],
  \label{Eq:SDipole} 
\end{equation}
where $q$ is the source-dipole strength.
Higher-order solutions can be constructed
by combining lower multipoles.
A general flow field
solving the Stokes equations
can be expressed as a sum of all relevant force and source singularities \cite{Pozrikidis1992,Kim2013,Spagnolie2012,Mathijssen2015}.

The flow singularities describe the far field of colloids moving in a Newtonian fluid.
To fulfill the no-slip boundary condition at the surface of 
a colloid,
they have to be combined.
For example, the exact flow field for a
sphere of radius $R$ 
sedimenting
at velocity $\mathbf{u}$
is a combination of 
the
stokeslet and source-dipole flow field.
In 
the laboratory frame,
where the sphere 
moves with velocity $\mathbf{u}$,
the flow field reads \cite{Dhont1996}
\begin{equation}
\mathbf{v}(\mathbf{r}) =   \textbf{\textsf{S}}(\mathbf{r} - \mathbf{r}_0)\mathbf{u}
\label{Eq:Sphere1}
\end{equation}
with
\begin{equation}
\textbf{\textsf{S}}(\mathbf{r}) = \frac{3}{4} \frac{R}{r} \left( \mathbf{1}
 + \hat{\mathbf{r}}\otimes\hat{\mathbf{r}}     \right)  + \frac 1 4  \frac{R^3}{r^3}  \left( \mathbf{1}
 - 3 \hat{\mathbf{r}}\otimes\hat{\mathbf{r}}     \right).
\end{equation}
It
is shown in figure~\ref{Fig:flow}(d).
Figure~\ref{Fig:flow}(g) illustrates the
flow field around a fixed sphere in a uniform background flow 
$\mathbf{v}(\mathbf{r}) = \mathbf{u}$.
It is the same as
in equa\-tion~(\ref{Eq:Sphere1})
but with a constant $\mathbf{u}$ added.

A colloid surrounded by a field gradient moves since some phoretic mechanism establishes a slip-velocity field at its surface
\cite{Anderson1989}. The resulting
flow field is often that of a source dipole
given in equation~(\ref{Eq:SDipole}) \cite{Anderson1989}.
Also, for active emulsion droplets and Janus colloids the source dipole field is dominant \cite{Thutupalli2011}, or at least present \cite{Spagnolie2012,Bickel2013}.
Interestingly,
the flow field of equation~(\ref{Eq:SDipole}),
taken in the frame of a colloid moving with velocity $\mathbf{u} = 2q/(8\pi \eta R^3) \mathbf{e}$,
is not only valid in the far field but often agrees with the
slip-velocity field at the particle surface
and thereby 
also
determines the hydrodynamic near field \cite{Anderson1989}.
In figures~\ref{Fig:flow}(f,i) we show the
source-dipole
flow field 
of a
colloid,
initiated by some
some phoretic 
mechanism,
either in the lab frame 
 [see equation~(\ref{Eq:SDipole})] 
or
in the co-moving particle frame, 
where the colloid velocity $\mathbf{u}$ has been subtracted.

We will discuss the flow fields around force-dipole swimmers 
illustrated
in figures~\ref{Fig:flow}(e,h) in section~\ref{Sec:Prescribed}.

\subsection{Propulsion by near-surface flows: self-phoresis and Marangoni propulsion}
\label{Sec:Propulsion}

Biological microswimmers typically have to perform a non-reciprocal deformation of their cell body
in order
to swim \cite{Purcell1977,Lauga2009a,Elgeti2015b}.
In contrast, active colloids are able to move autonomously without
changing their shape periodically.
Instead, they create
tangential fluid flow near their surfaces.
Different
mechanisms for 
initiating such a self-phoretic
motion exist but 
still
the details 
have not been fully
understood
yet
(see, for example, the discussions in \cite{Brown2014}).

One mechanism for self-propulsion is the following: Janus particles have two 
distinct
faces.
Typically, one of them catalyzes a reaction of molecules in the surrounding fluid.
Since reactants and products interact differently with the particle surface, a pressure gradient along the
surface is created, which drives fluid flow,
as we will discuss in  the following.
Janus colloids can 
have
different shapes. The most common examples are half-coated spheres 
\cite{Golestanian2005a,Golestanian2007a,Howse2007a,Palacci2010,Jiang2010b,Ke2010,Volpe2011,Ebbens2011,Baraban2012a,Gao2012,Zheng2013a},
bimetallic rods \cite{Paxton2004b,Fournier-Bidoz2005b,Wang2006,Sundararajan2008},
or dimers, which consist of two linked spheres with different 
chemical properties of their surfaces
\cite{Rueckner2007,Valadares2010}.

In contrast, also
colloidal particles with initially uniform surface properties
are sometimes able to move autonomously by 
spontaneous
symmetry breaking 
\cite{DeBuyl2013a,Schmitt2013,Michelin2013,Schmitt2016}.
Prominent examples are active emulsion droplets, where Marangoni stresses at the surface
drive
a slip velocity
field
and 
thereby initiate
self-propulsion
\cite{Sumino2005,Toyota2009,Thutupalli2011,Schmitt2013,Izri2014,Schmitt2016}.
Swimming dimer droplets are formed when two active emulsion droplets are connected with a surfactant bilayer 
\cite{Thutupalli2013}.

\begin{figure}
\includegraphics[width=\columnwidth]{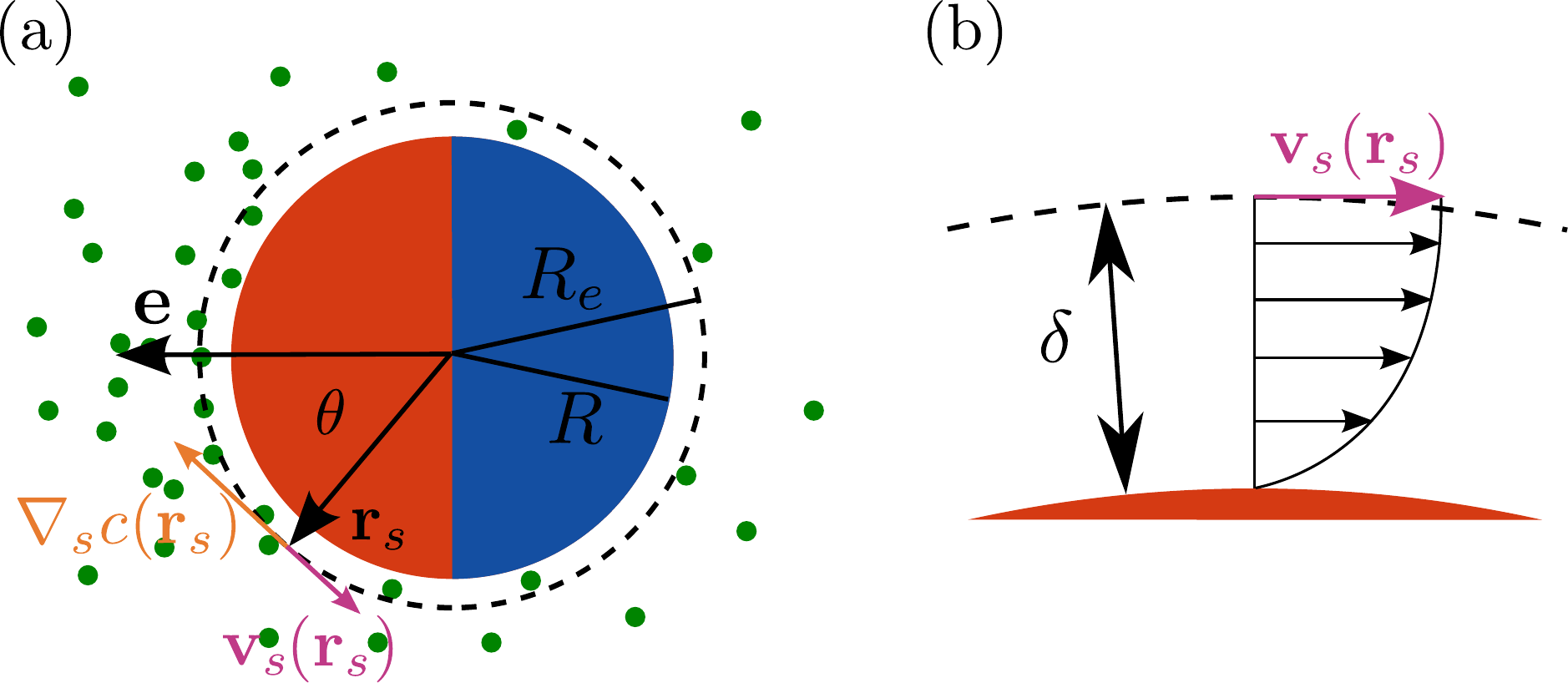}
\caption{(a) Sketch of a 
Janus particle of radius 
$R$
with a catalytic cap (red)
moving by self-diffusiophoresis. The particle creates
a local  concentration gradient 
of a solute,
$\boldsymbol{\nabla}_sc(\mathbf{r}_s)$, in tangential direction at 
position $\mathbf{r}_s$  
and at
a distance 
$\delta=R_e - R$
away from the particle surface.
The surface velocity $\mathbf{v}_s$ is proportional to the
concentration gradient.
(b) Within the diffusive boundary layer of thickness $\delta$ 
a surface velocity
field
$\mathbf{v}_s$ 
develops,
which depends on the position $\mathbf{r}_s$ at the surface.}
\label{Fig:phoresis}
\end{figure}

\subsubsection{Experimental realisations of active colloids}
\label{Sec:ExpReal}
At present a 
variety
of different  active colloidal systems have been realized experimentally.
Paxton \etal \cite{Paxton2004b} and Fournier-Bidoz \etal \cite{Fournier-Bidoz2005b} were the first 
to create micrometer-sized 
active
bimetallic 
rods.
They
propel themselves
by consuming fuel such as $\textrm{H}_2\textrm{O}_2$ 
at one end of the rod due to the different surface chemistry \footnote{Self-propulsion powered by $\textrm{H}_2\textrm{O}_2$
has first been observed for $\mathrm{mm}$ sized plates moving at the air-liquid interface \cite{Ismagilov2002}.}.
Here, a local ion gradient 
creates an electric field,
which
initiates fluid flow close to the rod and hence 
electrophoretic self-propulsion\footnote{The electric field acts on local charges and thereby produces body forces in the fluid.}.
Spherical active colloids \cite{Howse2007a,Palacci2010}
and sphere-dimers \cite{Valadares2010}
are able to swim 
by self-diffusiophoresis, where a 
gradient in
chemical species close to the surface
generates
a pressure gradient
along the surface
and thereby fluid flow
\cite{Golestanian2005a}.

Water
droplets
can become active and
move in oil by consuming bromine fuel supplied inside the droplet. 
The bromination of mono-olein surfactants at the water-oil interface induces 
gradients in surface tension and thus
Marangoni stresses, 
which initiate surface flow and thereby
self-propulsion
 \cite{Thutupalli2011,Schmitt2013,Thutupalli2013,Herminghaus2014}.
Even pure
water droplets
are 
able to move by spontaneous symmetry breaking of the surfactant concentration
at the interface  \cite{Izri2014} as do liquid-crystal droplets \cite{Herminghaus2014}.
In both cases, surfactant micelles are crucial to initiate the spontaneous symmetry breaking
\cite{Schmitt2016}.
%

When swimmers depend on fuel, they
stop moving after it has been consumed.
Instead of 
supplying
fuel in the solution, 
permanent illumination with
laser light as a power source 
can sustain
self-propulsion for an 
arbitrarily
long time.
For example, 
in Refs.\ \cite{Volpe2011,Buttinoni2012b} the authors suspend Janus particles with a golden cap in a water-lutidine mixture.
Heating the cap with laser light demixes the binary liquid and initiates self-diffusiophoretic motion.
Recent theoretical calculations 
have
been performed in Refs.~\cite{Wurger2015,Samin2015}.

Finally,
the release of bubbles behind 
an
active colloid \cite{Gibbs2009a,Manjare2012}
or from cone-shaped microtubes \cite{Sanchez2010a,Solovev2010}
can also lead to self-propulsion.
%
%
%
Moreover,
self-propelled colloids 
were
designed 
that are powered
by ultrasound \cite{Rao2015a}.

Currently the number of 
experimentally realized 
colloidal swimmers is increasing fast.
We refer to recent reviews, which also discuss  
more
details
of the different mechanisms for generating self-locomotion
 \cite{Ozin2005,Paxton2006,Hong2010d,Ebbens2010a,Mirkovic2010,Sengupta2012,Kapral2013,Wang2013a,Abdelmohsen2014}.

\subsubsection{Theoretical description and modeling of 
self-phoretic swimmers}
\label{Sec:SwimTheory}

We discuss in more detail the concept of \textit{self-phoresis}, by which 
solid
active 
colloids move forward, and
also comment on the \textit{Marangoni-propulsion} of active emulsion droplets.

The gradient of an external field $X(\mathbf{r})$ induces pho\-re\-tic transport of a passive 
colloid of radius $R$
since the field
determines the interaction between the colloidal surface and
the surrounding fluid
\cite{Anderson1989}.
Within a thin layer of thickness 
$\delta \ll R$ the
field gradient $\boldsymbol{\nabla}X(\mathbf{r})$ induces a tangential near-surface flow, 
which 
increases from zero
in radial direction
and saturates at
an effective radius $R_e=R+\delta$
to a constant value
[see figure~\ref{Fig:phoresis}(b)].
Typically, the
external field $X(\mathbf{r})$ is 
an electric potential $\phi(\mathbf{r})$, a chemical concentration field $c(\mathbf{r})$, or
a temperature field $T(\mathbf{r})$ and the respective colloidal transport mechanisms are called \textit{electrophoresis},  \textit{diffusiophoresis}, and \textit{thermophoresis}.

In contrast, 
active colloids 
create 
the local
field gradient $\boldsymbol{\nabla}X(\mathbf{r})$ and 
the resulting
near-surface flow by themselves.
In the simplest case, 
half-coated Janus spheres 
are used
as sketched in figure~\ref{Fig:phoresis}(a).
The catalytic cap (red) catalyzes a chemical reaction of the fuel towards a chemical product (shown as green dots),
which diffuses around.
In 
steady state a non-uniform concentration profile around the particle is established
\cite{Golestanian2005a,Golestanian2007a,Golestanian2009a,Popescu2009,Popescu2010,Ebbens2012,Sharifi-Mood2013,Michelin2014,deGraaf2015,Yariv2015}.
Since
the Janus colloid produces the diffusing chemical by itself, the propulsion mechanism is called 
self-diffusiophoresis.
Recent works showed how
details of the surface coating and ionic effects  determine the
swimming speed and efficiency 
of
catalytically driven active colloids \cite{Sabass2010,Sabass2012,Sabass2012a,Brown2014,Ebbens2014,deGraaf2015,Brown2015b}.

Janus colloids coated with metals such as 
gold  are able to move by heating the metal cap \cite{Jiang2010b,Bickel2013,Baraban2013b,Qian2013,Bregulla2014}.
The resulting local
temperature gradient
induces
an effective slip velocity, which is known as \textit{Soret effect}, and
the active colloid moves by
self-thermophoresis.
Furthermore,
bimetallic nano- and microrods move by self-electrophoresis \cite{Paxton2005,Moran2010,Yariv2011}.
Here an ionic current near the surface 
drags fluid with it
and
thereby induces
an effective slip velocity at the surface.

All the systems mentioned above 
create a tangential slip velocity
$\mathbf{v}_s$ close to the 
particle surface
and
proportional to
the field gradient  
$\boldsymbol{\nabla}_s X(\mathbf{r})$ along the surface \cite{Anderson1989},
\begin{equation}
\mathbf{v}_s(\mathbf{r}_s) = - \kappa(\mathbf{r}_s) \left. 
\boldsymbol{\nabla}_s  X(\mathbf{r})\right|_{|\mathbf{r}|=R_e} \, .
\label{Eq:slipv}
\end{equation}
The slip-velocity coefficient
$\kappa$ depends on the specific phoretic mechanism and material properties
of the particle-fluid interface, which vary with the location $\mathbf{r}_s$.
For
spherical active colloids with axisymmetric surface coating,  $\mathbf{v}_s(\theta)$ depends on the 
polar
angle $\theta$  
of the position vector $\mathbf{r}_s$ relative to
the orientation vector $\mathbf{e}$
[see figure~\ref{Fig:phoresis}(a)].

To determine the swimming speed and the flow field around the active colloid, one  
takes
the slip velocity at the effective radius 
$R_e=R+\delta$.
In concrete, one solves
the homogeneous Stokes equations 
for $ r>R_e$ with
the boundary condition 
$\mathbf{u}(\mathbf{r}_s) = \mathbf{v}_s + v_0\mathbf{e} + \boldsymbol{\Omega} \times \mathbf{r}_s$,
where $v_0$ is the constant swimming speed
and $\boldsymbol{\Omega}$ is the angular velocity of the active colloid.
Since phoretic
transport is force- and torque free 
\cite{Anderson1989}, 
both the hydrodynamic force and torque of
equations
(\ref{Eq:FMstress1}) and (\ref{Eq:FMstress2})
should be zero when evaluated 
along the surface with
radius $R_e$.
This gives the translational
and angular velocity of a spherical (self-)phoretic colloid \cite{Anderson1989},
\begin{eqnarray}
\label{Eq:VColloid}
  \mathbf{V} &= v_0\mathbf{e} = - \langle \mathbf{v}_s  \rangle  \\
\label{Eq:OmColloid}
   \boldsymbol{\Omega} &=  - \frac{3}{2R} \langle \mathbf{v}_s  \times \mathbf{n} \rangle \, ,
\end{eqnarray}
where $\langle \cdots \rangle$ 
denotes the average taken over the surface
with 
normal $\mathbf{n}$.
Note that for axissymmetric  surface velocity profiles, $\boldsymbol{\Omega} = \mathbf{0}$ and the active colloid 
moves on a straight line.
This is, for example, true for half-coated Janus spheres.
Nevertheless, rotational Brownian motion 
reorients the swimmer and 
it performs a persistent random walk, as we will discuss
in section \ref{Sec:Walk}.

To gain a better 
insight into self-phoretic motion
on the microscopic level,
explicit hydrodynamic mesoscale simulation techniques 
have
been used to study 
self-diffusiophoretic 
and self-thermophoretic 
 motion.
In particular, reactive multi-particle colision dynamics (R-MPCD)  is a method to explicitely simulate
the full hydrodynamic flow fields and chemical fields around 
a
swimmer.
It also includes
chemical reactions between 
solutes
and 
thermal noise \cite{Tucci2004}.
The method
was
used to simulate the self-phoretic motion of spherical Janus colloids \cite{DeBuyl2013,Yang2014}
and 
self-propelled
sphere dimers \cite{Rueckner2007,Tao2008a,Tao2009a,Tao2010,Valadares2010,Thakur2011,Thakur2012,Yang2011a,Yang2014,Colberg2014a}.
The simulated flow field of a sphere-dimer swimmer agrees with that of a force dipole \cite{Yang2014d,Reigh2015},
which has recently 
been confirmed by an analytic calculation \cite{Reigh2015}.
Dissipative particle dynamics was used to determine the effect of particle shape on 
active
 motion \cite{Lugli2012}, and to calculate the flow fields 
initiated by self-propelled Janus colloids 
for different fluid-colloid interactions \cite{Fedosov2015}.

In active emulsion droplets spontaneous symmetry breaking generates gradients in the density of surfactant molecules 
at the droplet-fluid interface and thus gradients in surface tension $\sigma$. They drive \textit{Marangoni flow} at the interface
\cite{Thutupalli2011,Schmitt2013,Michelin2013,Izri2014,Schmitt2016}.
The slip velocity field does not simply follow from an equivalent
of equation~(\ref{Eq:slipv}), which links slip velocity to $\boldsymbol{\nabla}_s \sigma(\mathbf{r}_s)$.
Instead, one has to solve the Stokes equations inside and outside the droplets and match the difference in tangential viscous stresses by $\boldsymbol{\nabla}_s \sigma(\mathbf{r}_s)$ \cite{Schmitt2016}. So, in contrast to colloidal particles fluid flow also 
evolves inside the droplet \cite{Sumino2005,Toyota2009,Ohta2009,Thutupalli2011}.

%
%
%
%

\subsubsection{Swimming with prescribed surface velocity}
\label{Sec:Prescribed}
We have seen that the 
active
motion of self-phoretic colloids is mainly determined by the 
surface velocity
field
$\mathbf{v}_s$ independent of the underlying physical or chemical mechanisms
to realize $\mathbf{v}_s$.
In the simplest case 
one assumes
a
\textit{prescribed} 
surface velocity field
 without taking the underlying mechanism 
into account.
Such
model swimmers are 
useful to study 
how
hydrodynamic flow fields 
influence the 
(collective)
dynamics of microswimmers.

\begin{figure}
\begin{center}
\includegraphics[width=0.8\columnwidth]{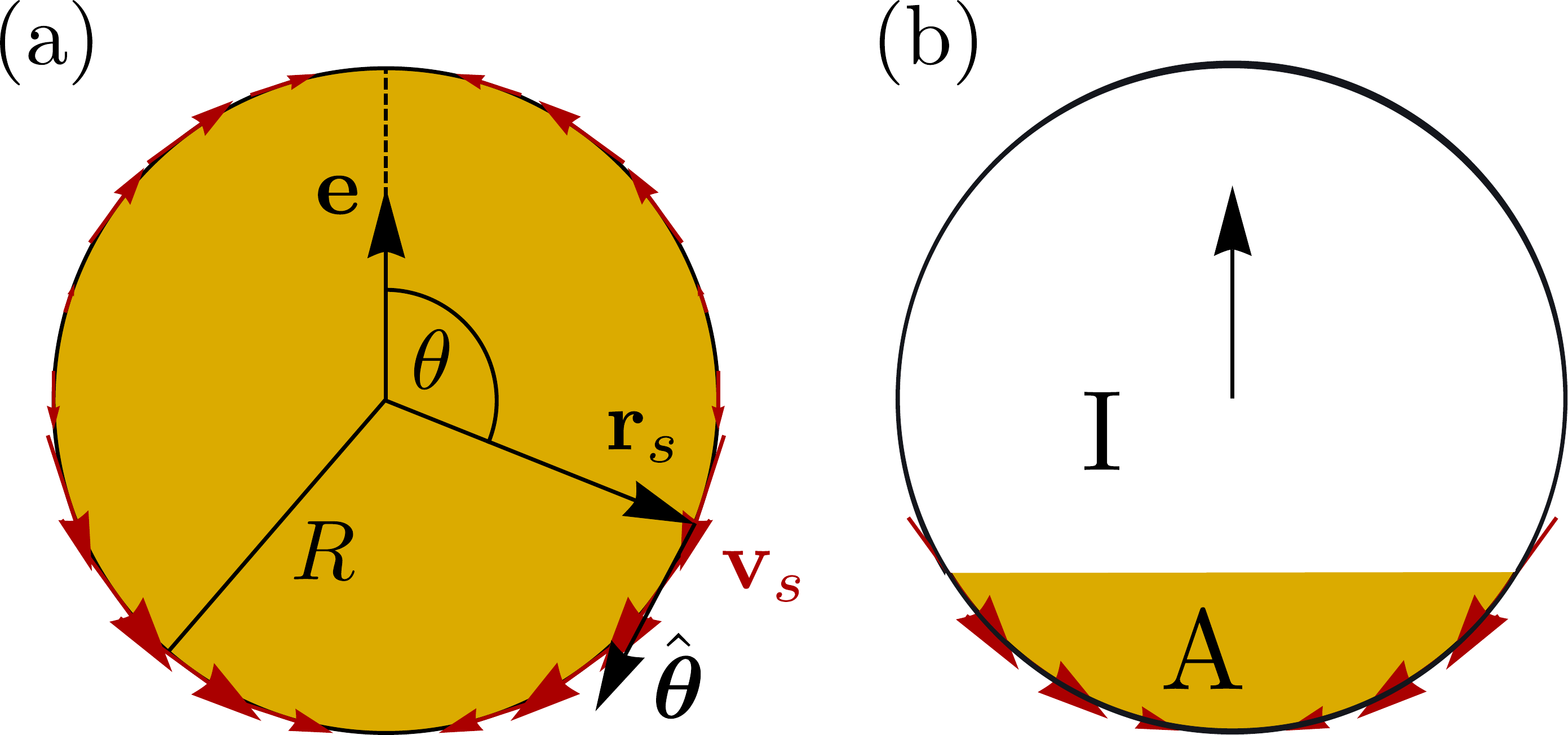}
\end{center}
\caption{(a) Sketch of a spherical squirmer of radius $R$ swimming along direction $\mathbf{e}$. The axisymmetric surface 
velocity field $\mathbf{v}_s$ 
points along the
tangential 
polar
direction $\hat{\boldsymbol{\theta}}$ 
and
depends on the position $\mathbf{r}_s$ on the surface, i.e., on the 
polar
angle $\theta$.
(b) Sketch of a 
simple model of an
active Janus sphere, which is inert (I) 
at
the front part and active (A) 
at
the rear part.}
\label{Fig:prescribed}
\end{figure}

One prominant example 
is the so-called \textit{squirmer} \cite{Lighthill1952,Blake1971a}.
It was originally proposed to model ciliated microorganisms such as
\textit{Paramecium} and \textit{Opalina}.
Nowadays,
the squirmer
is frequently used as a simple model microswimmer.
In its 
most common
form, the squirmer is a 
solid
spherical particle of radius $R$
with
a prescribed 
tangential surface velocity field
\cite{Ishikawa2006},
which
is expanded into an appropriate set of basis functions involving spherical harmonics
\cite{Pak2014,Schmitt2016}.
In
the important case of 
an axisymmetric velocity field with only polar
velocity component $v_{\theta}$, 
the expansion uses
derivatives of Legendre 
polynomials
$P_n(\cos\theta)$ 
\cite{Lighthill1952,Blake1971a,Ishikawa2006},
\begin{eqnarray}
v_{\theta} &= \sum_{n=1} B_n \frac{2}{n(n+1)}\sin\theta \frac{d P_n(\cos\theta)}{d \cos\theta } \\
         &= B_1 \sin \theta ( 1+ \beta \cos\theta) + \dots
\end{eqnarray}
where the 
coefficients
$B_n$ 
characterize
the surface velocity modes and
$\mathbf{v}_s=v_{\theta}\hat{\boldsymbol{\theta}}$
is directed along the longitudinals [see figure~\ref{Fig:prescribed}(a)].
We
also introduced the dimensionless \textit{squirmer parameter} $\beta = B_2/B_1$, which defines the strength
of the second squirming mode 
relative
to the first one.
Without loss of generality we set $B_1>0$
and the squirmer swims into the positive $z$ direction as we will demonstrate below.
Typically,
only the first two modes are considered and $B_n=0$ for $n \ge 3$ \cite{Ishikawa2006,Downton2009a}.
The reason is that 
$B_1$ and $B_2$ 
already 
capture the basic features of microswimmers.
If $B_1$ is the only non-zero coefficient,
the
resulting
flow field 
in the bulk fluid
is exactly that of a source dipole 
and illustrated in figures~\ref{Fig:flow}(f).
The coeffcient
$B_1$ is linearly related to the source-dipole strength $q$ in
equation~(\ref{Eq:SDipole}), $q= 8\pi \eta B_1 R^3/3$.
If we add the second mode, $B_2 \ne 0$,
the far field of the swimmer is that of a force dipole  
shown in
figure~\ref{Fig:flow}(b).
It decays more slowly with distance
than the source dipole field. 
To match the boundary condition at the swimmer surface,
an additional term $\propto r^{-4}$
is 
needed
\cite{Ishikawa2006}.
Figure~\ref{Fig:flow}(e) illustrates the complete
squirmer flow field 
for a pusher with $\beta=-3$.
The second mode coefficient
$B_2$ is related to the force dipole strength $p$
introduced in equation~(\ref{Eq:Dipole})
by
$p=-4\pi R^3\eta B_2$.
So, 
$B_2 \propto \beta<0$ describes 
a pusher ($p>0$),
while for $B_2 \propto \beta>0$ the squirmer realizes
a puller ($p<0$).

We note that 
a coordinate-free expression
of
the surface velocity 
field
reads \cite{Ishikawa2006}
\begin{eqnarray}
\label{Eq:VsSquirmer1}
\mathbf{v}_s &= \sum_{n=1} B_n \frac{2}{n(n+1)}[(\mathbf{e}\cdot \hat{\mathbf{r}})\hat{\mathbf{r}} - \mathbf{e}]
 \frac{d P_n(\mathbf{e}\cdot \hat{\mathbf{r}})}{d (\mathbf{e}\cdot \hat{\mathbf{r}} ) } \\
\label{Eq:VsSquirmer2}
         &= B_1 ( 1 + \beta \mathbf{e}\cdot \hat{\mathbf{r}}_{s} )   [ (\mathbf{e}\cdot \hat{\mathbf{r}}_{s} )
 \hat{\mathbf{r}}_{s} - \mathbf{e}  ] + \dots \, ,
\end{eqnarray}
where we introduced the
the swimming direction $\mathbf{e}$ and the unit vector 
$\hat{\mathbf{r}} = \mathbf{r} / r$.
We calculate the
velocity of the squirmer 
by
inserting equation~(\ref{Eq:VsSquirmer1}) into  equation~(\ref{Eq:VColloid}) and 
obain
$v_0 = 2B_1/3$.
Hence, the swimming velocity
only depends on the first mode $B_1$.

A simple model of an active Janus particle 
also uses
a prescribed surface velocity
field
[see figure~\ref{Fig:prescribed}(b)].
While
the inert part (I)
of the particle surface
fulfills the boundary condition of a passive colloid ($\mathbf{v}_s=0$),
the chemically active cap (A) generates nearby slip flow and $\mathbf{v}_s\neq 0$
\cite{Spagnolie2012}.

Noteworthy, numerical simulations 
were also
performed with ellipsoidal and rodlike 
swimmers
driven by a surface flow
\cite{Ishikawa2006a,Saintillan2007,Saintillan2008,Kanevsky2010,Saintillan2012,Spagnolie2012,Lushi2013,Lushi2014}.

\subsection{Colloidal surfers and rollers}
\label{Sec:RollSurf}

Some 
colloidal objects
need bounding sur\-faces to move forward.
For example, mag\-ne\-tically actuated colloidal doublet rotors perform non-reciprocal motion in the presence of solid surfaces,
which leads to directed motion \cite{Tierno2008,Tierno2008a}.
Bricard \textit{et al.}
 applied a static electric field 
between two conducting glass slides,
while
the lower 
slide was
populated with sedimented insulating spheres \cite{Bricard2013}.
At sufficiently high field strength the particles start to rotate
about
a random 
direction
parallel to the surface.
This
is known as \textit{Quincke rotation},
where the polarization axis rotates away from the field direction and thereby
spontaneously breaks the axial symmetry
\cite{Quincke1896,Jones1984,Das2013,Bricard2013}.
The well-known translational-rotational coupling of a spherical colloid 
close to
a no-slip surface then 
induces
directed motion \cite{Happel2012,Jakli2008}.
Its
well-defined velocity 
depends on the electric field strength, $v_0 \sim [(E/E_Q)^2-1]^{1/2}$, where $E_Q$ is the critical field strength
for the Quincke rotation to set in.
Note that this \textit{rolling} motion was already observed and explained by J\'akli \etal \cite{Jakli2008}, 
who 
investigated
Quincke rotation in liquid crystals \cite{Liao2005,Jakli2008}.

Palacci \textit{et al.} studied the motion of \textit{colloidal sur\-fers} \cite{Palacci2013,Palacci2014}.
They consist of a photoactive material (e.g.~hematide) 
embedded in polymer spheres,
which are dispersed
in a  
solvent
containing $\textrm{H}_2\textrm{O}_2$.
After 
sedimenting
to the 
lower
substrate, 
the particles are illuminated by UV light. 
The
photoactive material triggers the dissoziation of 
$\textrm{H}_2\textrm{O}_2$, which generates osmotic flow along the substrate.
This rotates the active part towards the substrate and the particles start 
moving.
Although the exact swimming mechanism is not known yet, the substrate and the osmotic flow are  necessary 
for self-propulsion \cite{Palacci2013,Palacci2014}.

\subsection{Active Brownian particles}
\label{Sec:ABP}
Modeling the motion of biological and artificial microswimmers has become a largely growing field in statistical physics, hydrodynamics, and soft matter physics.
A minimal model
for microswimmers and other active individuals are \textit{active Brownian particles} \cite{Erdmann2000,Schweitzer2003,Romanczuk2012}.  
They
only capture 
the
very basic features
of active motion:
(i) overdamped dynamics of particle position $\mathbf{r}$ and particle orientation  $\mathbf{e}$, (ii) 
motion with
intrinsic particle velocity $v_0$ along the direction  $\mathbf{e}$, and (iii)
thermal (and non-thermal) translational
as well as
rotational noise, $\boldsymbol{\xi}$ and $\boldsymbol{\xi}_r$.
The motion of spherical (see, 
for example,
\cite{VanTeeffelen2008,TenHagen2011,TenHagen2011a,Zheng2013a,Babel2014,Bialke2012,Buttinoni2013e,
Henkes2011,Fily2012,Redner2013b,Ni2013a,Mognetti2013,Fily2014,Yang2014e,Wysocki2014,Zottl2014,Stenhammar2014}),
elongated (see, 
for example,
\cite{Peruani2006,Baskaran2008,Baskaran2008a,Wensink2008,Elgeti2009,TenHagen2011,Wensink2012,Wensink2012b,Kaiser2012}), and
active Brownian particles 
with more complex shapes
 \cite{Wittkowski2012,Kummel2013,Wensink2014,TenHagen2014} 
has
extensively been studied.

\begin{figure}
\begin{center}
\includegraphics[width=0.9\columnwidth]{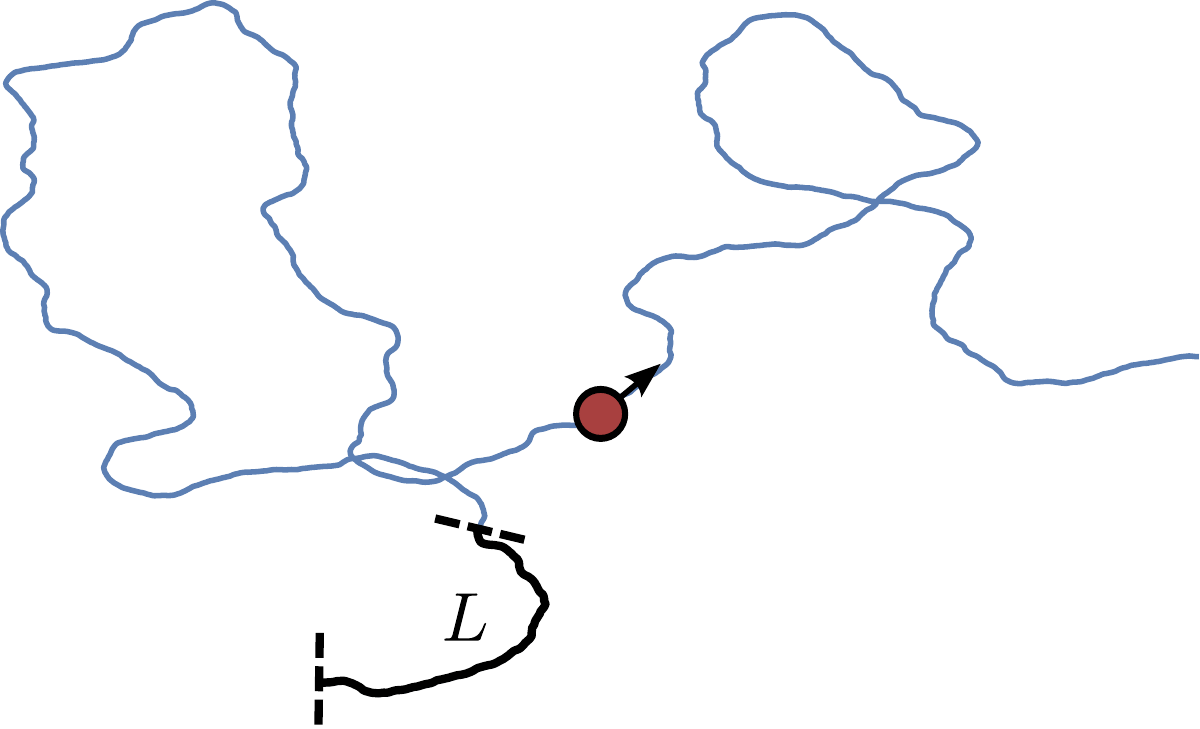}
\end{center}
\caption{Example trajectory of active Brownian motion. The persistence length $L$ defines the typical 
distance over which the active particle has lost information about its initial orientation.
In units of the swimmer diameter it is called persistence number, see equation~(\ref{Eq:Per}).
}
\label{Fig:MSD}
\end{figure}

The
most  important example is the overdamped  motion of an active Brownian sphere.
The 
dynamics for its position $\mathbf{r}(t)$ and 
orientation $\mathbf{e}(t)$  
is determined
by the Langevin equations
\begin{eqnarray}
\label{Eq:LangevinR}
\dot{\mathbf{r}} &= v_0\mathbf{e} + \sqrt{2D}\boldsymbol{\xi} \\
\label{Eq:LangevinE}
\dot{\mathbf{e}} &= \sqrt{2D_r}\boldsymbol{\xi}_r \times \mathbf{e} \, ,
\end{eqnarray}
where $D$ is the translational and $D_r$ the rotational diffusion constant of a sphere.
The rescaled stochastic force $\boldsymbol{\xi}$ and torque $\boldsymbol{\xi}_r$  
fulfill the conditions for delta-correlated
Gaussian white noise,
\begin{equation}
 \langle \boldsymbol{\xi}(t)  \rangle = 0 \enspace, \enspace
 \langle\boldsymbol{\xi}(t)\otimes\boldsymbol{\xi}(t')      \rangle = \textbf{\textsf{1}}\delta(t-t')
\label{Eq:NoiseT2}
\end{equation}
and   
\begin{equation}
 \langle \boldsymbol{\xi}_r(t)  \rangle = 0 \enspace , \enspace
 \langle\boldsymbol{\xi}_r(t)\otimes \boldsymbol{\xi}_r(t')      \rangle = \textbf{\textsf{1}}\delta(t-t') \,.
\label{Eq:NoiseR2}
\end{equation}
The unit vector $\mathbf{e}$ 
performs 
a random walk
on  the unit sphere (3D) or unit circle (2D).
A typical 
spatial
trajectory in 2D is shown in figure~\ref{Fig:MSD}.

\section{Generic Features of Single Particle Motion}
\label{Sec:Single}

In the following we discuss generic features of active motion, which already reveal themselves
for single active colloids.

\subsection{Persistent Random Walk}
\label{Sec:Walk}
Beside their directed swimming motion, active
colloids 
also
perform 
random motion due to the presence of noise as discussed in section~\ref{Sec:ABP}.
This leads to a persistent random walk, which is described by an effective diffusion constant on sufficiently large 
length scales. 
We summarize the basic theory in the following.
We use the Langevin
equations~(\ref{Eq:LangevinR}) and (\ref{Eq:LangevinE})
of an active Brownian particle that
moves with constant speed $v_0$ and changes its orientation $\mathbf{e}(t)$ stochastically.
Together with 
equations\ (\ref{Eq:NoiseR2}) characterizing rotational noise,
the orientational time correlation function can 
be computed \cite{Dhont1996},
\begin{equation}
\langle \mathbf{e}(t) \cdot \mathbf{e}(0) \rangle = e^{-t/\tau_r} \, ,
\label{Eq:Cep}
\end{equation}
where $\tau_r$ is the \textit{orientational correlation time} or \textit{persistence time}.
It
quantifies the time
during which the swimmer's orientations are correlated.
The persistence time is proportional to the inverse rotational diffusion  constant $D_r$ and for
a spherical particle 
becomes
\begin{equation}
\tau_r = \frac{1}{(d-1)D_r} \, ,
\end{equation}
where  $d$ is the 
spatial
dimension 
of the rotational motion.
For pure thermal 
motion  $D_r = k_BT/(8\pi\eta R^3)$ and strongly depends on the radius $R$ of the particle.
Note that the translational diffusion constant of a sphere is $D=k_BT/(6\pi\eta R)$ \cite{Dhont1996}.

To quantify the importance of translational and rotational diffusion compared to the 
directed
swimming motion,  
we 
define
the \textit{active P\'eclet number} $\textrm{Pe}$ 
 and the \textit{persistence number} $\textrm{Pe}_r$,
respectively.
To do so, we introduce the
ballistic time scale 
$t_0=2R/v_0$,
which the microswimmer needs
to swim its own diameter $2R$,
and
the diffusive time scale 
$t_D = 4R^2/D$,
which it
needs to diffuse its own size.
The P\'eclet number compares the two time scales, 
\begin{equation}
\textrm{Pe} = \frac{t_D}{t_0} = \frac{2Rv_0}{D}.
\label{Eq:Pe}
\end{equation}
For $\textrm{Pe} \ll 1$ the particle motion is mainly governed by translational diffusion, such as for passive Brownian particles,
while for
$\textrm{Pe} \gg 1$ 
diffusive transport is negligible 
compared to active motion.

The persistence number compares the persistence time $\tau_r$ to the ballistic time scale $t_0$,
\begin{equation}
\textrm{Pe}_r = \frac{\tau_r}{t_0} = \frac{v_0}{2RD_r} = \frac{v_0\tau_r}{2R} = \frac{L}{2R},
\label{Eq:Per}
\end{equation}
where $L= v_0 \tau_r$ is the \textit{persistence length}, i.e., the distance
a swimmer moves approximately in one direction.
Thus,
$\textrm{Pe}_r$ 
measures the persistence length in units of the particle diameter
 \cite{Friedrich2008,Li2009,Goetze2010,Taktikos2011}. 
Note that for purely thermal noise,
\begin{equation}
 D_r = \frac{3D}{4R^2} \quad \mathrm{and} \quad \textrm{Pe}_r = \frac{\textrm{Pe}}{3} \, .
\label{Eq:DDr}
\end{equation}
Non-thermal rotational noise, however, can significantly decrease the  persistence of the swimmer \cite{Howse2007a,Jiang2010b}
such that $\textrm{Pe}_r \le \textrm{Pe}/3$.
Typical values for $\textrm{Pe}$ and  $\textrm{Pe}_r $ for spherical micron-sized
 active colloids are  of the order 
$10^0-10^2$ \cite{Howse2007a,Jiang2010b,Palacci2010,Buttinoni2012b,Palacci2013}.
In figure~\ref{Fig:MSD}  a typical persistence length of the random walk of a spherical microswimmer in 2D is indicated.

We now discuss the mean square displacement of the active particle, which defines 
an
effective diffusion 
coefficient
$D_\mathrm{eff}$.
The persistent random walk of active particles was first formulated by  
R.~F\"{u}rth in 1920
and  also experimentally  verified   with Paramecia and an unknown species \cite{Fuerth1920}.
Using equations (\ref{Eq:LangevinR})--(\ref{Eq:NoiseR2})
the mean square displacement $\langle \Delta r^2(t) \rangle = \langle |\mathbf{r}(t) - \mathbf{r}(0)|^2 \rangle$
 can be calculated 
as
\cite{Howse2007a,Downton2009a,TenHagen2011}
\begin{equation}
\langle \Delta r^2(t) \rangle = 2dDt + 2v_0^2\tau_rt - 2v_0^2\tau_r^2(1-e^{-t/ \tau_r}) \, .
\label{Eq:MSDS1}
\end{equation}
For times small compared to the persistence time,  $t \ll \tau_r$,
equation~(\ref{Eq:MSDS1}) reduces to
\begin{equation}
\langle \Delta r^2(t) \rangle
 = v_0^2t^2 + 2d D t,
\label{Eq:MSDS2}
\end{equation}
which 
includes contributions
from the ballistic motion of the swimmer
and 
from translational diffusion.
However, 
for $t \gg \tau_r$
the swimmer 
orientation decorrelates from
its initial
value
and the active colloid
performs diffusive motion,
\begin{equation}
\langle \Delta r^2(t) \rangle =  (2d D + 2v_0^2\tau_r) t =  
2d D_{\mathrm{eff}} t,
\label{Eq:MSDS3}
\end{equation}
with the \textit{effective diffusion constant} 
\begin{equation}
 D_{\mathrm{eff}} = D + 
 \frac{1}{d}
 v_0^2 \tau_r
 =
D + \frac{v_0^2}{d (d-1) D_r} \, .
\label{Eq:MSDS3b}
\end{equation}
For pure thermal noise,  using equations~(\ref{Eq:DDr})
and $d=3$,
it can be 
rewritten either
with the persistence number $\textrm{Pe}_r$
or the P\'eclet number $\textrm{Pe}$,
\begin{equation}
 D_{\mathrm{eff}} =  D(1+2\textrm{Pe}_r^2) = D(1+\frac{2}{9}\textrm{Pe}^2) 
\label{Eq:MSDS4}
\end{equation}
Hence, $D_{\mathrm{eff}}$
quadratically increases with $\textrm{Pe}$.
Note that $D_{\mathrm{eff}}$ can also be calculated via the velocity-autocorrelation function using
the corresponding Green-Kubo relation \cite{Kapral2013}.

\subsection{Motion under gravity}
\label{Sec:Gravity}
When the density of the active colloid $\rho_c$ is larger than the density of the surrounding medium $\rho_f$,
it sediments in the gravitational field $\mathbf{g} = - g \hat{\mathbf{z}}$. The active colloid
experiences a sedimentation force $\delta m \mathbf{g} \sim (\rho_c-\rho_f) \mathbf{g} = const$,
where $\delta m$ is the buoyant mass of the active colloid
and
$g=9.81m/s$.
The
Langevin equation
for the position of the sedimenting active colloid reads 
\begin{equation}
\dot{\mathbf{r}} = v_0\mathbf{e} + (\delta m/\gamma)\mathbf{g} + \sqrt{2D}\boldsymbol{\xi}_1 \, .
\end{equation}
Alteratively, one formulates the Smoluchowski equation for the 
propability distribution $\rho(\mathbf{r},\mathbf{e},t)$
for position and orientation
\cite{Enculescu2011,Wolff2013},
\begin{equation}
\frac{\partial \rho}{\partial t} + \nabla \cdot \mathbf{J}^{\mathrm{trans}} + \mathcal{R} \cdot \mathbf{J}^{\mathrm{rot}} = 0 \, ,
\label{Smoluchowski}
\end{equation}
with the respective translational and rotational currents
\begin{equation*}
\mathbf{J}^{\mathrm{trans}} = -D \nabla \rho  + \left(v_0 \mathbf{e} + \delta m \mathbf{g} / \gamma \right)   \rho
\enspace , 
\enspace
\mathbf{J}^{\mathrm{rot}}  = -D_r \mathcal{R} \rho \,. 
\label{eq.Smolu}
\end{equation*}
The latter is purely diffusive with the rotational operator $\mathcal{R} = \mathbf{e} \times \nabla_{\mathbf{e}}$.

A multipole expansion of $\rho$ gives an exponential sedimentation profile for the spatial density, as for passive particles,
\begin{equation}
\rho_0(z) = \rho_0(0) \exp(-z / \delta_{\mathrm{eff}}) \, .
\end{equation}
However, the
sedimentation length
  \cite{Tailleur2008,Tailleur2009,Palacci2010,Enculescu2011,Ginot2015}
\begin{equation}
\delta_{\mathrm{eff}} =
\delta_0 
\left( 1 +  \frac{2\textrm{Pe}^2}{9} \right) =
 \frac{\delta_0 D_{\mathrm{eff}} }{D}
\end{equation}
is increased
compared to the passive case with $\delta_0 = k_B T / (mg)$. This can also be rationalized by introducing an increased
effective temperature $T_{\mathrm{eff}} = T D_{\mathrm{eff}} / D$ \cite{Palacci2010}. Clearly, the larger sedimentation
length is due to the larger effective diffusion of active particles. However, the deeper reason is that they develop 
polar order by aligning against the gravitational field \cite{Enculescu2011}.

Gravitaxis of spherical active colloids with asymmetric mass distribution 
was
observed 
in experiments
\cite{Campbell2013} and described theoretically \cite{Wolff2013}.
Interestingly, the interplay of
gravity and
intrinsic circling motion 
of asymmetric L-shaped active colloids \cite{Kummel2013} leads to various complex swimming paths
and gravitaxis, induced by the shape asymmetry 
\cite{TenHagen2014}.

\subsection{Taxis of active colloids}
\label{Sec:Taxis}
Taxis means the ability of
biological microswimmers such as bacteria or sperm cells
to move along field gradients. For example,
gradients of 
a chemical field or a light intensity field
can initiate
\textit{chemotaxis} \cite{Berg1972} or \textit{phototaxis} \cite{Witman1993}, respectively.
We already mentioned that colloids also react to field gradients, which generate phoretic transport.
Combined with activity, one obtains artificial microswimmers that can mimic biological taxis.
Here we discuss how 
active colloids move in 
field gradients  $\boldsymbol{\nabla}X(\mathbf{r})$
that are
(i) externally applied such as in classical phoretic transport, (ii) generated by other self-phoretic active colloids, or (iii) generated
by the active colloid itself 
and therefore known as
\textit{autochemotaxis}.

\subsubsection{Motion in external field gradients}
\label{Sec:FieldGradients}
\begin{figure}
\begin{center}
\includegraphics[width=\columnwidth]{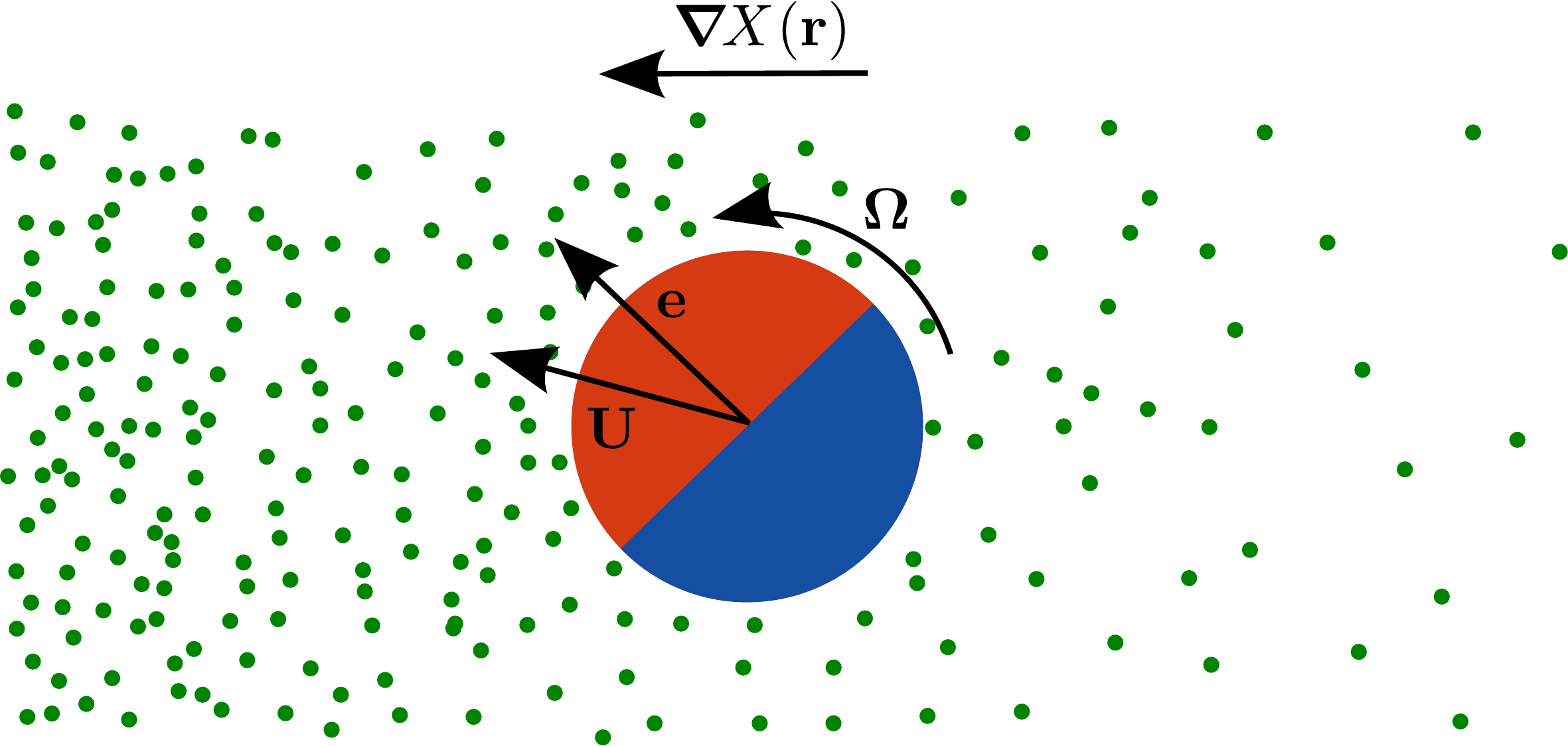}
\end{center}
\caption{An active colloid moving in an external 
field
gradient 
$\boldsymbol{\nabla}X(\mathbf{r})$.
Due to phoresis its
velocity $\mathbf{U} = v_0\mathbf{e} + \mathbf{U}_C$ is not parallel to its intrisic symmetry axis $\mathbf{e}$ 
and it 
rotates
with 
angular velocity
$\boldsymbol{\Omega}\neq \textbf{0}$. 
}
\label{Fig:phoresis1}
\end{figure}

Self-phoretic active colloids are surrounded by a cloud of 
self-generated field gradients 
but
also respond to external gradients 
$\boldsymbol{\nabla}X(\mathbf{r})$.
Again, this can be 
realized
by diffusio-, electro-, or thermophoresis.
Then, spherical 
colloids
at position $\mathbf{r}$, either passive or active,
translate with
velocity $\mathbf{U}_C$ and 
rotate
with angular velocity $\boldsymbol{\Omega}_C$ \cite{Anderson1989},
\begin{eqnarray}
\mathbf{U}_C &= 
\Big[
\langle \kappa (\mathbf{r}_s)\rangle \mathbf{1} 
- \frac 1 2 \langle (3\mathbf{n}\otimes \mathbf{n} - \mathbf{1})\kappa (\mathbf{r}_s)  \rangle   
\Big]
\boldsymbol{\nabla}X(\mathbf{r}) \, , \\
\boldsymbol{\Omega}_C &= \frac{9}{4R}\langle \mathbf{n}\kappa (\mathbf{r}_s) \rangle \times 
\boldsymbol{\nabla}X(\mathbf{r}) \, ,
\end{eqnarray}
where
$\kappa (\mathbf{r}_s)$ is the slip-velocity coefficient introduced in equation (\ref{Eq:slipv})
and
$\langle \ldots \rangle$ is again the average taken over the particle surface
with normal $\mathbf{n}$.
Finally, the
total 
colloidal
velocity reads
 $\mathbf{U} = v_0\mathbf{e} + \mathbf{U}_C$, and the angular velocity
is simply $\boldsymbol{\Omega} = \boldsymbol{\Omega}_C$.
This is  sketched in figure~\ref{Fig:phoresis1}.
Note, however, that the velocity $v_0$ will in general depend on the local concentration $X(\mathbf{r})$ \cite{Saha2014a}.

Experimentally, the chemotactic response of active Janus colloids in a concentration gradient $\boldsymbol{\nabla}c(\mathbf{r})$
 of $\textrm{H}_2\textrm{O}_2$ 
 was
 observed
 \cite{Hong2007a,Ibele2009a,Sen2009c,Hong2010c,Solovev2013a,Baraban2013a,Duan2013a,Wang2013a}.
A
self-propelled 
nano-dimer motor 
moving in
a chemical gradient 
was
simulated in Ref.~\cite{Thakur2011a}.
Recent studies on interacting self-phoretic active colloids (see 
chapter~\ref{Sec:Collective})
considered phoretic attraction
and
repulsion
but also
reorientation of a particle 
due to field gradients produced by 
others
\cite{Theurkauff2012,Palacci2013,Saha2014a,Pohl2014,Bickel2014a,Pohl2015,Liebchen2015}.

\subsubsection{Autochemotaxis}
The chemical field 
$c(\mathbf{r},t)$,
which 
a self-diffusiophoretic particle 
creates,
diffuses 
on the characteristic time scale
$t_c \propto D_c^{-1} R^2$,
where $D_c$ is the diffusion constant of the chemical species
and $R$ is the particle radius.
Since the chemical diffuses fast, one typically has 
$t_c \ll \tau_r$, where $\tau_r$ is the orientational persistence time  
introduced in section\ \ref{Sec:Walk}. 
%
%
The particle interacts
with its own chemical trail,
which
is known as \textit{autochemotaxis}.
Thus, the chemical field couples back to the motion of the microswimmer, which can drastically modify its 
ballistic and diffusive movement.
The implications
were mainly
studied 
in the context of biological autochemotactic systems
\cite{Tsori2004,Grima2005,Sengupta2009a,Taktikos2011,Kranz2015}
and recently discussed for active colloids \cite{Golestanian2009a}.

In Ref.\ \cite{Golestanian2009a} the interaction of an active colloid with its self-generated
chemical cloud 
was investigated.
Since the active colloid performs rotational diffusion
and 
the
released product 
molecules
at the chemically active surface diffuse 
on
a finite time $t_c$,
a non-stationary concentration profile around the particle evolves.
As a result,
the  mean square displacement 
of equation\ (\ref{Eq:MSDS1})
is modified 
and shows anomalous diffusion at intermediate time scales and a modified effective diffusion in the 
long-time limit \cite{Golestanian2009a}.

\subsection{Motion in external fluid flow}
\label{Sec:FluidFlow}
Biological microswimmers have to respond to fluid flow in their natural environments \cite{Rusconi2015}.
In addition, we expect that in the near future new experiments and simulations on the motion of active colloids in microchannel flows will be performed.
Finally, one has
the vision that articial microswimmers may be used in the future to navigate through human blood vessels \cite{Nelson2010}.

Here we  discuss
the
basic physical mechanisms for microswimmers such as  active colloids
moving in fluid
 flow.
We consider an axisymmetric active particle with orientation $\mathbf{e}$ 
that swims
in a background flow
field
$\mathbf{u}(\mathbf{r},t)$ at low Reynolds number.
This
problem
was first formulated for biological microswimmers by J~O\ Kessler and T~J\ Pedley (see, e.g., \cite{Pedley1992} for a review).
If variations of the flow field on the size of the particle are small, 
a passive particle 
simply acts
as a \textit{tracer particle} 
and follows
the stream lines of the flow.
Thus its
velocity 
is 
$\mathbf{u}(\mathbf{r}(t))$,
where $\mathbf{r}(t)$
denotes
the center-of-mass position of the particle.
For an active particle one has to add the swimming velocity and the total particle velocity becomes
$\mathbf{V} = v_0\mathbf{e} +  \mathbf{u}(\mathbf{r}(t))$.
We
assume a constant intrisic particle speed $v_0$, which is not altered by the presence of the flow.
The particle
orientation $\mathbf{e}$, however, 
changes continuously in fluid flow.
For a spherical particle 
Fax\'ens second law
determines the angular velocity as
$ \boldsymbol{\Omega} = \frac 1 2 \boldsymbol{\nabla} \times \mathbf{u}$ 
proportional to the local vorticity of $\mathbf{u}(\mathbf{r})$
\cite{Dhont1996}.
For an elongated ellipsoid of length $L$ and width $W$ 
the angular velocity
is modified to
\begin{equation}
 \boldsymbol{\Omega} = \frac 1 2 \boldsymbol{\nabla} \times \mathbf{u} + G \mathbf{e} \times \textsf{E} \, \mathbf{e} \, ,
\label{Eq:FaxenR}
\end{equation}
where $\textsf{E}$ is the 
strain 
rate
tensor and $G=(\gamma^2-1)/(\gamma^2+1)\in [0,1)$ 
is a geometrical factor related to
the aspect ratio of the swimmer,
$\gamma=L/W$.
At low Reynolds number the equations of motions are overdamped \cite{Pedley1992}
and
they read
\begin{equation}
\label{Eq:EOMFlow1}
 \dot{\mathbf{r}} =  v_0\mathbf{e} +  \mathbf{u} \quad \mathrm{and} \quad
 \dot{\mathbf{e}}  = 
\boldsymbol{\Omega} 
\times \mathbf{e} \, .
\end{equation}
The solutions of 
the second equation
for the 
deterministic
rotation of 
an
axisymmetric passive particle in shear flow
are known as \textit{Jeffery orbits} named after G~B~Jeffery \cite{Jeffery1922}.
The
orientation vector $\mathbf{e}$
moves on a periodic orbit
on the unit sphere,
where the
specific solution
$\mathbf{e}(t)$
depends on the \textit{Jeffery constant}, which is a constant of motion 
of the second equation in (\ref{Eq:EOMFlow1}).

\begin{figure}
\includegraphics[width=\columnwidth]{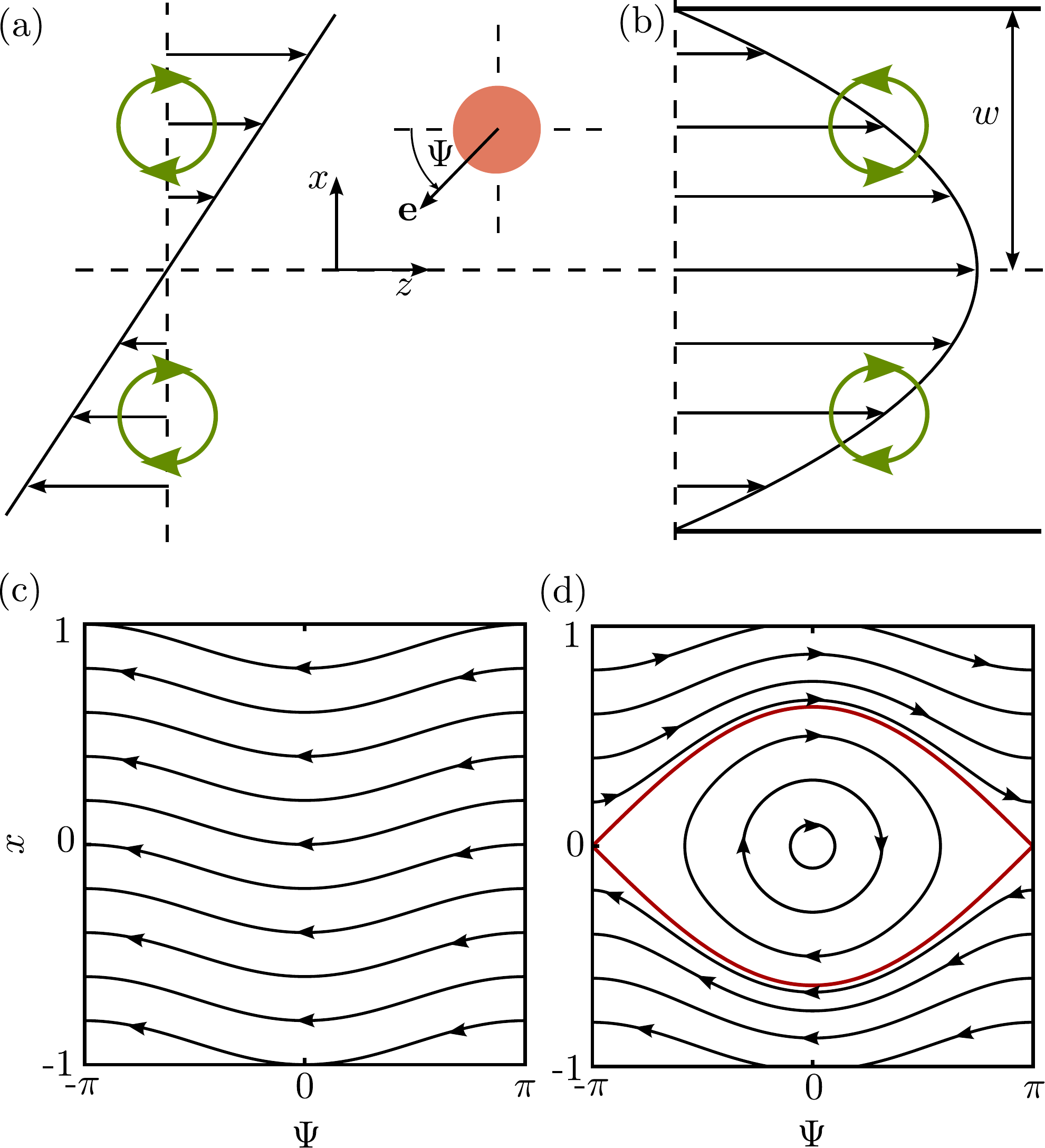}
\caption{
A swimmer with lateral position $x$ and orientation angle $\Psi$ moves along planar trajectories
in (a) simple shear flow and (b) Poiseuille flow.
The flow vorticities are indicated in green.
The corresponding phase portraits or trajectories in
$x$-$\Psi$ phase space are shown in (c) for 
shear flow and in (d) for 
Poiseuille flow.
} 
\label{Fig:Flow}
\end{figure}

To demonstrate  basic features of the
problem,
we focus here on a 
spherical active particle 
without noise
moving in 
two dimensions
in a steady unidirectional flow 
field
$\mathbf{u} = u_z(x)\hat{\mathbf{z}}$.
As sketched in figures~\ref{Fig:Flow}(a,b), the swimmer moves in the $x$-$z$ plane 
and its orientation is described by the angle $\Psi$.
The equations of motion (\ref{Eq:EOMFlow1}) 
then simplify to 
\cite{Pedley1992,Zilman2008,TenHagen2011a,Zottl2012a,Zottl2013}
\begin{eqnarray}
\label{Eq:EOMShear1}
 \dot{\Psi} &=  - \frac 1 2 \partial_x u_z(x), \\
\label{Eq:EOMShear2}
 \dot{x}  &= -v_0\sin\Psi, \\
\label{Eq:EOMShear3}
 \dot{z}  &= -v_0\cos\Psi + u_z(x).
\end{eqnarray}
Since only 
equations
(\ref{Eq:EOMShear1}) and (\ref{Eq:EOMShear2}) are coupled and  do not depend on $z$, 
one first considers
the solutions in $x$-$\Psi$ space.
Once
$x(t)$ and $\Psi(t)$ are known, they can be inserted into 
equation (\ref{Eq:EOMShear3})
to determine
$z(t)$ 
by integration.

For example, in simple shear flow, $\mathbf{u} = \dot{\Gamma}x\hat{\mathbf{z}}$,  where $\dot{\Gamma}=\textrm{const}$ 
is the shear rate [see also figure~\ref{Fig:Flow}(a)],
a spherical swimmer tumbles 
with a constant angular velocity $\dot{\Psi} = -\dot{\Gamma}/2$
and 
it moves on a
\textit{cycloidic} 
trajectory \cite{TenHagen2011a}.
The 
corresponding phase portrait in
$x$-$\Psi$ space 
is
shown in figure~\ref{Fig:Flow}(c).
Since the angular velocity $\dot{\Psi}$ of the swimmer  is independent of $x$, all starting positions $x(0)$ 
initiate the same cycloidic trajectory.

The situation is more complex in pressure-driven 
planar
Poiseuille flow in a channel of width 
$2w$.
The flow field becomes
$\mathbf{u} = v_f(1 - x^2/w^2) \hat{\mathbf{z}}$,
where $v_f$ is the flow speed in the center of the channel [see also figure~\ref{Fig:Flow}(b)].
Here the angular velocity 
$\dot{\Psi} = v_f x /w^2$
of the swimmer
is linear in
the 
lateral
position $x$ in the channel.
When the swimmer starts sufficiently far away from the centerline
or oriented downstream,
it does not cross the centerline and again tumbles in flow
but with non-constant angular velocity.
In contrast, when the swimmer starts oriented upstream 
and
sufficiently close to the centerline,
flow vorticity rotates it towards the center of the channel, it crosses the centerline, the flow vorticity changes sign, and 
the swimmer changes its sense of rotation, and is again 
directed
towards the centerline.
This results in an upstream-oriented swinging motion around the centerline.
For small amplitudes the frequency becomes $\omega = (v_0v_f / w^2)^{1/2}$ \cite{Zottl2012a}. 
It depends on the swimming speed and flow curvature $2v_f/w^2$.
The dependence on flow speed $v_f$ has recently been confirmed in experiments with a biological
microswimmer called \textit{African trypanosome} \cite{Uppaluri2012}.

In figure\ \ref{Fig:Flow}(d) swinging
and tumbling trajectories are 
represented
in $x$-$\Psi$ 
phase space.
They are divided by 
a
separatrix shown in red.
The equations of motion (\ref{Eq:EOMShear1}) and (\ref{Eq:EOMShear2}) for swimming in Poiseuille flow
are formally the same as for the mathematical pendulum.
In particular, they can be combined to $\ddot \psi + \omega^2 \sin \psi = 0 $ with $\omega$ given above.
Thus the
oscillating
 and 
circling
solutions
of the pendulum
correspond to the swinging and tumbling 
trajectories
of the swimmer, respectively.

Due to the formal correspondence with the pendulum, the
system also possesses a Hamiltonian $H$.
It consists of
a \textit{potential energy}, depending on the angle $\Psi$,
and a \textit{kinetic energy}, depending on the position $x$, which formally plays the role of a velocity \cite{Zottl2012a}.
The existence of a Hamiltonian 
implies
time-reversal symmetry 
and
reflects the fact that the trajectories do not converge to stable solutions,
\emph{i.e.},\ active particles do not cluster
at specific lateral positions
or focus in flow.
Only when 
time-reversal symmetry is broken, for example, in the presence of swimmer-wall hydrodynamic interactions
\cite{Zottl2012a}, bottom-heaviness \cite{Kessler1985a}, phototaxis \cite{Garcia2013,Jibuti2014},
flexible body shape \cite{Tournus2014},
or viscoelastic flows \cite{Mathijssen2015a},
does the dynamics 
become
\textit{dissipative} in the sense of dynamical systems and particles aggregate
at specific locations in the flow.
Although swinging and tumbling trajectories in microchannel Poiseuille flow have been observed experimentally with
biological microswimmers \cite{Uppaluri2012,Rusconi2014a},
they have not
yet
been confirmed 
in experiments with active colloids.
Spherical and elongated
swimmers moving in three dimensions in channels with elliptic cross sections still show
swinging-like and tumbling-like trajectories, 
which can become quasi-periodic \cite{Zottl2013}  or even chaotic\ \cite{Zottl2016}.

We also note that the simple model equations (\ref{Eq:EOMFlow1}) 
only 
capture
the basic features of swimming active particles in flow.
For example, adding thermal noise destroys the periodicity of the solutions and 
swinging and tumbling trajectories 
become stochastic
\cite{Rusconi2014a,Ezhilan2015a}.
Swimmers are
even
able to cross the separatrix
and switch between the two swimming modes
 \cite{Zottl2012a}.
Furthermore, the dynamics is altered when
the flow field
couples to
the chemical field,
which
a self-phoretic particle
generates around itself
\cite{Tao2010,Frankel2014,Uspal2015a}.
This becomes even more complicated in 
the presence of
bounding surfaces \cite{Zottl2012a,Chilukuri2014,Uspal2015a,Palacci2015}, where
active colloids can swim stable against the flow and show \textit{rheotaxis} \cite{Uspal2015a,Palacci2015}.
Finally, deformable
active particles such as active droplets in flow also show complex swimming trajectories \cite{Tarama2013}.

\subsection{Motion near surfaces}
\label{Sec:Surface}
Recently, 
studying the motion of active colloids in the vicinity of flat or curved surfaces
came into focus
\cite{Volpe2011,Spagnolie2012,Kreuter2013,Elgeti2013,Takagi2014a,Uspal2015,Spagnolie2015,Brown2015,Schaar2015,Wysocki2015a,Elgeti2015a,Mozaffari2015,Ibrahim2015a,Das2015}.
Confinement is usually implemented in experiments by using flat walls  \cite{Volpe2011,Kreuter2013},
walls with edges \cite{Das2015},
patterned surfaces \cite{Volpe2011},
or passive colloids, which act as curved walls \cite{Takagi2014a,Brown2015}.

\begin{figure}
\includegraphics[width=\columnwidth]{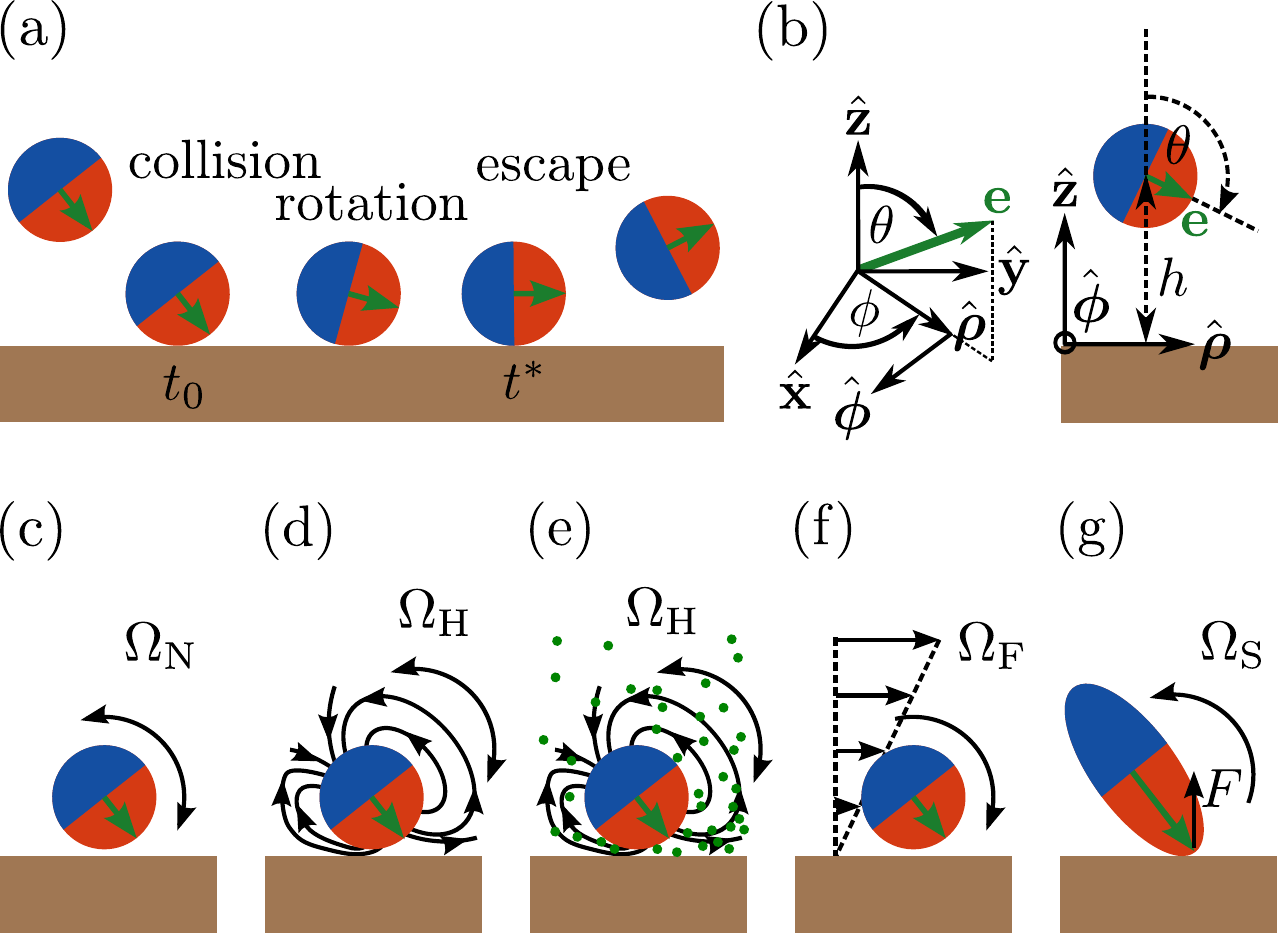}
\caption{(a) Collision of an active colloid with a surface: Collision at time $t_0$, reorientation close to the wall, 
and escape from the wall at time $t^{\ast}$.
(b) Definition of the coordinate system for an active colloid swimming in front of a wall: distance from the wall $h$,
and orientation to the wall $\theta$.  (c)-(g) Reorientation mechanisms close to the wall contributing to its angular velocity
$\Omega_{W,\phi}=\dot{\theta}$: Rotational noise $\Omega_N$ (c), 
hydrodynamic swimmer-wall interactions $\Omega_H$ (d) which can depend on the chemical field $c(\mathbf{r})$ (e),
 external fluid flow $\Omega_F$ (f), and steric interactions $\Omega_S$ (g).
}
\label{Fig:Wall}
\end{figure}

A key feature of active particles is that they accumulate 
at
surfaces even in the absence of electrostatic or other 
attractive
swimmer-wall interactions, in contrast to passive particles.
The accumulation of biological swimmers confined between two walls 
was first
observed by Sir Rothschild with living sperm cells \cite{Rothschild1963},
and later by other researchers 
using swimming bacteria \cite{Berke2008,Li2009,Molaei2014a}.
Active
particles accumulate 
at walls,
since
they need some time, after colliding with a surface,
to reorient away from the surface until they can escape 
\cite{Li2009,Enculescu2011,Elgeti2013,Schaar2015}.
This is
sketched in figure \ref{Fig:Wall}(a).

An
axisymmetric 
Brownian
microswimmer 
moves
in bulk 
along its director $\mathbf{e}$.
In the absence of any intrinsic or extrinsic torques 
acting on $\mathbf{e}$, the swimming direction only rotates due to
rotational noise $\boldsymbol{\Omega}_N = \sqrt{2D_r}\boldsymbol{\xi}_r$ and its dynamics is determined by
the Langevin equations (\ref{Eq:LangevinR}) and  (\ref{Eq:LangevinE}).
The presence of 
a
surface 
alters the dynamics of the swimmer,
which can be summarized by adding a wall-induced velocity $\mathbf{v}_W$ and a
wall-induced angular velocity $\boldsymbol{\Omega}_W$ to  
equations (\ref{Eq:LangevinR}) and  (\ref{Eq:LangevinE}), respectively.
A spherical swimmer in front of a flat wall and the relevant 
polar
coordinate system 
to express
the orientation
vector
$\mathbf{e} = e_{\rho}\hat{\boldsymbol{\rho}} + e_{\phi}\hat{\boldsymbol{\phi}} + e_z \hat{\mathbf{z}}$
are
shown in figure~\ref{Fig:Wall}(b).
Noise
alters all components of the orientation and position vectors of the swimmer 
according to the modified Langevin equations
(\ref{Eq:LangevinR}) and  (\ref{Eq:LangevinE}),
as before.
Due to the axial symmetry,
deterministic contributions to the drift velocities
$\mathbf{v}_W$  and $\boldsymbol{\Omega}_W$  only depend on the distance $h$ of the swimmer from the wall
and on the polar angle $\theta$ between the swimmer orientation
and the wall normal. 
So, one has
$\mathbf{v}_W = v_{W,\rho}(h,\theta)\hat{\boldsymbol{\rho}} + v_{W,z}(h,\theta)\hat{\mathbf{z}} $ and
$\boldsymbol{\Omega}_W = \Omega_{W,\phi}(h,\theta)\hat{\boldsymbol{\phi}}$.

The reasons for the wall-induced 
translational
and angular velocities depend on the specific properties of the swimmer, 
the wall, and the fluid.
In figure~\ref{Fig:Wall}(c)-(g) we sketch important mechanisms for active colloids 
to reorient close to
the wall 
with angular velocity
$\Omega_{W,\phi}=\dot{\theta}$.

First, rotational noise 
causes the angular velocity
$\Omega_N = \sqrt{2D_r}\xi_r +  D_r \cot\theta$,
where $\xi_r$ is again Gaussian white noise [see equation~(\ref{Eq:NoiseR2})], 
and
$D_r$ is 
the rotational diffusion constant close to the wall, which in general depends on $h$ \cite{Brenner1961}.
The effective rotational drift term
$D_r \cot\theta$ 
results  from the Stratonovich interpretation of equation (\ref{Eq:LangevinE}) with its multiplicative noise 
and ultimately appears
since $\mathbf{e}$ performs random motion on the unit sphere \cite{Berne1976,Raible2004,Schaar2015}.
The distribution of active Brownian particles confined between two parallel plates 
was
studied
by solving the corresponding 
Smoluchowski
equation \cite{Elgeti2013}
and
also in combination with gravity \cite{Enculescu2011,Wolff2013}.

Second, in order to fulfill the  
appropriate
boun\-dary condition at the wall,
the fluid flow around the microswimmer is modified compared to the bulk solutions discussed in section~\ref{Sec:LowRe}.
The resul\-ting hydrodynamic swimmer-wall interactions are res\-pon\-sible for a hydrodynamic attraction/repulsion
quantified by $\mathbf{v}_W$
and 
the
reorientation rate
$\Omega_H$ of the microswimmer in front of a slip- or no-slip wall
 \cite{Berke2008,Drescher2011b,Spagnolie2012,Schaar2015,Mathijssen2015} [see figure~\ref{Fig:Wall}(d)].
In particular, for generic pusher and pullers explicit expressions for $\mathbf{v}_W$ and $\Omega_H$ exist
based on their flow 
fields
given in equation (\ref{Eq:Dipole}).
While pushers have a stable orientation parallel to the wall, pullers are oriented perpendicular to 
bounding surfaces
\cite{Berke2008,Schaar2015}.
Hydrodynamic interactions of the
model swimmer squirmer (introduced in section~\ref{Sec:Prescribed})
with a wall 
have
recently been studied by several research groups
\cite{Llopis2010,Crowdy2010,Crowdy2011,Spagnolie2012,Zottl2012a,Zhu2013,Crowdy2013,Wang2013,Ishimoto2013,Li2014a,Papavassiliou2015,Lintuvuori2015}. 
In contrast to far-field hydrodynamic swimmer-wall interactions, we observed that
in lubrication approximation the force-dipole contribution of squirmer pullers 
to $\Omega_H$
behaves like the 
reorientation
rate induced by a generic pusher (and vice versa)\footnote{The source dipole contribution,
however, shows the same behavior 
as in far-field hydrodynamics.}.
We mentioned this fact
in Ref.~\cite{Schaar2015} and in the Supplemental Material of Ref.~\cite{Zottl2014} based on 
calculations from Ref.~\cite{Ishikawa2006}.
We note that 
bounding surfaces also alter
the concentration of chemical fields around self-phoretic swimmers
since
 the no-flux boundary condition at the wall
 has to be fulfilled.
 This 
 changes
the concentration gradient near the active colloid and 
thereby the driving
surface velocity 
field,
which 
in turn
determines the 
flow field and hence the hydrodynamic swimmer-wall interactions
\cite{Uspal2015,Mozaffari2015,Ibrahim2015a} [see figure~\ref{Fig:Wall}(e)].

Third, external fluid flow such as shear flow close to the wall induces an additional
reorientation rate $\Omega_F$
as illustrated in
figure~\ref{Fig:Wall}(f).
This rotates microswimmers
preferentially
against the flow 
resulting in a net upstream motion near  walls
in combination with 
hydrodynamic swimmer-wall interactions
 \cite{Hill2007,Nash2010,Costanzo2012,Zottl2012a,Chilukuri2014}.
Finally,
external flow 
can also alter the 
driving
surface velocity 
field of active particles
and the concentration of 
chemicals
used to propel them
\cite{Frankel2014,Uspal2015a}.

Forth, elongated swimmers 
that
collide with a bounding surface
tend to align with 
it
due to steric interactions \cite{Wensink2008,Li2009,Elgeti2009}. 
Upon collision a force acts from the wall
on the front of the particle.
The resulting
torque 
rotates the swimmer 
with an angular velocity $\Omega_S(\theta)$ 
until it is parallel to the wall
\cite{Li2009} [see figure~\ref{Fig:Wall}(g)].

In the absence of noise  stable swimming close 
to
the wall is possible, either with a 
 fixed orientation
angle $\theta_S$ 
\cite{Lauga2006,Berke2008,Uspal2015,Schaar2015}
or 
performing
(meta-)stable  oscillations in $h$-$\theta$ 
configuration space
close to the surface \cite{Or2009,Crowdy2010,Ishimoto2013,Li2014a}.
It is also possible that 
a
swimmer orients perpendicular to the wall and gets stuck there 
as noiseless generic pullers \cite{Berke2008,Schaar2015} or 
active 
Janus
particles
with large caps
 \cite{Uspal2015,Mozaffari2015} 
 would do.
However, a real
microswimmer
always experiences
thermal (and non-thermal) translational and rotational noise 
and 
therefore
can escape from the wall as sketched in figure~\ref{Fig:Wall}(a).

As a simple example we consider an active Brownian sphere with radius $R$.
Its
dynamics  in front of a wall
is governed by equations~(\ref{Eq:LangevinR}) and (\ref{Eq:LangevinE}),
where hydrodynamic and other swimmer-wall interactions are neglected.
In addition, its dynamics is restricted to 
$z\ge R$.
The particle hits the wall at an incoming 
orientation
angle $\theta_0$ at time $t_0$ [see also figure~\ref{Fig:Wall}(a)].
At the micron scale 
the P\'eclet number is often much larger than one meaning that
translational 
diffusion
is 
negligible against the active
swimming 
motion
and the active particle swims at the surface with $h(t) \approx R$.
It
can only escape from the wall when rotational noise  drives the particle's orientation to an escape angle $\theta^{\ast}=\pi/2$,
where it can swim away from the wall at time $t^{\ast}$.
The total time $t=t^{\ast}-t_0$ the active sphere stays at the surface is 
then called \textit{detention time} \cite{Schaar2015}, \textit{retention time} \cite{Wysocki2015a},
  \textit{residence time} \cite{Spagnolie2015},  \textit{escape time} \cite{Drescher2011b},  \textit{trapping time} \cite{Takagi2014a},
or \textit{contact time} \cite{Volpe2011}  and is a stochastic variable.
Its distribution $f(t|\theta_0)$, 
which we call \textit{detention time distribution} 
in
\cite{Schaar2015},
depends on the incoming angle $\theta_0$
and on the rotational diffusion constant $D_r$ near the wall. 
It
can be calculated 
using the theory of first passage times.
The mean detention time $T=\int_0^{\infty}t f(t|\theta_0) dt$ of the active Brownian particle at the surface decreases 
linearly with $D_r$ \cite{Schaar2015},
\begin{equation}
  T^{\mathrm{ABP}} = \frac{1}{D_r} \ln (1-\cos \theta_0).
\label{Eq:MFPT}
\end{equation}

\begin{figure}
\begin{center}
\includegraphics[width=0.9\columnwidth]{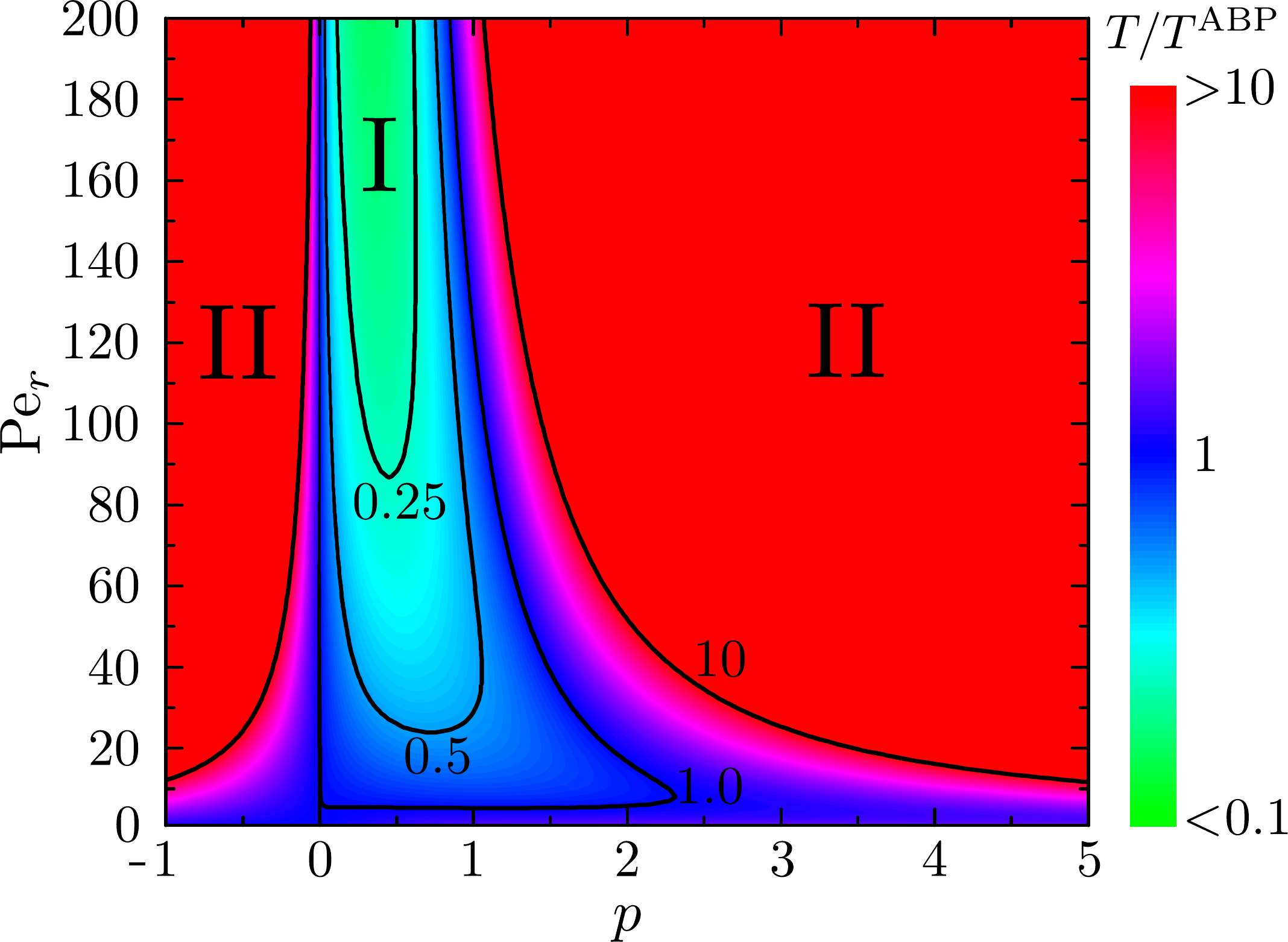}
\end{center}
\caption{Mean detention times $T/T^{\mathrm{ABP}}$ for pushers ($p>0$) and pullers ($p<0$) depending on the force 
dipole strength $p$ and the persistence number $\mathrm{Pe}_r$.
The incoming angle is
$\theta_0=3\pi/4$.
In region I $T/T^{\mathrm{ABP}}<1$, wheras in region II $T/T^{\mathrm{ABP}}\gg 1$.
(Adapted from Ref.~\cite{Schaar2015}; copyright (2015) American Physical Society). 
}
\label{Fig:WallPushPull}
\end{figure}

Simple model pushers and pullers are hydrodynamically trapped 
at
surfaces but can  escape through rotational noise.
Their mean detention times $T(p,\textrm{Pe}_r)$ compared to $T^{\mathrm{ABP}}$ are shown in figure~\ref{Fig:WallPushPull}.
For large persistence numbers $\mathrm{Pe}_r$ and sufficiently large force dipole strength $|p|$ (region II in 
figure~\ref{Fig:WallPushPull}),
the
escape from the surface can be mapped onto the escape of a particle over a large
potential barrier $\Delta U$ (in our case the hydrodynamic torque potential)
and the mean detention times 
are
approximated by Kramers-like formulas,
$T\sim e^{\Delta U / D_r}$ \cite{Drescher2011b,Schaar2015}.
Interestingly, for sufficiently small dipole strength $p$ a pusher 
escapes
more quickly
from a surface than an active Brownian particle \cite{Schaar2015},
as indicated by region I of figure~\ref{Fig:WallPushPull}. 
This is due to the wall-induced reorientation rate.

\subsection{Motion in complex environments}
\label{Sec:Complex}
In section~\ref{Sec:Surface} we discussed the interaction of an active colloid with a flat surface
to reveal the
basic physics of swimmer-wall interactions.
However, active colloids 
may also
move in more complex environments,
with curved bounding surfaces, 
for example,
in the presence of obstacles.

Recent experiments studied the dynamics of elongated and spherical
active colloids moving in the vicinity of passive spherical colloids \cite{Takagi2014a,Brown2015,Kummel2015}.
Elongated rods get hydrodynamically trapped 
by
spherical colloids,
orbit around the colloid,
and 
escape with the help of rotational diffusion \cite{Takagi2014a,Spagnolie2015,Wysocki2015a}.
An active colloid moving in a dense hexagonally packed monolayer of spherical colloids
also shows this orbiting around larger passive colloids
\cite{Brown2015}.
Interestingly, the 
orbiting
speed oscillates periodically due to the presence of the
six nearest neighbors and the orbiting phase can be used to 
analyze the hydrodynamic flow field created by the swimmer \cite{Brown2015}.
Active colloids can also merge and compress colloidal clusters or even locally melt colloidal crystals \cite{Kummel2015}.
Active colloids moving in circular confining
geometries 
also aggregate
at curved interfaces \cite{Volpe2011}
and the curvature radius
compared to the swimmer persistence length determines the 
density distribution
at the bounding surface
\cite{Fily2014b,Fily2015,Wysocki2015a}.

Wedge-like obstacles 
are
used to rectify the transport of biological microswimmers \cite{Galajda2007,Wan2008} or to
 trap active particles  \cite{Kaiser2012}.
Periodic
arrays of spherical or ellipsoidal obstacles 
enhance
the directed swimming
of active colloids as demonstrated experimentally \cite{Volpe2011,Brown2015}.
Finally, in theory
the rectification of active particles with different ratchet mechanisms 
was
demonstrated
\cite{Ghosh2013,Pototsky2013}.

\section{Collective Dynamics}
\label{Sec:Collective}
Up to now we characterized the dynamics of non\-interacting active colloids under various conditions 
(see chapter~\ref{Sec:Single}). In the following we review 
their emergent collective dynamics,
when they interact.

In biological active matter 
a 
large
variety of emergent patterns has been reported.
For example, \textit{bacterial turbulence} 
was
found in 
concentrated 
suspensions
of swimming bacteria \cite{Dombrowski2004,Wensink2012b,Aranson2013b}.
Other intriguing examples from biology are polar patterns in dense suspensions
of actin filaments driven by molecular motors \cite{Schaller2010} or
flowing active nematics formed by concentrated microtubules together with kinesin motors  \cite{Sanchez2012}.
Also,
swimming sperm cells can form
self-organized vortex arrays 
\cite{Riedel2005}
as do
 collectively moving microtubules \cite{Sumino2012}.
The collective migration of cells on a substrate 
depends on physical forces and stresses \cite{Trepat2009}.
Populations of swimming bacteria are able to phase-separate 
\cite{Schwarz-Linek2012,Chen2015c} or form \textit{active crystals} \cite{Petroff2015}.

Recently, 
the collective motion of artificial microswimmers such as active colloids has been studied, both experimentally and theoretically.
These
systems are very attractive for two reasons.
First, 
they 
can help to understand the main underlying physical 
principles governing
the collective motion of active microscopic 
individuals, which share
similar physical 
properties, namely,
self-propulsion, rotational diffusion, low Reynolds number hydrodynamics, 
as well as
short-range and 
phoretic
interactions
\cite{Poon2013a}.
Second, the 
self-assembly of 
interacting
active colloids 
offers
the possibility of 
forming
a new class of \textit{active materials}
with novel and tunable properties
\cite{Wang2015,Boekhoven2015}.






\subsection{Experimental observations of collective motion}
\label{Sec:Collective:Exp}

\subsubsection{Bound states and self-assembly of active colloids}
Active particles
in a (semi-)dilute suspension collide much more frequently compared to passive particles
with a rate that
increases linearly with colloidal density
and, in particular, with the P\'eclet number
\cite{Tailleur2008,
Redner2013b,
Stenhammar2013,Bialke2013}.
Therefore, 
self-propulsion, also in combination with
swimmer shape and 
interactions, can 
generate
self-assembled bound 
states of active colloids,
which autonomously translate and rotate 
depending on the 
specific 
shape of the
assembled 
clusters
\cite{Wang2009,Ebbens2010b,Wang2013b,Gao2013a,Solovev2013a,Wykes2015}.
Interestingly, recent theoretical 
works indeed
suggest the importance of swimmer shape, surface chemistry, and hydrodynamic interactions
for
the structures 
formed by
self-assembled active colloids \cite{Thakur2010,Soto2014a,Soto2015,Pandey2014,Sharifi-Mood2015,Bayati2015}.

\begin{figure*}
\includegraphics[width=\textwidth]{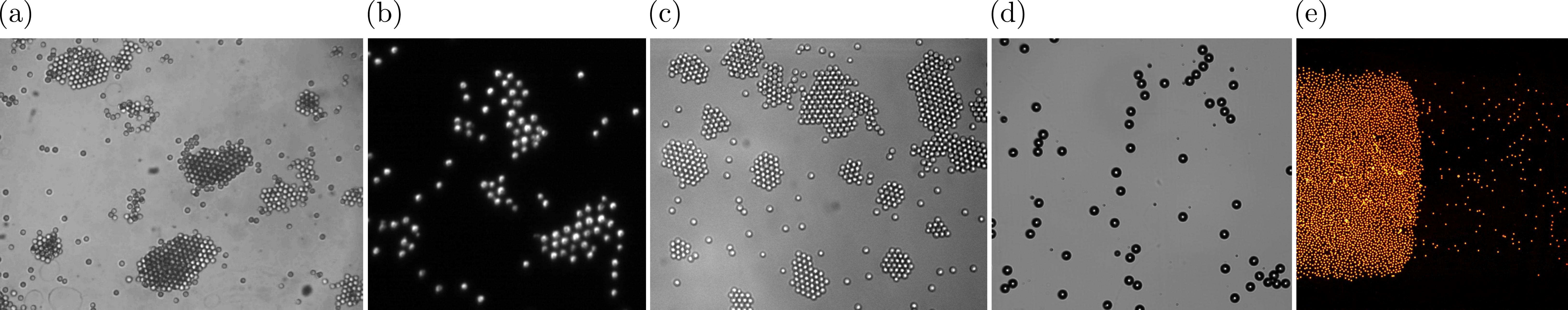}
\caption{Experiments on the collective dynamics of spherical active colloids.
(a) 
Phase separation of a suspension of 
active
Janus particles 
activated by locally demixing a critical binary fluid \cite{Buttinoni2013e}
(adapted with permission from Ref.~\cite{Buttinoni2013e};
copyright (2013)
American Physical Society).
(b) Formation of dynamic clusters of self-phoretic active colloids, which interact via chemotaxis \cite{Theurkauff2012} 
(courtesy of C\'ecile Cottin-Bizonne).
(c) Formation of living crystals 
by
light-activated colloidal surfers consisting of hematide cube attached to a polymer sphere 
\cite{Palacci2013} (courtesy of J\'er\'emie Palacci).
(d) Collective motion of a monolayer of active emulsion droplets \cite{Thutupalli2011} (courtesy of Shashi Thutupalli).
(e) Formation of a band of colloidal Quincke rollers in a racetrack geometry due to hydrodynamic and electrostatic interactions
\cite{Bricard2013} 
(courtesy of Denis Bartolo).}
\label{Fig:Experiments}
\end{figure*}


\subsubsection{Dynamic clustering and phase separation}
\label{Sec:ExpMIPS}
Passive hard-sphere colloidal systems phase-separate at relatively large densities and in a narrow density region 
into a fluid and crystalline phase favored by an increase in entropy\ \cite{Anderson2002}.
However, if one turns on activity in active particles,
phase-separation can already occur
at lower densities.
%
%
%
This so-called \textit{motility-induced phase separation} 
was first
discussed in the context of run-and-tumble bacteria \cite{Tailleur2008}
but seems to be generic for 
active particles, 
which slow down in the presence of other particles
\cite{Fily2012,Redner2013b,Cates2013,Cates2015}.
Typically, they
phase-separate into a 
gaslike and a fluid-/solidlike state at sufficiently high density and swimming velocity.
This
happens even in the absence of any aligning mechanism \cite{Fily2012,Redner2013b}.

Simple model systems to experimentally study the influence of activity on the phase behavior of self-propelled particles
are active colloidal suspensions
confined to a
monolayer,
which is
either sandwiched between two 
bounding
plates 
\cite{Thutupalli2011,Buttinoni2013e}
or sedimented 
on
a substrate
 \cite{Theurkauff2012,Palacci2013,Bricard2013}.
These
quasi-two-dimensional
systems 
allow
a relatively easy tracking of particle positions and 
thereby the
monitoring 
of particle dynamics.

Probably the most simple experimental system 
showing motility-induced phase separation 
was introduced
by Buttinoni \etal 
using light-activated spherical 
Janus particles 
that swim in a binary fluid close to the critical demixing transition
(see also section~\ref{Sec:ExpReal}).
Tuning
the laser intensity, the swimming speed -- and therfore the P\'eclet number -- can be adjusted.
Since phoretic and electrostatic interactions between colloids seem to be negligible
in these experiments,
the 
Janus particles serve as a simple experimental model for active hard spheres swimming in a low Reynolds number fluid. 
At
sufficiently low  P\'eclet number and areal density, the
particles 
do
not form clusters but rather 
behave
as an \textit{active gas}.
However,
for higher particle density and swimming speed, 
densely packed clusters 
emerge, which
coexist 
with a gas of active particles (see figure~\ref{Fig:Experiments}(a)).
Increasing
the  P\'eclet number, the size of these clusters grow, as predicted earlier by simulations and 
theory
(see chapter~\ref{Sec:Collective:Simu} and \ref{Sec:Collective:Phys}).

Very dynamic particle clusters form in the experiments of Theurkauff \etal with
self-phoretic spherical colloids
half-coated with platinum,
which 
become active when
adding $\textrm{H}_2\textrm{O}_2$ \cite{Theurkauff2012}  [see figure~\ref{Fig:Experiments}(b)].
Effective particle interactions exist, both attractive and repulsive, which in combination
can produce pronounced
dynamic clustering already at very low 
area fractions of a few percent
\cite{Theurkauff2012,Pohl2014,Pohl2015}.
%
%
These 
effective
forces arise from diffusiophoresis, which the colloids experience in non-uniform chemical fields 
produced by their neighbors while consuming $\textrm{H}_2\textrm{O}_2$
(see also chapter~\ref{Sec:Taxis}).
Therefore, such artificial systems can mimic chemotactic processes found in biological systems.
Aggregation of chemotactic active colloids 
was also reported in
further
experiments~\cite{Hong2007a,Sen2009c,Baraban2013a,Palacci2013,Xu2015a}.

Most recently,
the sedimentation profile of interacting self-phoretic 
colloids under gravity 
was
studied in detail \cite{Ginot2015}.
Here, far away from the bottom surface the density of swimmers is very 
small
and shows an exponential decay as described in section~\ref{Sec:Gravity}.
Closer to the bottom clusters are formed due to phoretic interactions at semi-dilute particle 
suspensions.
The cluster formation could be mapped to an adhesion process of a corresponding equilibrium system.

Palacci \etal 
investigated
the collective motion of \textit{colloidal surfers} (see section~\ref{Sec:RollSurf}) on a substrate.
Here again, the combination of self-propulsion and steric and chemical interactions triggers the formation of 
clusters \cite{Palacci2013},
which
show a well-defined crystalline structure  [see figure~\ref{Fig:Experiments}(c)].
These \textit{living crystals} are highly dynamic; they form, rearragange, and break up quickly.

All of these systems show the emergence of \textit{positional order} 
through
the formation of clusters, but no significant \textit{orientational order} 
in the swimming direction
has been reported.
Nevertheless, even for spherical active particles
\textit{polar order} 
can emerge
in the presence of hydrodynamic, electrostatic, or other interactions between nearby particles.
They lead
to a local alignment of swimmer orientations, as we report in the following section.

\subsubsection{Swarming and polarization}
\label{Sec:Swarm}
Local mechanisms for aligning 
active particles give rise to new collective phenomena.
We summarize them
here.

Quite surprisingly, even 
for spherical active 
col\-loids,
where an aligning mechanism is not obvious,
polar order 
was
reported in a dense suspension of swarming active emulsion droplets \cite{Thutupalli2011}
 [see figure~\ref{Fig:Experiments}(d)].
The 
observed
large-scale structures and swirls show some behavior reminiscent of biological swarms \cite{Vicsek2012}.
Hydrodynamic interactions between active droplets,
due to the flow fields they create,
are expected to play an important role.
However,
their
details
are not 
fully clear.
Marangoni flow at the 
droplet surfaces
cause
hydrodynamic flow fields 
in the bulk fluid
(see section~\ref{Sec:SwimTheory}),
but this Marangoni flow
might be 
modified by the presence of other active droplets.
It
remains 
to be investigated
how this effects 
the collective
dynamics. 
Noteworthy, the system reported in  Ref.~\cite{Thutupalli2011} is strongly confined 
in the plane
by  a curved petri dish, which seems to 
be important
for observing
the 
swarming
dynamics.
Similar behavior was noted
for vibrated 
granular matter \cite{Narayan2007,Kudrolli2008,Deseigne2010}. 
Finally, swimming liquid crystal droplets form crystalline rafts that float
above the substrate
\cite{Herminghaus2014}.

Bricard \etal studied the collective motion of Quincke rollers (see section~\ref{Sec:RollSurf}) in a racetrack geometry \cite{Bricard2013}.
Here, both electrostatic and hydrodynamic interactions between the rollers determine their collective motion.
For constant
strength of the applied 
electric field, 
the collective motion can be tuned by modifying the area fraction of the particles.
While at sufficiently low densities the particles form an apolar active gas,
at a critical density 
propagating polar bands 
emerge
[see figure~\ref{Fig:Experiments}(e)].
At even higher densities a polar liquid 
is stabilized,
where all particles move in the same direction.
Interestingly, when the 
confining
geometry is changed to a square or to a circular disc, 
a single macroscopic vortex 
forms
\cite{Bricard2013,Bricard2015}.
Similar vortices 
occur
in circularly confined 
bacterial suspensions \cite{Wioland2013,Lushi2014}.

In a very recent work,
Nishiguchi and Sano observed \textit{active turbulence} in a monolayer of swimming spherical 
colloids \cite{Nishiguchi2015}. Here, the Janus colloids 
sediment on
a substrate and
start to
swim 
when an
AC electric field
is applied.
While both hydrodynamic and electrostatic interactions 
seem
to play a role 
for generating turbulence,
the detailed mechanisms
are not well understood yet.
Noteworthy, up to now active turbulent states have only 
been observed for non-spherical active particles.

\subsection{Modeling and analysing the collective motion of active colloidal suspensions}
\label{Sec:Collective:Simu}

\subsubsection{Active Brownian particles}
The most basic realization 
for studying
interacting active colloids
in theory and simulations are
active Brownian particles (ABPs) (see also section~\ref{Sec:ABP}).
The equations of motion for a single, free ABP are given in equations~(\ref{Eq:LangevinR}) and (\ref{Eq:LangevinE}), 
and the solution is a persistent random walk (see section~\ref{Sec:Walk}).


The 
dynamics of interacting
ABPs can be implemented in 
different ways
using Brownian dynamics or even kinetic Monte Carlo  simulations \cite{Berthier2014}.
To account for the fact that particles cannot 
interpenetrate
each other 
due to steric repulsion,
hard or soft potentials between the ABPs are typically used.
For example, the Weeks-Chandeler-Anderson (WCA) potential $V(r)$ is 
employed
frequently to model 
soft-core or (almost) hard-core
repulsion between two 
particles.
It is a 
Lennard-Jones potential acting between spherical particles of radius $R$ and 
cut off at distance $d^\ast$, where the Lennard-Jones Potential has its minimum
\cite{Weeks1971},
\begin{equation}
 V(r) = 
\left\{ \begin{array}{rl}
 4\epsilon \left[ {\left( \frac{\sigma}{r} \right)}^{12} - {\left( \frac{\sigma}{r}\right)}^{6} \right]  + \epsilon,
  & \textnormal{for } r < d^\ast \\
0, & \textnormal{for } r \ge d^\ast.  \end{array} \right. 
\label{Eq:WCA}
\end{equation}
Here, $d^\ast = 2^{1/6}\sigma$, $\epsilon$ is the potential strength, 
and  $\sigma=2R/2^{1/6}$
is chosen
such that the interaction force $\mathbf{F} = -\boldsymbol{\nabla} V$ 
becomes non-zero at $d^\ast < 2R$, \textit{i.e.},
when two ABPs overlap.
A simple
alternative 
to implement hard-core interaction
is the following \cite{Palacci2013,Pohl2014}:
whenever
particles overlap during a simulation, one separates them along the line connecting their centers.

Then, in
a system consisting of $N$ active particles 
the equation of motion for the $i$-th particle 
reads
\begin{equation}
 \dot{\mathbf{r}}^{(i)} = v_0\mathbf{e}^{(i)} + \mu  \sum_{j \ne i}\mathbf{F}^{(ij)}  + \sqrt{2D}\boldsymbol{\xi} \, ,
\label{Eq:IA}
\end{equation}
where the index $j$ runs over all other $N-1$ particles, 
$\mathbf{F}^{(ij)}$
is the interaction force 
from particle $j$ on $i$,
and $\mu=(6\pi\eta R)^{-1}$ is the mobility coefficient of the ABP.
Together with the equation for the stochastic reorientation of the particle orientations $\mathbf{e}^{(i)}$ [see equation~(\ref{Eq:LangevinE})]
equation~(\ref{Eq:IA}) is solved, e.g., by Brownian dynamics simulations.
Additional attractive and/or repulsive
interactions may be included  \cite{Redner2013,Mognetti2013}.

\begin{figure}
\begin{center}
\includegraphics[width=0.85\columnwidth]{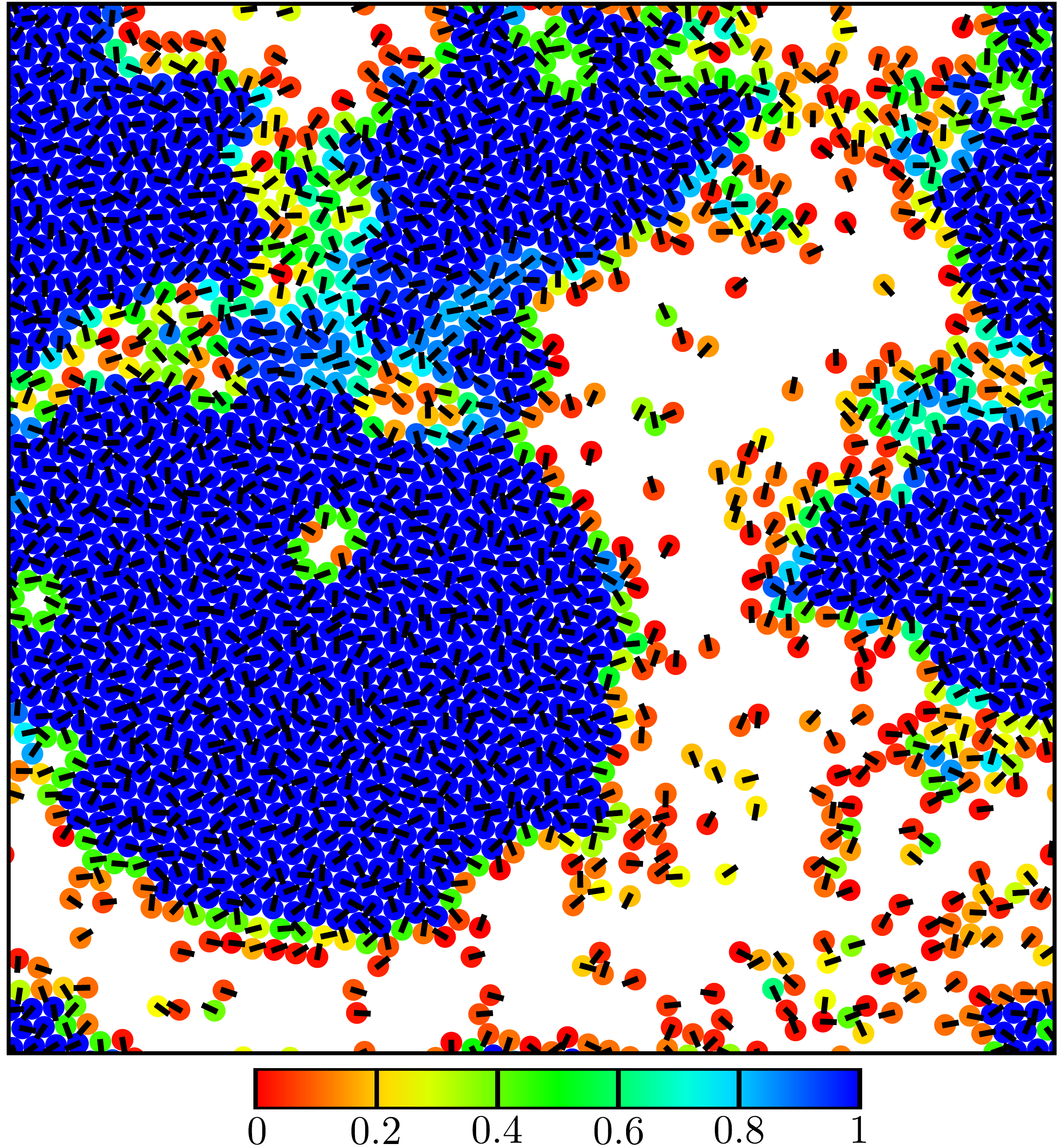}
\end{center}
\caption{Snapshot of a phase-separated suspension of 
active Brownian disks (area fraction $\phi=0.64$, P\'eclet number $\textrm{Pe}=360$).
The (almost) hard-core repulsion was implemented by using a WCA potential [equation~(\ref{Eq:WCA})] with $\epsilon=1000$.
The color indicates the local 6-fold bond-orientational order 
measured by
$|q_6|^2$ 
from
equation~(\ref{Eq:q6}).
 }
\label{Fig:BDsnap}
\end{figure}

While elongated ABPs tend to align with neighbors after steric collisions \cite{Peruani2006,Wensink2008,Baskaran2008},
active Brownian spheres and disks lack any 
intrinsic
aligning mechanism.
Nevertheless, different mechanisms
such as aligning, phoretic, 
or
hydrodynamic
interactions
alter the particle orientations $\mathbf{e}^{(i)}$ 
and
can be included into
equation~(\ref{Eq:LangevinE}) for the orientation 
vector
(see, for example, \cite{Szabo2006,Henkes2011a,Palacci2013,Pohl2014,Hennes2014,VanDrongelen2015}).

In theory and simulations,
the 
simplest
system 
for studying
collective motion of active Brownian particles consists of 
active Brownian disks, which interact only via hard-core repulsion  in two dimensions (2D).
This minimal model has 
frequently
been used 
to 
investigate
the collective dynamics of active particles 
\cite{Fily2012,Bialke2012,Redner2013b,Buttinoni2013e,Redner2013,Stenhammar2013,Bialke2013,Zottl2014,Stenhammar2014,Yang2014e,Mallory2014,Takatori2014,Speck2014a,Levis2014a,Farage2015}.
Bialk\'e \etal \cite{Bialke2012}, Fily \etal \cite{Fily2012}, and Redner \etal \cite{Redner2013b}
were the first
to explore dynamic structure formation within this model.
The two relevant parameters in the system are the areal density $\phi$ and the P\'eclet number $\textrm{Pe}$.
The latter
is linear in
the persistence number $\textrm{Pe}_r$
for pure thermal noise
[see equation~(\ref{Eq:DDr})].
While at low densities the ABPs form an active gas, 
they start to phase-separate into a gaslike and a crystalline phase at $\phi \gtrsim 0.3$ and sufficiently large $\textrm{Pe}$ 
 \cite{Fily2012,Redner2013b,Buttinoni2013e,Redner2013,Stenhammar2013,Bialke2013,Zottl2014,Stenhammar2014,Bialke2014}.
Figure~\ref{Fig:BDsnap} shows a
typical snapshot of the  phase-separated state for $\phi=0.64$ and $\textrm{Pe}=330$.
The crystalline structure is not perfect but 
also contains
defects.
For sufficiently small P\'eclet numbers 
the dense cluser phase is 
rather
liquidlike \cite{Bialke2012,Redner2013b}.

In
three dimensions (3D) the system also phase-separates.
However,
the cluster phase does not have crystalline order but is 
rather
fluidlike and the local density can reach the random-close-packing limit \cite{Wysocki2014a}.
Furthermore,
the coarsening dynamics of the clusters
clearly differ in 2D and 3D.
While in 3D the 
mean
domain size grows as $\sim t^{1/3}$ in time  \cite{Stenhammar2014},
similar to equilibrium coarsening dynamics, 
a cluster coarsens more slowly
in 2D
according to
$\sim t^{0.28}$ \cite{Redner2013,Stenhammar2013,Stenhammar2014}.

We note that the minimal model can be extended in 
several
ways leading to new emergent 
collective
behavior.
For example, introducing 
polydispersity in
the ABPs, results in active glassy behavior \cite{Berthier2013a,Ni2013a,Berthier2014,Fily2014,Farage2014,Szamel2015}.
Additional attractive interparticle forces lead to gel-like structures \cite{Redner2013} or 
very dynamic crystalline clusters
at low densities \cite{Palacci2013,Mognetti2013,Pohl2014}.
The collective motion of
self-propelled 
Brownian rods 
was
studied extensively
\cite{Peruani2006,Wensink2008,Baskaran2008,Baskaran2008a,Kudrolli2008,Yang2010,Ginelli2010,Kudrolli2010,McCandlish2012,Wensink2012}.
Due to the local alignment of the rods, one observes the formation of dynamic swarming clusters
\cite{Peruani2006}, 
moving bands \cite{Ginelli2010}, 
and even
turbulent states \cite{Wensink2012}.
Finally,
the self-assembly of 
active particles with more complex shapes 
was investigated
\cite{Wensink2014,Nguyen2014}.

To summarize,
the simple model of ABPs is
able to qualitativly reproduce important aspects of the observed emergent behavior in active colloids such as 
motility-induced phase separation (see section~\ref{Sec:ExpMIPS}) by only accounting for self-propulsion and steric 
hindrance.
However,
it neglects the effect of 
flow fields 
generated by
active colloids and 
does not include
phoretic interactions, which we will discuss in sections~\ref{Sec:CollectiveHI} and \ref{Sec:CollectiveChemo}.

\subsubsection{Microswimmers with hydrodynamic interactions}
\label{Sec:CollectiveHI}
As we have discussd in section~\ref{Sec:HowSwim}, active colloids moving in a Newtonian fluid create a flow field 
around 
themselves.
In the following, 
we discuss 
how these
flow fields 
determine
the collective dynamics of microswimmers.

\paragraph{Hydrodynamic interactions between active colloids}
We shortly introduce here the basic principles of hydrodynamic interactions between microswimmers 
(see also Refs.~\cite{Ishikawa2006,Pooley2007,Lauga2009a,Ishikawa2009,Poon2013a,Hennes2014,Elgeti2015b}).
Active colloids at positions $\mathbf{r}_i$ and swimming in bulk with velocities $\mathbf{U}_i=v_0\mathbf{e}_i$
create undisturbed flow fields $\mathbf{v}_i(\mathbf{r};\mathbf{r}_i,\mathbf{e}_i)$ around themselves, when they are
far apart from each other. The flow fields depend on the swimmer type, as discussed in section~\ref{Sec:HowSwim}.
In leading order of hydrodynamic interactions, each microswimmer is advected by the undisturbed flow fields from their
neighbors and the colloidal velocity becomes $\mathbf{U}_i' = \mathbf{U}_i + \mathbf{U}_i^{\mathrm{HI}}(\{
\mathbf{r}_j,\mathbf{e}_j \})$ with $\mathbf{U}_i^{\mathrm{HI}} = \sum_{j \ne i} \mathbf{v}_j(\mathbf{r}_i;\mathbf{r}_j,\mathbf{e}_j)$.
In addition, the vorticities of the undisturbed flow fields add up to determine the angular velocities 
$\boldsymbol{\Omega}_i' = \boldsymbol{\Omega}_i^{\mathrm{HI}}(\{\mathbf{r}_j,\mathbf{e}_j\})$ in leading order.
When particles come closer together, the undisturbed flow fields do not satisfy the no-slip boundary condition at the
surfaces of neighboring particles. They have to be modified and thereby the hydrodynamic contributions,
$\mathbf{U}_i^{\mathrm{HI}}(\{\mathbf{r}_j,\mathbf{e}_j \})$ and 
$\boldsymbol{\Omega}_i^{\mathrm{HI}}(\{\mathbf{r}_j,\mathbf{e}_j\})$,
to the colloidal velocities change.

Still in
the dilute limit, where the distances between the swimmers are much larger than their radii,
one can improve on the
\textit{far-field hydrodynamic interactions}
using Fax\'en's law \cite{Dhont1996}.
For example, for spherical particles of radius $R$
the hydrodynamic contributions $\mathbf{U}_i^{\mathrm{HI}}$ and $\boldsymbol{\Omega}_i^{\mathrm{HI}}$ to
the colloidal velocities
read 
\begin{eqnarray}
\label{Eq:HIU}
\mathbf{U}_i^{\mathrm{HI}} &= \sum_{j \ne i} 
\left(1  + \frac{R^2}{6} 
\nabla_i^2 
\right) \mathbf{v}_j(\mathbf{r}_i;\mathbf{r}_j,\mathbf{e}_j), \\
\label{Eq:HIOm}
\boldsymbol{\Omega}_i^{\mathrm{HI}} &= \frac 1 2 
\boldsymbol{\nabla}_i
\times \sum_{j \ne i} \mathbf{v}_j(\mathbf{r}_i;\mathbf{r}_j,\mathbf{e}_j) \,
\end{eqnarray}
where  $\mathbf{v}_j(\mathbf{r}_i;\mathbf{r}_j,\mathbf{e}_j)$ is
again
the undisturbed flow field created by swimmer $j$
and evaluated at the position $\mathbf{r}_i$ of swimmer $i$.
Similar expressions hold for ellipsoidal particles \cite{Happel2012,Kim2013}.

In more dense suspensions of active colloids the 
approximation 
of
far-field hydrodynamics
[equations~(\ref{Eq:HIU}) and (\ref{Eq:HIOm})] is not valid any more.
In contrast, $\mathbf{U}_i^{\mathrm{HI}} $ and $\boldsymbol{\Omega}_i^{\mathrm{HI}}$ are 
mainly determined by \textit{near-field hydrodynamic interactions}.
Their
calculation is 
much more complicated and strongly depends on the specific swimmer model.
Hence, they usually have to be determined via hydrodynamic simulations, which capture the hydrodynamic near fields correctly.
For
squirmers lubrication theory can be applied to calculate $\mathbf{U}_i^{\mathrm{HI}} $ and $\boldsymbol{\Omega}_i^{\mathrm{HI}}$,
but it only holds for interparticle distances 
$d \ll R$
\cite{Ishikawa2006}.

\paragraph{Collective motion of squirmers}
The squirmer model introduced in section~\ref{Sec:Prescribed} is a simple model to study 
how
hydrodynamic interactions 
influence the collective motion of
active colloids.
Ishikawa and Pedley were the first 
to use
the 
boundary element method
and Stokesian 
dynamics simulations 
for investigating
the
collective motion
of squirmers
in bulk fluids 
\cite{Ishikawa2007b,Ishikawa2007,Ishikawa2007a,Ishikawa2008a,Ishikawa2008,Ishikawa2010,Ishikawa2012,Ishikawa2014}.
Typically, in
a suspension of squirmers the reorientation rates $\boldsymbol{\Omega}_i^{\mathrm{HI}}$ 
chaotically evolve
in time and hydrodynamic interactions 
thus modify rotational diffusion and an increased effective diffusion constant
$D_r^{\mathrm{HI}}$
results
\cite{Ishikawa2007b,Molina2013}.
For 
squirmer
pullers polar order can emerge\cite{Ishikawa2008a}, which 
was
also 
quantified further in 
Refs.~\cite{Evans2011,Alarcon2013a,Delmotte2015}.
This is in contrast to hydrodynamic simulations of self-propelled rods, which show local polar order for 
generic pushers but not for pullers (see 
the last paragraph of this section).
The reason for the 
different behavior of
squirmer and active rod suspensions might be due to different 
types of
near-field hydrodynamic interactions
as dicussed in section~\ref{Sec:Surface}.


Noteworthy, in large suspensions of  squirmer pullers 
temporal
density 
variations
 emerge, where a large cluster 
periodically 
forms and breaks
apart \cite{Alarcon2013a}.
The 
time
correlation function of these density fluctuations show oscillatory behavior with a well-defined frequency.
Recently, also the 
dynamics of 
many
squirmers confined between two hard walls has been studied \cite{Lambert2013,Pagonabarraga2013a,Zottl2014,Li2014a,Oyama2015}.
For a separation distance
much larger 
than
the swimmer size,
again a huge dynamically evolving cluster emerges.
It
travels between the walls and has been interpreted as a propagating sound wave
\cite{Oyama2015}.

Recent investigations 
studied
the
collective motion of squirmers moving 
either
in 2D \cite{Aguillon2012,MatasNavarro2014,MatasNavarro2015} or in quasi-2D \cite{Zottl2014} 
in order to 
reveal the influence of
hydrodynamic interactions
on the dynamics of
active 
colloidal
suspensions.
The quasi-2D geometry constrained the squirmers to a
monolayer similar 
to
experiments
in Ref.~\cite{Buttinoni2013e} (see section~\ref{Sec:Collective:Exp}).
2D simulations showed that long-range hydrodynamic interactions 
result in
strong reorientation rates $\boldsymbol{\Omega}_i^{\mathrm{HI}}$ 
that
are sufficient to entirely suppress motility-induced phase separation of squirmers \cite{MatasNavarro2014}.
Simpler
non-squirming  swimmers simulated with the 
lattice-Boltzmann
method 
in 2D
\cite{Ramachandran2006,Llopis2006} 
and with Stokesian 
dynamics in a monolayer in 3D
\cite{Mehandia2007} 
showed
some clustering.

\begin{figure}
\includegraphics[width=\columnwidth]{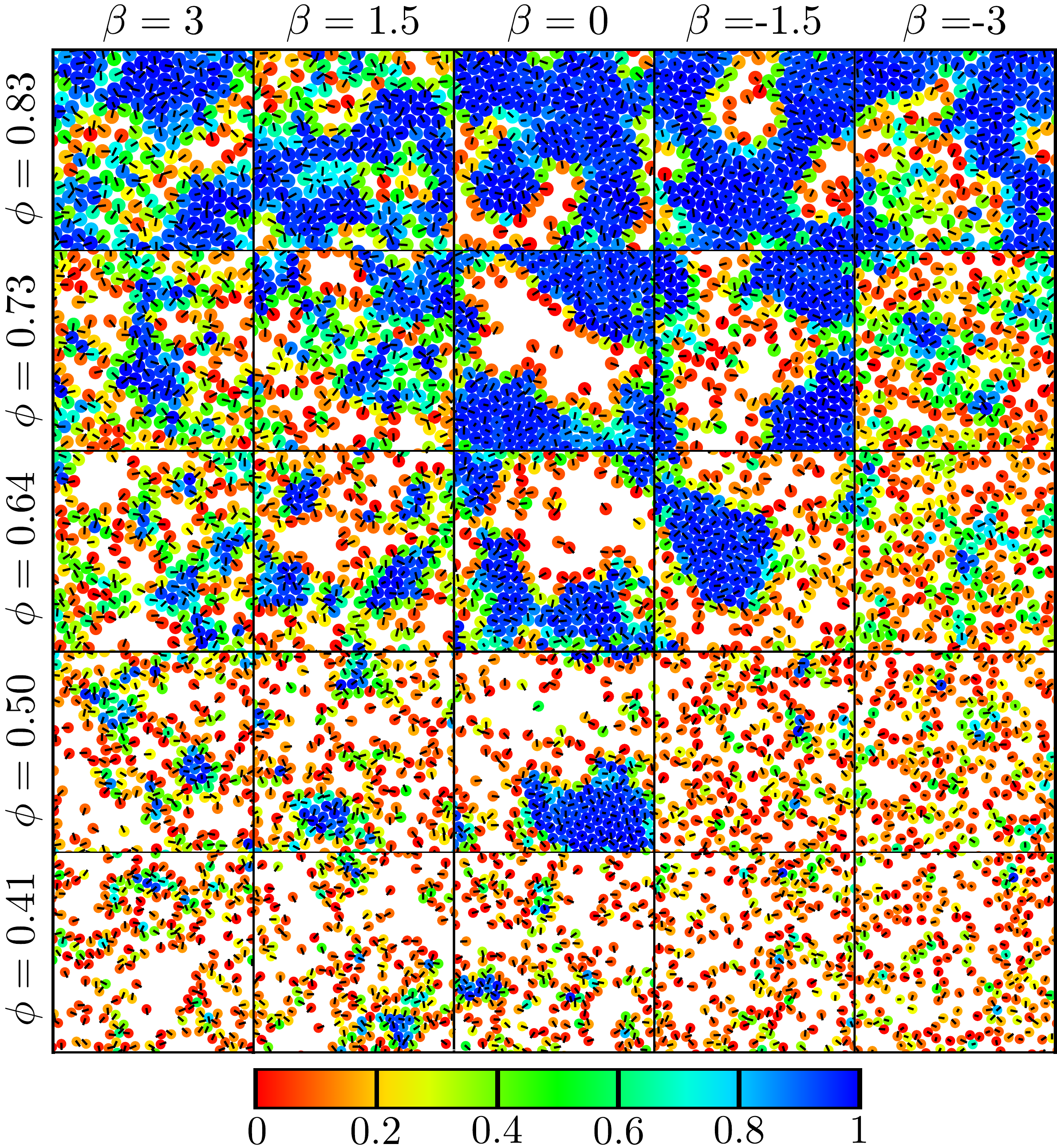}
\caption{Collective motion of squirmers in a quasi-2D geometry  depending on the 
area fraction $\phi$
and 
squirmer parameter $\beta$.
The color indicates the local 
6-fold bond-orientational order 
measured by
$|q_6|^2$ 
from equation~(\ref{Eq:q6}).
(Adapted from Ref.~\cite{Zottl2014};
copyright (2014) American Physical Society.)
}
\label{Fig:CollectiveSquirmer}
\end{figure}

We 
performed three-dimensional
MPCD simulations of squirmers 
and
strongly confined
them
between two parallel plates,
such that they could only move in a monolayer (quasi-2D geometry)
\cite{Zottl2014}.
The simulations were based on Refs.~\cite{Downton2009a,Goetze2010,Zottl2012a}, where
single squirmers were implemented with MPCD.
Our results showed
that 
the
phase behavior 
of squirmers
strongly depends on the swimmer type, characterized by the squirmer 
parameter $\beta$, and on areal density $\phi$  (see figure~\ref{Fig:CollectiveSquirmer}).
While neutral squirmers
($\beta = 0$)
and weak pushers 
($\beta < 0$)
phase-separate at a sufficiently high density,
pullers 
($\beta > 0$)
only form small 
and
short-lived clusters.
Strong pushers do not 
cluster at all and only develop one crystalline region at high areal densities.
They
tend to point 
with their swimming directions
perpendicular to the 
bounding
walls, which significantly reduces their 
in-plane persistent motion so that clustering does not occur.
Pullers are more oriented parallel to the walls. 
But
their rotational diffusivity is 
strongly
enhanced
so that the persistent motion is again too small to exhibit
phase separation.
In contrast, neutral squirmers and weak pushers also swim parallel to the walls and their hydrodynamic rotational diffusion $D_r^{\mathrm{HI}}$
is sufficiently small to 
allow
stable clusters
to form and hence 
they
phase-separate.
We
do not observe polar order in any of the studied systems.
This
could be due to the presence of 
the walls or
 thermal noise, which 
was
absent in all the other 
simulations on the collective dynamics of squirmers.
We currently perform large-scale simulations, which confirm all these findings and clearly demonstrate
the first-order phase transition associated with phase separation.

\paragraph{Collective motion of elongated microswimmers and actively spinning particles}
At present experiments have focussed on the emergent behavior of spherical active colloids, as described in sections~\ref{Sec:ExpMIPS} and \ref{Sec:Swarm}.
Nevertheless, we expect that in the near future novel experiments on the collective motion of self-propelled rod-shaped colloids
will be performed.

Aditi Simha and Ramaswamy were the first to study the
role of long-range hydrodynamic interactions 
in
the collective motion of 
swimming force dipoles
with polar or nematic order
using continuum field 
theory
 \cite{AditiSimha2002}.
The collective motion of hydrodynamically interacting 
active dumbbells, which are modeled as pairs of point forces,
was addressed
in Refs.~\cite{Hernandez-Ortiz2005,Underhill2008}
and later in Ref.~\cite{Hinz2015} using 
dissipative particle dynamics.
More refined, explict simulations of hydrodynamically interacting self-propelled rods 
were
performed by Saintillan and Shelley \cite{Saintillan2007,Saintillan2012},
Lushi \etal \cite{Lushi2013,Lushi2014}, and Krishnamurthy and Subramanian \cite{Krishnamurthy2015} 
based on
slender-body theory.
Unlike
squirmer suspensions, pusher rods and dumbbells show local polar 
(as well as
nematic) order and 
form
large-scale vortices, in qualitative agreement with experiments with pusher-type bacteria \cite{Dombrowski2004,Wensink2012b}.
Typically, in
these simulations 
only 
approximate
flow fields based on far-field 
hydrodynamics were implemented.
A recent simulation study 
with
active dumbbells
improved the resolution of hydrodynamic interactions between the swimmers using
the method of fluid particle dynamics
\cite{Furukawa2014}.
Motility-induced phase separation 
was observed, 
and it was shown that hydrodynamic interactions enhanced cluster formation.
Yang \etal studied the
hydrodynamics of collectively swimming flagella 
and
observed the formation of dynamic jet-like clusters of 
synchronously beating
flagella \cite{Yang2010}.
Finally, 
recent simulations 
investigated the collective motion of actively rotating disks, which tend to form crystalline structures \cite{Goto2015,Yeo2015}.

\subsubsection{External fields: gravity and traps}   \label{sec.fields}

As discussed in section~\ref{Sec:Gravity}, a suspension of non-interacting active colloids under gravity
shows an exponential sedimentation profile \cite{Palacci2010} and develops polar order against gravity \cite{Enculescu2011}.
The
presence of hydrodynamic interactions in a dilute suspension of sedimenting run-and-tumble swimmers does not 
significantly modify 
such a
sedimentation profile \cite{Nash2010}.
However, when active Brownian particles are strongly bottom-heavy, they collect in a layer while swimming against the upper
boundary of a confining cell. This layer is unstable, when hydrodynamic interactions are included, and the particles
move downwards in plumes similar to observations made in bioconvection \cite{Hennes2013,Pedley1992}.
%
%
%

\begin{figure}
\includegraphics[width=\columnwidth]{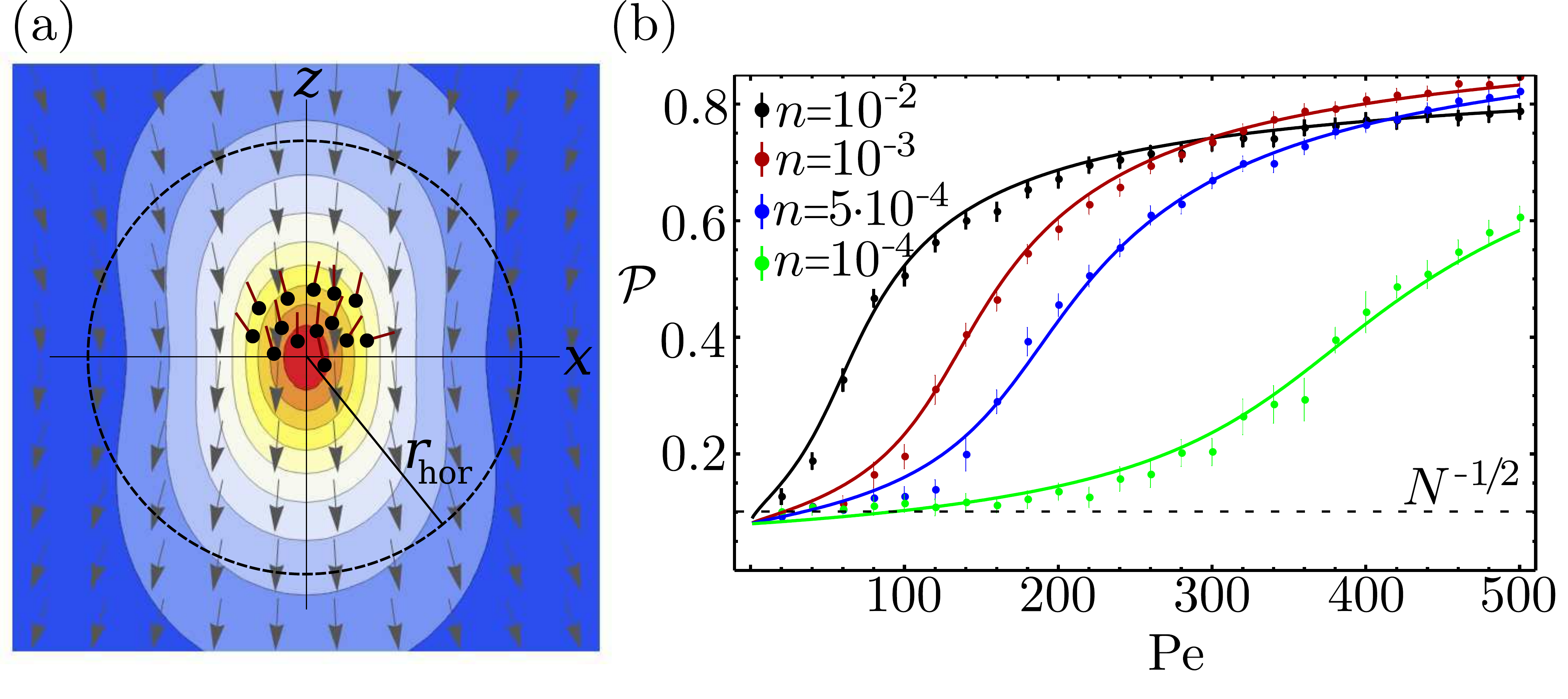}
\caption{
Pump formation of hydrodynamically interacting active particles 
(a) The formation of a macroscopic pump induces a regularised Stokeslet flow. 
(b) Global polar order parameter $\mathcal P$ versus P\'eclet number $\mathrm{Pe}$
for different volume fractions
(adapted from Ref.~\cite{Hennes2014}; copyright (2014) American Physical Society).
}
\label{Fig:Pump}
\end{figure}

A similar instability occurs for
run-and tumble swimmers \cite{Nash2010} and active Brownian particles of radius $R$ \cite{Hennes2014},
when they move in a harmonic trap potential while interacting hydrodynamically. The microswimmers
form a macroscopic pump state 
that
breaks the rotational symmetry
of the trap.
The radial force $\mathbf{F} = -k \mathbf{r}$ $(k>0)$ 
confines
the particles to a spherical shell of radius
$r_{\mathrm{hor}} = R \mathrm{Pe}/ \alpha$,
where $\alpha= k R^2/k_BT$ is the trapping P\'eclet number \cite{Hennes2014}.
When started 
with random positions and orientations,
the
active particles first accumulate at the horizon
at radial distance $r_{\mathrm{hor}}$, where they
point 
radially
outwards  \cite{Nash2010,Hennes2014}.
Due to the externally applied 
trapping
force, each particle 
initiates the
flow field
of a stokeslet,
in leading order.
However, a
uniform 
distribution
of stokeslets
on the horizon sphere, all
pointing radially outward, 
is
not stable
against small perturbations.
In particular, particles' swimming directions are rotated by nearby stokeslets. The particles move towards each other
creating denser regions, which are advected towards the center.
At sufficiently large swimmer density 
the active particles align in their own flow field and thereby generate
a macroscopically 
ordered state [see figure~\ref{Fig:Pump}(a)],
quantified by the global polar order parameter $\mathcal P$ (see also section~\ref{Sec:CorrFunctOrder}).
The coarse-grained flow field, created by the swimmers, has  the form of a regularised stokeslet
and 
pumps fluid along the polar axis to the center of the trap.
The pump formation
does not occur, if
the P\'eclet number,
the trapping strength, or the density
is too small,
as shown in figure~\ref{Fig:Pump}{(b).

Interestingly, the form of the  distribution $\Phi(\theta)$ of orientation angles $\theta$ 
measured
against the pump axis 
can be calculated analytically from the corresponding Smo\-luchowski equation.
One obtains
$\Phi(\theta) \sim \exp(A \cos\theta)$,
reminiscent of dipoles aligning in a field $A$, which is produced by the swimmers themselves. In simulations we find
$A \sim (\mathrm{Pe} \mathcal P)^{\gamma}$, with $\gamma \approx 1$ for low densities and decreasing for 
higher densities.
Thus, the
mean flow field created by 
all the
swimmers
and the associated 
polarization
$\mathcal P$ 
play
the same 
respective
role
as the 
mean
magnetic field and the magnetization 
in Weiss' theory 
of ferromagnetism
\cite{Hennes2014}.

In Ref.\ \cite{Pototsky2012} the
collective behavior of 
purely}
repulsive active particles in two-dimensional traps 
was mapped on a system of passive particles with modified trapping potential and then formulated
as a dynamical density functional theory
\cite{Pototsky2012}.
In very good agreement with the numerical solution of the corresponding Langevin equations,
one could show
that the 
radial
distribution in the trap 
including
packing effects 
strongly depends
on the P\'eclet number.

\subsubsection{External flow and rheology}
\label{Sec:Rheology}
We have discussed the response of a single microswimmer to an externally applied flow field $\mathbf{u}$ in 
section~\ref{Sec:FluidFlow}. In turn, a suspension of microswimmers is able to modify  $\mathbf{u}$ and the 
rheological properties of the fluid.

In pure Newtonian fluids the shear stress tensor $\boldsymbol{\uptau} $ is linear to the strain rate tensor,  
$\boldsymbol{\uptau} = \eta_0 [ \boldsymbol{\nabla}\otimes\mathbf{u} + (\boldsymbol{\nabla}\otimes\mathbf{u})^t ] $,
where 
$\eta_0$ is the shear viscosity.
Adding a dilute suspension of passive colloids 
with
volume fraction $\phi \ll 1$
is known to increase the \textit{effective viscosity} $\eta$. In 
leading
order 
in $\phi$
the increase is linear,
$\eta = \eta_0(1 + \frac 5 2 \phi)$, as calculated by Einstein \cite{Einstein1906,Einstein1911}.
In contrast, adding a suspension of swimming bacteria \cite{Sokolov2009,Gachelin2013,Lopez2015} or algae \cite{Rafai2010a,Mussler2013}
to water can lead to both an increase and a decrease  of the viscosity;
an effect which was first predicted in Ref.~\cite{Hatwalne2004}.
%
%
Strikingly, at high shear rates  
$\eta$ approaches zero
for swimming \textit{E.~coli} suspensions \cite{Lopez2015},
which thus becomes a \emph{superfluid}. This
was
also 
predicted before for sheared active gels \cite{Cates2008}.

While up to date no experimental studies exist 
that measure
the effective viscosity $\eta$ of active colloidal suspensions, numerical simulations 
were
performed using the squirmer model.
The rheological properties of 
squirmer suspensions 
with hydrodynamic interactions were
simulated  using Stokesian 
dynamics
\cite{Ishikawa2007b} and
lattice
Boltzmann simulations \cite{Pagonabarraga2013a}.
In Ref.~\cite{Ishikawa2007b} a semi-dilute suspension of squirmers in unbounded shear flow 
show almost the same, but slightly reduced, apparent viscosity as for a semi-dilute passive colloidal suspension.
In wall bounded shear flow, as studied in Ref.~\cite{Pagonabarraga2013a}, 
the viscosity of squimer suspensions 
depends
strongly on the applied shear rate $\dot{\gamma}$ and on the swimmer type.
While for strong $\dot{\gamma}$ the viscosity of all squirmer suspensions is comparable to the effective viscosity of passive suspensions,
it can be strongly decreased or increased at lower $\dot{\gamma}$.
Namely, for all squirmer suspensions with sufficiently small $|\beta| \lesssim 20$ the effective viscosity is increased ($\eta / \eta_0 > 1$),
but for apolar pushers ($\beta \rightarrow -\infty$) and apolar pullers ($\beta \rightarrow \infty$) the effective viscosity
is strongly decreased ($\eta / \eta_0  < 0$) and almost vanishes for very small $\dot{\gamma}$.


\subsubsection{Chemotactic active colloids}
\label{Sec:CollectiveChemo}

As discussed in section~\ref{Sec:FieldGradients}, 
an active colloid 
that experiences a
nonuniform background concentration field $X(\mathbf{r})$
moves with an additional phoretic
velocity   $\mathbf{U}_C$ and 
angular velocity $\boldsymbol{\Omega}_C$,
which both 
are linear in
the local field 
gradient
$\boldsymbol{\nabla}X(\mathbf{r})$.
Self-phoretic
colloids 
act as sinks and sources in a chemical field
and
respond to 
local 
field
gradients 
created by 
neighboring
particles.
This 
introduces an effective interaction, which can either be attractive or repulsive. In particular, active colloids can orient parallel
or against a field gradient and thereby either swim towards or away from other particles.


The collective behavior of chemotactic active particles has been studied numerically 
with particle-based models using Langevin dynamics
simulations
\cite{Taktikos2012,Palacci2013,Pohl2014,Pohl2015},
and theoretically 
by
continuum models \cite{Saha2014a,Pohl2014,Liebchen2015}.
Depending on 
whether the active colloids are chemoattractive or chemorepulsive, or
whether they tend to rotate towards or against a chemical gradient, 
qualitatively different collective behavior emerges \cite{Taktikos2012,Palacci2013,Saha2014a,Pohl2014,Pohl2015,Liebchen2015}.

To capture the basic effects, we follow Ref.~\cite{Pohl2014} and 
include diffusiophoretic attraction/repulsion and reorientation into equations~(\ref{Eq:LangevinR}) and (\ref{Eq:LangevinE}).
Then the equations of motion for $N$ interacting particles ($i=1,\dots,N$) read \cite{Pohl2014,Pohl2015}
\begin{eqnarray}
\label{Eq:PhoreticR}
\dot{\mathbf{r}}_i &= v_0\mathbf{e}_i - \zeta_{\mathrm{tr}}\boldsymbol{\nabla}c(\mathbf{r}_i)+ \sqrt{2D}\boldsymbol{\xi} \\
\label{Eq:PhoreticE}
\dot{\mathbf{e}}_i &= - \zeta_{\mathrm{rot}}(\mathbf{1} 
- 
\mathbf{e}_i \otimes \mathbf{e}_i)\boldsymbol{\nabla}c(\mathbf{r}_i)
 + \sqrt{2D_r}\boldsymbol{\xi}_r \times \mathbf{e}_i \, ,
\end{eqnarray}
where $\zeta_{\mathrm{tr}}$ is the translational diffusiophoretic parameter, which quantifies chemoattraction 
($\zeta_{\mathrm{tr}}>0$) and chemorepulsion ($\zeta_{\mathrm{tr}}<0$),
and $\zeta_{\mathrm{rot}}$ is the rotational diffusiophoretic parameter, which describes reorientation 
along
($\zeta_{\mathrm{rot}}>0$) and 
against
($\zeta_{\mathrm{rot}}<0$) chemical gradients.
Here, the chemical field $c(\mathbf{r}) $ created by the swimmers is assumed to be 
a stationary sink, which moves around with the particles.
In leading order it decays linearly with the distance from the swimmer \cite{Golestanian2007a} but is screened due to the presence 
of other particles \cite{Palacci2013,Pohl2014}.
Dipolar contributions and higher multipoles are neglected.
We discuss here the 
two-dimensional
motion of relatively dilute systems (area fraction $5\%$) similar 
to
experiments \cite{Theurkauff2012,Palacci2013}.

\begin{figure}
\includegraphics[width=\columnwidth]{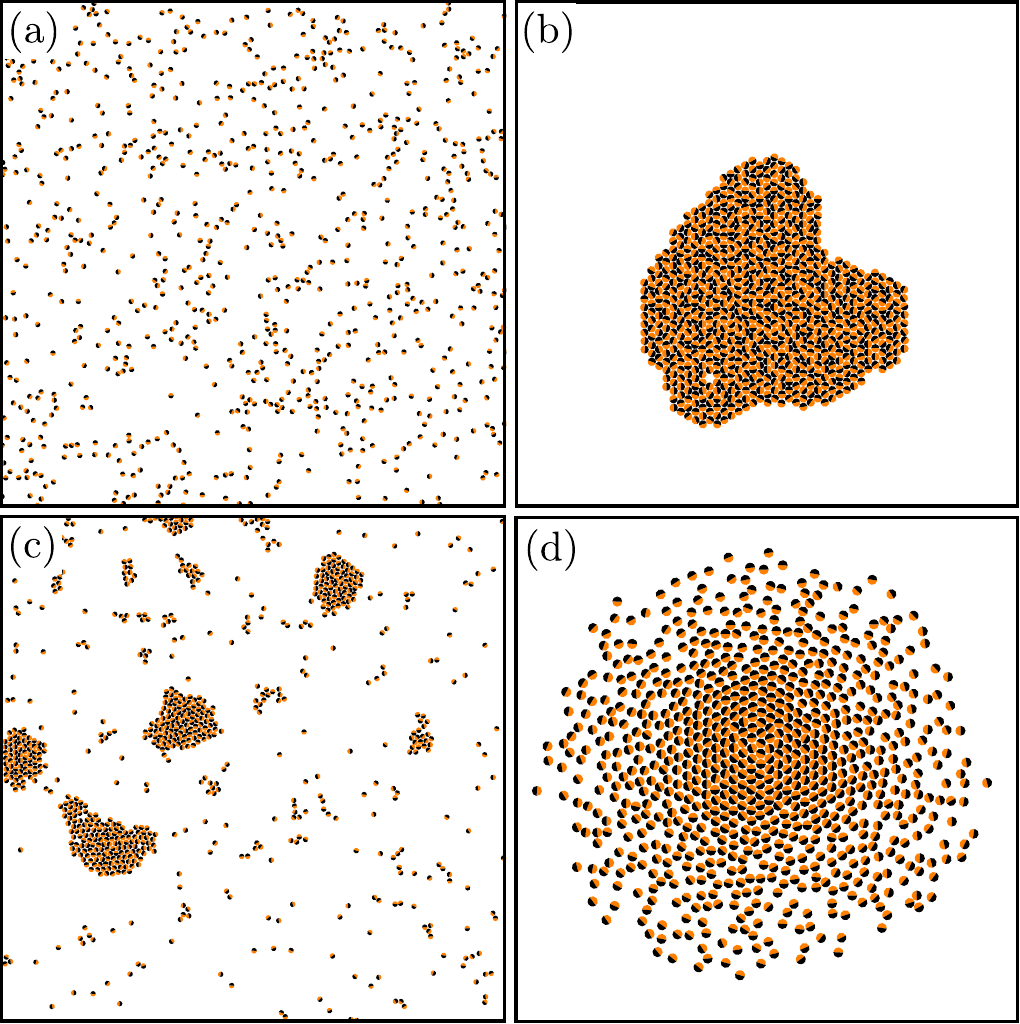}
\caption{
Collective motion of chemotactic active colloids at $5\%$ area fraction  
depending on the diffusiophoretic parameters $\zeta_{\mathrm{tr}}$ and $\zeta_{\mathrm{rot}}$.
(a) Gaslike state for small $\zeta_{\mathrm{tr}}$ and $\zeta_{\mathrm{rot}}$.
(b) Collapsed single-cluster state for sufficiently large $\zeta_{\mathrm{tr}}$.
(c) Dynamic clustering for $\zeta_{\mathrm{rot}}<0$ and sufficiently large $\zeta_{\mathrm{tr}}>0$.
(d) Core-corona state for chemorepulsion ($\zeta_{\mathrm{tr}}<0$).
[(a)-(c) adapted from Ref.~\cite{Pohl2014}; copyright (2014) American Physical Society. (d) adapted from Ref.~\cite{Pohl2015}; with kind permission of The European Physical Journal  (EPJ).]
}
\label{Fig:Pohl}
\end{figure}

First, for sufficiently small $\zeta_{\mathrm{tr}}$ and $\zeta_{\mathrm{rot}}$ the particles are in a gaslike state,
where 
they hardly cluster
[see figure~\ref{Fig:Pohl}(a)], similar to active Brownian disks which interact purely due to steric hindrance \cite{Fily2012,Redner2013b}.
Second, for sufficiently strong attraction all the particles end up in a collapsed state and 
form a single stable cluster \cite{Taktikos2012,Pohl2014,Saha2014a}
reminiscent of the chemotactic collapse obtained from the Keller-Segel model \cite{Keller1970} [see figure~\ref{Fig:Pohl}(b)].
Third, for $\zeta_{\mathrm{rot}}<0$ and sufficiently large $\zeta_{\mathrm{tr}}>0$ dynamic clustering is observed 
[see figure~\ref{Fig:Pohl}(c)]
in good agreement with experiments \cite{Theurkauff2012,Palacci2013}.
Dynamic clusters form since chemoattraction is balanced by the reorientation of the swimmers such that they try to swim
away from each other.
The cluster-size distribution
(see also section~\ref{Sec:StructProp})
 follows a power-law decay.
Fourth, 
in the opposite case,
for chemorepulsion  ($\zeta_{\mathrm{tr}}<0$) but alignment 
against
the chemical gradient  ($\zeta_{\mathrm{tr}}>0$)
an oscillating state 
can emerge,
where all particles periodically collapse and dissolve \cite{Pohl2015}.
Further decreasing $\zeta_{\mathrm{tr}}$,
the particles
ultimately
arrange in a stable core-corona state as shown in figure~\ref{Fig:Pohl}(d),
which also occurs when screening is turned off.
Oscillating states reminiscent of \textit{plasma oscillations} and aster formation  
were
also 
observed 
in
a continuum model \cite{Saha2014a}.
Different scenarios for pattern formation in chemorepulsive active colloids have recently been explored in Ref.~\cite{Liebchen2015}.

The collective behavior of a dilute suspension of 
light-heated
active colloids
interacting by \textit{thermoattraction} and \textit{thermorepulsion}
was
analyzed in Ref.~\cite{Golestanian2012a}.
While thermorepulsive colloids show a depletion zone in the center 
of a confining region,
thermoattractive colloids aggregate in the center and collapse at sufficiently high density \cite{Golestanian2012a}.
Using Brownian dynamics simulations, the formation of a dense cometlike swarm,
which consists
of interacting thermally active colloids moving towards an external light source,
was reported \cite{Cohen2014}.
Finally,
Ref.~\cite{Bickel2014a} discusses the
orientation of thermally active Janus particles in a temperature gradient, as well as the effect of P\'eclet number on aggregation and dispersion of the particles.

The collective motion or pattern formation of active particles on the liquid-air interface of a thin liquid film
was studied in Ref.\ \cite{Pototsky2014}, while passive but chemically active particles at the interface were treated 
in Ref.\ \cite{Masoud2014}.
A 2D continuum approach to couple chemical signaling and flow fields created by pushers and pullers 
was
used 
in
\cite{Lushi2012a}.

A first 
work by Thakur and Kapral includes
full hydrodynamic and phoretic interactions, which is computationally very expensive
\cite{Thakur2012}.
They studied the collective dynamics of up to 10 
dimer motors in a quasi-2D geometry in the presence of thermal noise
and observed dynamic cluster formation,
which is mainly triggered by phoretic interactions but altered by hydrodynamic interactions
 \cite{Thakur2012}.

\subsubsection{Mixtures of active and passive colloids}
Sten\-hammar \etal investigated the
structure and dynamics of a monodisperse mixture of active and passive Brownian disks 
for a broad range of P\'eclet numbers Pe, total area fractions $\phi_0$, and fraction of active particles $0 \le x_A \le 1$
\cite{Stenhammar2015b}.
They observed phase separation induced by the active particles.
However,
in contrast to motility-induced phase separation of 
purely active Brownian disks ($x_A=1$), cluster formation, cluster dynamics, and cluster break-up are much more dy\-namic for 
active-passive particle mixtures ($x_A<1$).
At sufficiently high Pe phase separation is al\-rea\-dy possible for a relatively low amount of active particles ($x_A>0.15$).
In the phase-separated state the active particles are mainly concentrated at the border of the clusters.
This
has recently been confirmed 
in experiments
already with a very small  amount ($x_A <3 \%$)
of active colloids
\cite{Kummel2015}.
In contrast, 
above the threshold density for the crystallization of passive colloids, active particles accumulate
at 
grain boundaries. When passing through crystalline domains, active particles
are able to melt them locally and create defects along their trajectory \cite{Kummel2015}.
Finally, by
performing event-driven Brownian dynamics simulations,  Ni \etal \cite{Ni2014b} 
realized that a small amount of active particles helps to crystallize
passive polydisperse glassy hard spheres in three dimensions.

Various types of 
segregation
in a mixture of passive and self-propelled hard rods moving in 2D 
were
observed by McCandlish \etal \cite{McCandlish2012}, i.e.\
swarming, coherent clustering, and transient lane formation.
In dissipative particle dynamics simulations
a mixture of passive and active particles 
showed
the emergence of turbulence, polar order, and vortical flows \cite{Hinz2014}.




\subsection{Physical concepts of active suspensions}
\label{Sec:Collective:Phys}
In sections \ref{Sec:Collective:Exp} and \ref{Sec:Collective:Simu}
we have reviewed emergent behavior of collectively moving active colloids.
In the following we will present some important concepts and measures to 
capture the structure and dynamics of interacting active particles more
quantitatively.

\subsubsection{Motility-induced phase separation}
\label{Sec:MIPSTheory}
Motility-induced phase separation of active particles can mainly be understood as
a self-trapping mechanism:
When active particles  collide, they slow down due to steric hindrance
and a cluster can form for sufficiently high persistence 
number
\cite{Cates2013,Buttinoni2013e,Cates2015}.

In theory, the
separation of purely repulsive active Brownian particles into a 
liquidlike and a gaslike phase has recently been investigated 
by mapping the system to an equilibrium system,
for which
an effective free energy functional
was constructed
\cite{Tailleur2008,Cates2013,Stenhammar2013,Bialke2013,Speck2014a,Wittkowski2014a,Speck2015a}.
Here, the velocities of the particles in the presence of neighbors are typically approximated by
a function $v(\rho)$ which decays almost linearly with density $\rho$ \cite{Fily2012,Redner2013b,Stenhammar2013,Bialke2013,Stenhammar2014},
\begin{equation}
v(\rho) = v_0\left( 1- \frac{\rho}{\rho{^\ast}} \right) \, ,
\end{equation} 
where $\rho^{\ast} \approx \rho_{\mathrm{rcp}}$ is 
the density
close to random close packing  \cite{Cates2015}.
In the limit of large P\'eclet numbers, phase
coexistence 
occurs
when the following relation holds \cite{Tailleur2008,Cates2015},
\begin{equation}
\frac{dv(\rho)}{d\rho} < -\frac{v(\rho)}{\rho}.
\end{equation}
The
 coexistence of a gaslike and a 
liquidlike
phase emerges via spinodal decomposition;
similar as in equilibrium thermodynamics,
two-phase coexistence 
with the binodal densities $\rho_1$ and $\rho_2$
arises from 
the
tangent construction 
for an
effective free energy functional \cite{Tailleur2008,Cates2013,Stenhammar2013,Bialke2013}.
Interestingly, by linearizing the underlying hydrodynamic equations \cite{Bialke2013}
an effective Cahn-Hilliard equation for active phase separation can be constructed \cite{Speck2014a}.
Wittkowski \etal 
formulated
a scaler $\phi^4$ field theory for the  order parameter field $\phi(\mathbf{r},t)$
to study phase separation and its coarsening dynamics by an \textit{Active Model B} \cite{Wittkowski2014a},
which violates detailed balance. This procedure
has recently been extended to include hydrodynamics by coupling  $\phi(\mathbf{r},t)$ to 
the Navier-Stokes equations (\textit{Active Model H} \cite{Tiribocchi2015a}).


A kinetic approach 
for determining
a condition
for motility-induced phase separation 
was formulated
by Redner \etal \cite{Redner2013,Redner2013b}.
The
incoming flux $j_{\mathrm{in}}$ of active particles from the gas phase of density $\rho_g$ 
onto an existing
cluster
is compared with the outgoing flux $j_{\mathrm{out}}$ 
from
the border of a cluster 
into the gas phase.
While the incoming flux,  $j_{\mathrm{in}} \propto \rho_g \mathrm{Pe}$, 
is proportional to
particle speed and density,
the outgoing flux, $j_{\mathrm{out}} \propto \mathrm{Pe}_r^{-1}$, depends 
on the 
inverse
persistence number
and thus
on rotational dif\-fu\-sion.
Note that 
rotational diffusion
can be strongly 
enhanced
in the presence of hydrodynamic interactions \cite{Ishikawa2008a} and 
thereby
reduce or suppress phase separation \cite{Zottl2014,MatasNavarro2014}.
The condition for phase separation and the
fraction of particles in the cluster
are determined from
the steady state condition  $j_{\mathrm{in}} = j_{\mathrm{out}}$ \cite{Redner2013,Redner2013b}.

Finally, we 
comment on how to
quantify a phase-separated state obtained from  experimental or simulation data of active particles.
A phase-separated state 
consists
of a 
high-density
(liquid- or solidlike) and a 
low-density
(gaslike) region 
and has,
by definition, 
a bimodal distribution $p(\phi_l)$ of \textit{local} densities $\phi_l$.
The local density $\phi_l^{(i)}$ of particle $i$ at position $\mathbf{r}_i$ is typically defined via the size $A_i$ of its corresponding 
Voronoi cell  \cite{Okabe2009},
$\phi_l^{(i)} = A_0 / A_i$, where $A_0= \pi R^2$ in 2D or $A_0=4\pi R^3/3$ in 3D
and $R$ is the particle radius.


\subsubsection{Active crystals and structural properties}
\label{Sec:StructProp}
To characterize if active particles are in a gas-, 
liquid- or 
crystal-like
phase,
further
structural properties 
are determined from the particle
positions $\mathbf{r}_i(t)$.

\begin{figure}
\includegraphics[width=\columnwidth]{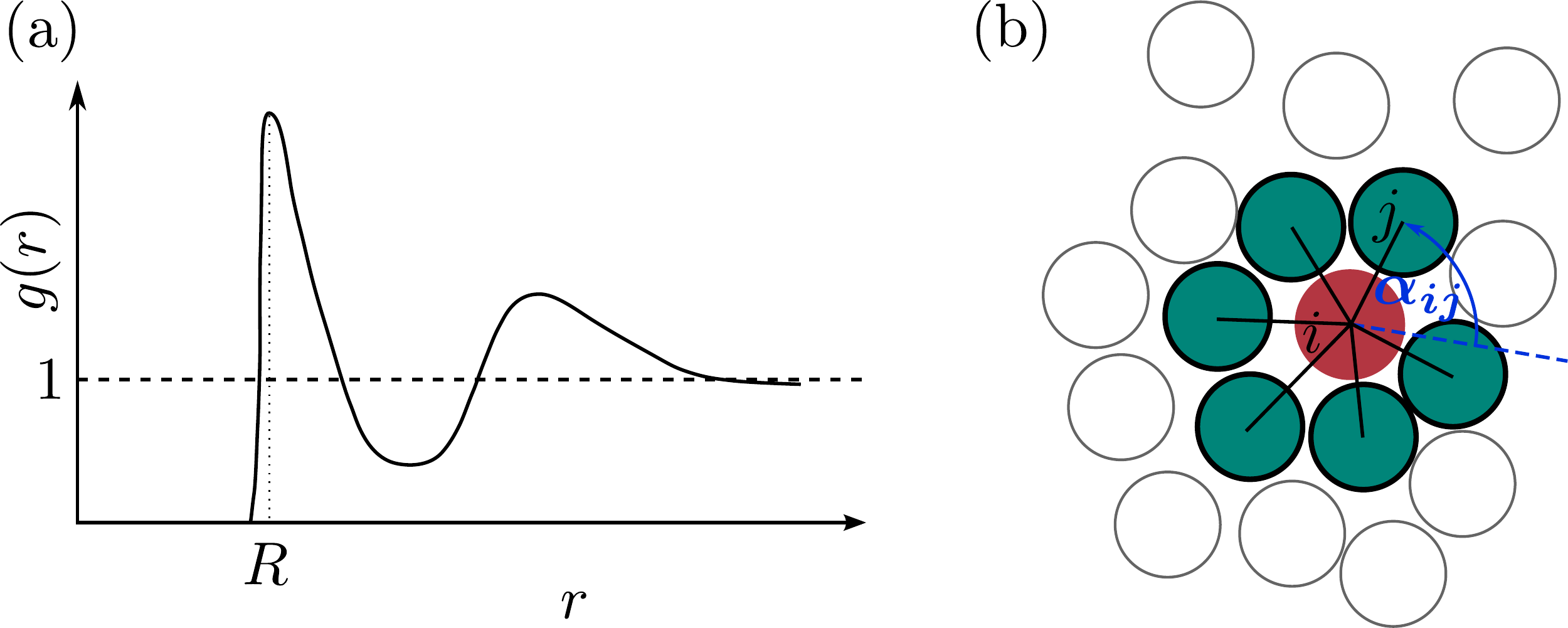}
\caption{
(a) Typical pair correlation function $g(r)$ of a liquid consisting of spherical particles of radius $R$.
(b) Sketch of local structure around particle $i$ (red), and definition of the angle $\alpha_{ij}$ from equation~(\ref{Eq:q6})
between particle $i$ and its six nearest neighbors (dark green).
The arbitrarily chosen axis is shown as a dashed blue line.
}
\label{Fig:Structure}
\end{figure}

\paragraph{Pair correlation function}
The pair correlation function (or radial distribution function) $g(r)$ measures the 
correlations of particles at distance $r$
and 
was
used to characterize the structure of collectively moving active colloids
 \cite{Ishikawa2008a,Thakur2012,Redner2013b,Buttinoni2013e,Alarcon2013a,Mognetti2013,Kummel2015}.
It is defined as the local density 
$\rho(r)$
of active particles in a spherical shell with radius $r$ around a central particle,
normalized by the mean particle density $\rho_0$,
thus $g(r)=\rho(r)/ \rho_0$.
For example, in 
two dimensions 
it can be calculated by
\begin{equation}
g(r) = 
\frac{1}{4\pi R^2 (N-1) \phi_0}  \left\langle \sum_j \sum_{i \neq j}\delta(|\mathbf{r}_i-\mathbf{r}_j|-r)      \right\rangle \, ,
\end{equation}
where $\phi_0$ and $N$ are the total area fraction and number of particles, respectively.
Peaks in $g(r)$ indicate favored interparticle distances.
While for a dilute gas no distances are favored and $g(r) \approx 1$,  
$g(r)$
shows a finite number of peaks with decreasing intensity for a liquid [see figure~\ref{Fig:Structure}(a)],
and pronounced peaks
for solids, where the actual peak positions 
indicate
the structure of the solid (e.g.,\ hexagonal).

\paragraph{Static structure factor}
The Fourier transform of the
pair correlation factor is the static structure factor $S(\mathbf{k})$, which
identifies periodic modulations
of a material 
with
wave vector $\mathbf{k}$.
It
is defined as
\begin{equation}
S(\mathbf{k}) = \frac{1}{N} \left\langle \sum_j \sum_ie^{i \mathbf{k} \cdot (\mathbf{r}_i - \mathbf{r}_j)}     \right\rangle \, ,
\end{equation}
and was
used to characterize the structure 
of
active colloidal clusters \cite{Fily2012,Theurkauff2012,Bricard2013,Redner2013b,Bialke2013,Ni2013a,Stenhammar2014}.

\paragraph{Bond orientational order}
Clusters of active colloids moving in a monolayer  
often
show 
hexagonal crystallike 
ordering,
see sections  \ref{Sec:ExpMIPS} and \ref{Sec:Collective:Simu},
and figures~\ref{Fig:BDsnap}, \ref{Fig:CollectiveSquirmer} and \ref{Fig:Pohl}.
The local 6-fold bond-orientational order   around particle $i$ is 
measured
by \cite{Steinhardt1983,Bialke2012,Redner2013,Zottl2014}
\begin{equation}
|q_6^{(i)}|^2 \quad \mathrm{with} \quad q_6^{(i)} = \frac 1 6 \sum_{j \in N_6^{(i)}} e^{i6\alpha_{ij}} \, ,
\label{Eq:q6}
\end{equation}
where $|q_6^{(i)}|^2 \in (0,1)$,  
the sum goes over the six nearest neighbors\footnote{Sometimes, it is more 
convenient to run the sum over all neighbors
within a small distance $\delta$ 
from
particle $i$ \cite{Redner2013}.}
of particle $i$, and the angle $\alpha_{ij}$ is measured between the distance vectors $\mathbf{r}_i - \mathbf{r}_j$ and 
an
arbitrarily
chosen axis [see figure~\ref{Fig:Structure}(b)].
For example, for particles in a cluster with perfect hexagonal order, $|q_6^{(i)}|^2=1$,
as 
indicated in figures~\ref{Fig:BDsnap} and \ref{Fig:CollectiveSquirmer}. 
The mean 
value,
$\langle|q_6|^2  \rangle$, 
can be used as an order parameter to measure how many particles are in a dense crystal-like cluster phase \cite{Zottl2014}.

Spatial correlations in
the 6-fold bond-orientational order
measured by
$\langle q_6^{\ast}(\mathbf{r}) q_6 (\mathbf{r}') \rangle$ \cite{Redner2013b}
and 
correlations of particle $i$ with
its six nearest neighbors,
 $\mathrm{Re} \frac 1 6 \sum_{j \in N_6^{(i)}} q_6^{(i)}q_6^{\ast (j)}  $, \cite{Lechner2008,Bialke2012}
was
used to distinguish between liquidlike, hexaticlike, and solidlike structures \cite{Bialke2012,Redner2013b}.
The global bond-orientational order parameter $|\langle q_6 \rangle|^2$ 
indicates an overall structural order in the system.
Note, however, that in general $|\langle q_6\rangle|^2$ depends on the system size.
Only for a perfect crystal, $|\langle q_6\rangle|^2=1$, 
independent of 
system size.

\paragraph{Cluster size distributions}
A cluster is typically defined by 
a
set of particles 
with a maximum distance $\epsilon$ between neighboring particles.
One characterizes the clusters
in a system of active colloids
by a cluster size distribution $p(N_c)$, where $N_c$ is the number of particles in the cluster 
\cite{Peruani2006,Theurkauff2012,Fily2012,Buttinoni2013e,Pohl2014,Fily2014,Kummel2015}.
In the gas phase $N_c=1$ (isolated particle) and $N_c=2$ (two colliding particles) 
mainly occur,
whereas 
clusters with $N_c \gg 1$ dominate in the
dense phase.
In phase-separating active colloidal suspensions the size of the largest cluster in the long-time limit 
was used
as an order parameter to study the onset of phase separation \cite{Buttinoni2013e}.

\subsubsection{Giant number fluctuations}
In thermal equilibrium
the number of particles $N$ in a volume $V$ in a grand canonical ensemble fluctuates with standard deviation 
$\Delta N = \sqrt{\langle N^2 \rangle - \langle N \rangle ^2} \sim \sqrt{N}$ 
\cite{Reichl2016}.
In contrast, self-propelled particle systems 
that
are 
out of equilibrium  often show \textit{giant number fluctuations} 
indicated by
$\Delta N \sim N^{\alpha}$ with $0.5 < \alpha \le 1$.
This 
was
predicted 
\cite{AditiSimha2002,Ramaswamy2003,Toner2005,Ramaswamy2010} and first demonstrated 
in experiments
with vibrated granular rods \cite{Narayan2007}
(see also the discussion in Ref.~\cite{Aranson2008a}).
Giant number fluctuations 
were also
reported in 
dynamic clustering of
active colloids \cite{Palacci2013}
and 
in simulations of
active Brownian particles \cite{Henkes2011a,Fily2012,Wensink2012}.
However,
electrostatic and hydrodynamic interactions suppress giant number fluctuations 
in  dense 
suspensions of polarly ordered Quincke rollers
\cite{Bricard2013}.


\subsubsection{Correlation functions and order parameters}
\label{Sec:CorrFunctOrder}
So far, in section \ref{Sec:Collective:Phys}, we have described quantitative measures 
using 
particle positions $\mathbf{r}_i$.
In the following, we 
consider
particle orientations $\mathbf{e}_i$ and velocities $\mathbf{v}_i$  to 
characterize
orientational order in the system.

\paragraph{Polar order}
To quantify 
global polar order in a suspension of active colloids, the order parameter 
\begin{equation}
P(t) 
= \left|\langle \mathbf{e}(t) \rangle\right| = \frac 1 N \left |\sum_i \mathbf{e}_i(t) \right |
\label{Eq:Polar}
\end{equation}
was
used
in squirmer suspensions  \cite{Ishikawa2008a,Evans2011,Alarcon2013a,Delmotte2015} or
for
active particles in a harmonic trap \cite{Hennes2014}.
In all these cases the particles are spherical and the aligning mechanisms 
have a pure hydrodynamic origin.
In the long-time limit $P(t)$ typically reaches a steady state 
with
$\mathcal{P} = \mathrm{lim}_{t \rightarrow \infty} P(t)$.
For fully developed polar order $\mathcal P=1$. 
However, even
an ensemble of randomly oriented particles 
shows a residual value for the order parameter:
$\mathcal P\approx N^{-1/2}$
\cite{Ishikawa2008a}.
Interestingly, 
as reviewed in section\ \ref{sec.fields}
and demonstrated for collectively moving Quincke rollers,
the 
formation
of polar order at a critical density or a critical P\'eclet number 
can be
mapped to the onset of
ferromagnetic order treated
within mean-field theory 
\cite{Bricard2013,Hennes2014}.
In the continuum limit the
polarisation
becomes
$\mathcal{P} = |\int_{\mathbb{S}_2} \Phi(\mathbf{e})\mathbf{e} dS|$ \cite{Bricard2013,Hennes2014},
where $dS$ is the surface element on the unit sphere $\mathbb{S}_2$, and $\Phi(\mathbf{e})$ the
orientational
distribution 
function
in 
steady state.

To measure the temporal correlations of the 
orientation vector
$\mathbf{e}$, the autocorrelation function 
$C_p=\langle \mathbf{e}(t) \cdot \mathbf{e}(0) \rangle$
from equation\ (\ref{Eq:Cep})
for 
a single particle 
is calculated and
then averaged over all particles, 
$\overline{C}_p = \overline{\langle \mathbf{e}(t) \cdot \mathbf{e}(0) \rangle}$.
If the decay of 
$\overline{C}_p$ 
is exponential, 
$\overline{C}_p = e^{-2D_r^{\mathrm{eff}}t}$,
$D_r^{\mathrm{eff}}$ defines an effective rotational diffusion constant.
For example,
due to hydrodynamic 
or phoretic
interactions
it can have a larger value than the coefficient
$D_r$ of an isolated swimmer \cite{Ishikawa2008a,Thakur2012,Zottl2014}.
Hence, $D_r^{\mathrm{eff}}$ 
quantifies
the
orientational persistence of interacting microswimmers.

Spatial correlations    
of particle orientations 
at
distance $r$ 
and
at 
time $t$  
are
described with the equal-time
polar pair correlation function \cite{Saintillan2007}
\begin{equation}
C_e(r,t) = \frac{\langle  \sum_{i \neq j}(\mathbf{e}_i(t) \cdot \mathbf{e}_j(t) \rangle^2)\delta(|\mathbf{r}_i-\mathbf{r}_j|-r)      \rangle}
         {\langle  \sum_{i \neq j}\delta(|\mathbf{r}_i-\mathbf{r}_j|-r)      \rangle}.
\label{Eq:Ce}
\end{equation}
The
average
is taken
over all 
pairs of swimmers $i,j$.
Note that the maximum distance $r$ in equation~(\ref{Eq:Ce}) is set by the systems size of the experiment or simulation.
In experiments it is often more convenient to measure 
particle
velocities $\mathbf{v}_i$ instead of 
their
orientations $\mathbf{e}_i$. 
Then, the spatial correlation function $C_v(r,t)$ is defined as $C_e(r,t)$
in equation~(\ref{Eq:Ce})
but with orientation vectors replaced by velocities.

From the
decay of the 
time-averaged
correlation functions 
$C_e(r) = 
\frac 1 T
 \int_0^TC_e(r,t) dt$ and  $C_v(r) = \frac 1 T \int_0^TC_v(r,t) dt$,
one can determine
correlation lengths $l_c$ 
for
orientations and velocities in the system
 \cite{Dombrowski2004,Saintillan2007,Ishikawa2008a,Ishikawa2008,Underhill2008,Hernandez-Ortiz2009,Thutupalli2011,Cisneros2011}.
Anticorrelations ($C_e<0$ or $C_v<0$) 
indicate
the formation of vortices, 
for example, 
in \textit{turbulent} bacterial suspensions \cite{Dombrowski2004,Dunkel2013b}.

To characterize the mobility of collectively moving active colloids, the dynamic order parameters
\begin{equation}
V_1(t) = \left|\langle \mathbf{v}(t) \rangle\right| = \frac 1 N \left |\sum_i \mathbf{v}_i(t) \right |
\label{Eq:V1}
\end{equation}
and 
\begin{equation}
V_2(t) = \langle \mathbf{v}(t) \cdot \mathbf{e}(t) \rangle = \frac 1 N  \sum_i \mathbf{v}_i(t) \cdot \mathbf{e}_i(t)
\label{Eq:V2}
\end{equation}
were
used.
While $V_1$ simply measures the mean speed of 
the
particles (see, e.g., \cite{Ishikawa2007a,MatasNavarro2014}),
 $V_2$ determines how fast they move along their 
intrinsic
directions $\mathbf{e}_i$ (see, e.g., \cite{Thakur2012,Zottl2014,MatasNavarro2014}). 
Possible deviations
of $V_2$ from the swimming velocity $v_0$ 
always indicate
colloidal interactions.

We note that 
equal-time correlations functions $C_u(r,t)$ and $C_{\omega}(r,t)$ are formulated for the
hydrodynamic flow
field $\mathbf{u}(\mathbf{r},t)$ and vorticity field $\boldsymbol{\omega}(\mathbf{r},t) 
= \nabla \times \mathbf{u}(\mathbf{r},t)$
created by the motion of
microswimmers
(see, e.g., \cite{Cisneros2010}).
Particularly interesting 
are
the Fourier
transforms
of the time-averaged correlation 
functions, $\int C_u(r,t) dt$ and $\int C_v(r,t) dt$,
which 
are 
used to define
 the energy spectrum $E(k)$ in 
an
  active fluid.
It 
was applied
to characterize \textit{active turbulence} in bacterial \cite{Wensink2012b} and active colloidal \cite{Nishiguchi2015} suspensions.
An order parameter for the strength of 
vorticity
in the fluid is the so-called \textit{enstrophy}, which is the spatial average of $\boldsymbol{\omega}^2$.


\subsubsection{Mean square displacement}
In order to quantify the 
temporal evolution of $N$ interacting active colloids by means of their trajectories $\mathbf{r}_i(t)$,
the particle-averaged mean square displacement 
\begin{equation}
\langle \Delta r^2(t) \rangle = \frac 1 N \sum_{i=1}^N |\mathbf{r}_i(t) - \mathbf{r}_i(0)|^2 \rangle
\label{Eq:MSDCollective}
\end{equation}
is
used \cite{Thakur2012,Fily2012,Redner2013b}.
Recall that for a single particle [equation~(\ref{Eq:MSDS1})] 
the motion  is ballistic ($\sim t^2$) at small times and diffusive ($\sim t$) at large times.
Collective
motion can introduce an 
intermediate superdiffusive regime $\sim t^{3/2}$ \cite{Thakur2012,Redner2013b}
and interactions between
active particles typically 
reduce
the long-time diffusion constant extracted from equation~(\ref{Eq:MSDCollective}) compared to
free swimmers.

\subsubsection{Active thermodynamics}

In \textit{equilibrium} 
thermodynamics macroscopic 
state variables such as temperature $T$ and pressure $p$ are well-defined quantities.
The values 
calculated from
their microscopic definitions coincide with the values obtained
by introducing them as first derivates of the
thermodynamic potentials.
Temperature and pressure are connected via an equation of state, for example, for an ideal gas 
one has
$pV=nRT$.

In \textit{nonequilibrium} thermodynamics it is in general not possible to define state variables such as $T$ and $p$ in a meaningful way.
Nevertheless, very recently quite a lot of effort has been made in 
formulating
a thermodynamic description of active matter systems
(see, e.g.,\ the discussions in \cite{Cates2015,Takatori2015,Marconi2015,Solon2015b,Ginot2015,Marchetti2015}).
For example, as already discussed in section\ \ref{Sec:MIPSTheory}, motility-induced phase separation can sometimes be 
addressed with
an effective free energy \cite{Cates2015}
 and the activity-induced collisions of the particles mapped to an effective attraction potential \cite{Farage2015}.
Another example is the controverse discussion about the concept of effective temperature in active particle systems
 \cite{Loi2008,Palacci2010,Wang2011,Szamel2014,Levis2015,Ginot2015}.
Indeed, sometimes 
one can assign an effective temperature to
the stochastic motion of active colloids.
Examples are the persistent random walk in bulk (see section~\ref{Sec:Walk}) or 
the sedimentation profile of non-interacting active particles in a gravitational field
(see section\ \ref{Sec:Gravity}).
In general, however, 
it is not possible to treat
the stochastic dynamics of microswimmers
with an effective temperature.

\paragraph{Active pressure}
Very recently some research groups 
have
started to work on defining 
pressure in active particle suspensions
\cite{Mallory2014,Fily2014b,Yang2014e,Takatori2014,Takatori2015,Fily2015,Ni2015,Solon2015,Solon2015b,Winkler2015,Yan2015,Ezhilan2015b,Bialke2015,Maggi2015,Speck2015b,Marchetti2015}
by using various definitions for pressure.
First, 
one views
pressure 
as 
the mechanical force,
which
an active particle system in a container
exerts per unit area
on bounding surfaces.
Second, 
in bulk 
it should be
the trace of the hydrodynamic stress tensor.
Third, pressure can be calculated as the derivative of a free energy with respect to volume.
Since,
in general,
a free energy does not exist 
for active particle systems, only the first two, microscopic definitions of pressure are usually accessible.

However, although all pressure definitions coincide in thermal equilibrium, they do not for active particle systems out of equilibrium.
Based on work from Refs.~\cite{Takatori2014,Yang2014,Mallory2014},
Solon \etal showed \cite{Solon2015} for 
interacting active Brownian disks that the mechanical pressure 
acting
on 
the
walls 
of
a container 
contains
the
sum of three terms 
including the \textit{active pressure} (also called \textit{swim pressure})  \cite{Takatori2014,Yang2014,Bialke2015,Solon2015}
\begin{equation}
p_A = \frac{\gamma v_0}{2V} \sum_i \langle \mathbf{r}_i \cdot \mathbf{e}_i \rangle \, ,
\end{equation}
where $v_0$ and $\gamma$ are particle speed and friction constant, 
respectively.
$V$ 
is
the 
area
(2D) or volume (3D)
of the simulation box
and the average $\langle \ldots \rangle$ is taken over time.
This expression
is independent of the swimmer-wall interaction potential and hence a state function \cite{Solon2015}.
However, in general,
pressure is not a state function
as demonstrated,
for example,
for active particles, which transfer torques to 
bounding
walls during collision \cite{Solon2015b}.

\section{Outlook and Perspectives}
\label{Sec:Outlook}

Studying the motion of self-propelled colloids is a relatively new field in physics
and many aspects have to be explored in more detail in the future.
As presented in section~\ref{Sec:HowSwim}, active colloids with fixed shape are able to move force-free by 
generating
fluid flow close to their
surfaces.
Nevertheless, even for the motion of a single active colloid all the physical and chemical mechanisms,
which lead to 
near-surface 
flow,
have not been
fully understood
yet.
Therefore, we expect that both in experiment and theory more studies on the locomotion mechanisms of
self-phoretic and Marangoni-flow driven active colloids will follow within the next years.
Progress has been made in understanding the motion of a single active colloid in flow,
close to surfaces, in field gradients, and under gravity, as discussed in section~\ref{Sec:Single}.
To
test the theoretical predictions, more experiments and simulations will be necessary 
in order
to gain a thorough understanding of the combination of self-propulsion and the aforementioned external stimuli.

A particular interest lies in understanding the emergent behavior in collectively moving active colloids,
as demonstrated in section~\ref{Sec:Collective}.
In contrast to biological microswimmers active colloids have a much simpler shape.
Therefore, they 
constitute
relatively simple model systems 
for studying
nonequilibrium features 
in the collective motion
of active particle systems.
For exampe, although
recent work 
has
shed 
some
light 
on
the mechanisms behind 
phase separation and dynamic clustering of active colloids,
all the  
principles for explaining
these observations
have not been
understood so far.
Using
modern microfluidic tools, we expect that novel experiments on the col\-lec\-tive behavior of active colloids in microchannels will 
be performed in the near future.
This includes
com\-plex microfluidic topologies or 
the controlled application of
fluid flow and external forces.

Novel simulations will be necessary to further explore the 
large
variety of collective behavior in active colloids, and novel theoretical approaches will help to understand the generic mechanisms 
behind 
it.
For example, detailed simulations of a large number of collectively moving active 
colloids,
which 
fully
resolve both 
hydrodynamic and phoretic interactions,
are not available at present.
We expect large-scale computer simulations to enable a detailed modeling
of such systems in the near future.
From 
a
theoretical point of view,  a general thermodynamic description of active matter systems is not 
available.
It
has to be explored, under which conditions such a description 
exists
and how to formulate it. 
For example, besides identifying an appropriate active pressure in a system
of active particles, one may also try to define surface tension solely 
induced by the activity of microswimmers.

All in all, understanding the generic properties of active systems and the collective motion of active colloidal suspensions 
poses fundamental and challenging questions of nonequilibrium statistical physics. As biomimetic systems they may also help to 
better understand biological processes such as chemotaxis or the aggregation of biological active matter.
Novel materials formed by active self-assembly may be constructed in the near future 
based on active microscopic constituents such as 
self-propelled colloids and thereby a new branch of material science is opened.
Finally, active colloids stimulate the imagination towards active cargo transport either in technological processes or in 
medical applications. Active colloidal systems are fascinating. They have already 
pathed the way to
new fields in different directions and 
there is much more to come.

\section*{Acknowledgments}
\label{Sec:Ack}
\addcontentsline{toc}{section}{\nameref{Sec:Ack}}
We thank 
Denis Bartolo,
Clemens Bechinger, C\'ecile Cottin-Bizonne, J\'er\'emie Palacci, and Shashi Thutupalli
for providing the images presented in figure~\ref{Fig:Experiments}.
We acknowledge funding from the Deutsche Forschungsgemeinschaft (DFG) within the 
project STA 352/10-1 and the priority program SPP 1726 "Microswimmers - From Single Particle Motion to Collective 
Behaviour" (STA 352/11).

\section*{References}
\label{Sec:Refs}
\addcontentsline{toc}{section}{\nameref{Sec:Refs}}

\end{document}